\documentclass[11pt,letterpaper]{article}
\usepackage{jheppub}

\usepackage{graphicx}
\graphicspath{{figures/}}
\usepackage{bbm}
\usepackage{amsmath}
\usepackage{amssymb}
\usepackage{mathtools}
\usepackage{amsthm}
\usepackage{mathrsfs}
\usepackage{marvosym}
\usepackage{dsfont}
\usepackage[labelformat=simple]{subcaption}
\usepackage{xcolor}
\usepackage{braket}
\usepackage{cleveref}
\usepackage{comment}
\usepackage{cancel}
\usepackage{float}
\usepackage{fancybox}
\usepackage{soul}
\usepackage[skins,theorems]{tcolorbox}
\tcbset{highlight math style={enhanced,
		colframe=black,colback=white,arc=0pt,boxrule=1pt}}

\allowdisplaybreaks

\definecolor{dark-gray}{gray}{0.20}
\definecolor{gray}{gray}{0.30}
\definecolor{light-gray}{gray}{0.80}
\definecolor{dark-red}{rgb}{0.7,0,0}
\definecolor{dark-green}{rgb}{0.1,0.4,0}
\definecolor{dark-blue}{rgb}{0.3,0.3,0.7}
\definecolor{light-blue}{rgb}{0.8,0.8,1}
\definecolor{swamp}{RGB}{240, 199, 197}

\newcommand{\be}{\begin{equation}}
	\newcommand{\ee}{\end{equation}}

\def\be{\begin{equation}}
	\def\ee{\end{equation}}
\def\bea{\begin{eqnarray}}
	\def\eea{\end{eqnarray}}

\newcommand{\beq}{\begin{equation}}  \newcommand{\eeq}{\end{equation}}
\newcommand{\bal}{\begin{aligned}}   \newcommand{\eal}{\end{aligned}}
\def\beqa{\begin{eqnarray}}
	\def\eeqa{\end{eqnarray}}

\newcommand{\KK}{\text{KK}}

\newcommand{\Spin}{\mathrm{Spin}}

\newcommand{\Tr}{\mathrm{Tr}}

\newcommand{\Lsp}{\ensuremath{\Lambda_\text{sp}}}
\newcommand{\Nsp}{\ensuremath{N_\text{sp}}}

\newcommand{\Mpld}{\ensuremath{M_{\text{Pl,}\, d}}}

\newcommand{\mKK}{\ensuremath{m_\text{KK}}}
\newcommand{\RKK}{\ensuremath{R_\text{KK}}}

\newcommand{\MBH}{\ensuremath{M_\text{BH}}}
\newcommand{\RBH}{\ensuremath{R_\text{BH}}}
\newcommand{\SBH}{\ensuremath{S_\text{BH}}}

\numberwithin{equation}{section}

\def\simleq{\; \raise0.3ex\hbox{$<$\kern-0.75em
		\raise-1.1ex\hbox{$\sim$}}\; }
\def\simgeq{\; \raise0.3ex\hbox{$>$\kern-0.75em
		\raise-1.1ex\hbox{$\sim$}}\; }

\newcommand{\figref}[1]{figure \ref{#1}}

\numberwithin{equation}{section}

\hypersetup{
	colorlinks=true,
	linkcolor=dark-blue,
	citecolor=dark-red,
	urlcolor=dark-green,
	linktoc=page
}

\theoremstyle{remark}

\newtheoremstyle{named}{}{}{\itshape}{}{\bfseries}{.}{.5em}{#3}
\theoremstyle{named}

\title{\centering Black Hole-Tower Correspondence:\\ Species backreaction in minimal black hole limits}

\author{Miquel Aparici$^{1,2,3}$,}
\author{Alvaro Herráez$^{3}$,}
\author{Joaquin Masias$^{3}$,}

\affiliation{$^{1}$Instituto de F\'{i}sica Te\'{o}rica IFT-UAM/CSIC,
C/ Nicol\'{a}s Cabrera 13-15, Campus de Cantoblanco, 28049 Madrid, Spain}
\affiliation{$^2$Departamento de F\'{i}sica Te\'{o}rica, Universidad Aut\'{o}noma de Madrid, Cantoblanco, 28049 Madrid, Spain}
\affiliation{$^3$Max-Planck-Institut f\"ur Physik, Boltzmannstrasse 8, 85748 Garching bei M\"unchen, Germany}

\emailAdd{miquel.aparici@estudiante.uam.es, aherraez@mpp.mpg.de, jmasias@mpp.mpg.de}

\abstract{The Black Hole-String Correspondence identifies the microstates of a minimal black hole with those of a highly excited string. In decompactification limits, where the light states are Kaluza--Klein modes, its proposed counterpart is the Black Hole-Tower Correspondence at the species scale. We extend it in three directions. First, we expand the free thermodynamic analysis to include towers with string-oscillators and Kaluza--Klein modes, and clarify whether the ensemble connects to a $d$-dimensional black hole or a wrapped black string. Second, we include gravitational backreaction in the pure Kaluza--Klein tower transition. With no winding-tachyon analogue, the interpolating configuration is the one-loop Kaluza--Klein gas, which reorganizes into higher-dimensional radiation and yields self-gravitating solutions whose instability points to the black string and whose thermodynamics match it at the correspondence point, $S\sim \Nsp$, where the size approaches the species length. This shows that the matching is robust under self-interactions. Third, we ask whether the two phases are continuously connected, revisiting and extending the worldsheet and topological analyses of the Black Hole-String case to include also Kaluza--Klein towers. Linear sigma model families interpolate smoothly in the heterotic string, while in type II the transition is obstructed. The cobordism classes of the two saddles agree in every structure we check, but their brane-charge lattices, given by bordism groups, differ, which is the source of the type II obstruction. This obstruction is thus perturbative, showing that the transition can only proceed through the non-perturbative, charge-violating processes of the kind predicted by the Cobordism Conjecture.}
\begin{document}
	\makeatletter
	\let\old@fpheader\@fpheader
	\renewcommand{\@fpheader}{  \vspace*{-0.1cm} \hfill IFT-UAM/CSIC-26-120\\ \vspace*{-0.1cm} \hfill MPP-2026-125}
	\makeatother
	
	\maketitle
	\setcounter{page}{1}
	\pagenumbering{roman}

	\hypersetup{
		pdftitle={},
		pdfauthor={},
		pdfsubject={}
	}	
	
\newcommand{\remove}[1]{\textcolor{red}{\sout{#1}}}

\newpage
\pagenumbering{arabic} 

\section{Introduction and Summary}\label{s:Intro}
The microscopic origin of black hole entropy is an open problem that must be addressed in any theory of quantum gravity. In the context of string theory, this question has been partially answered. For example, with extended $\mathcal{N}>1$ supersymmetry one can realize black hole microstates as bound states of wrapped D$p$-branes \cite{Strominger:1996sh}. Another partial answer in string theory comes from the celebrated Black Hole-String Correspondence \cite{Horowitz:1991cd, Susskind:1993ws, Horowitz:1996nw}, which identifies, at weak string coupling, the microstates of a black hole with those of a free, highly excited string. On one side of the correspondence one has a large black hole which shrinks as the string coupling is adiabatically reduced, until its horizon reaches the string scale. At this point, it transitions into a dense ball of highly excited strings. As $g_s$ approaches zero one then obtains a free string. The entropy, the temperature and the mass match parametrically on both sides of the transition, offering a possible explanation for black hole microstates, in terms of a weakly coupled microscopic description where the transition occurs. The intermediate regime, where gravitational self-interactions of the string become important, was also studied by Horowitz and Polchinski \cite{Horowitz:1996nw, Horowitz:1997jc}, who found a classical Euclidean saddle, the Horowitz-Polchinski (HP) solution, which interpolates between the free string and the black hole. This solution is constructed in the Euclidean time formalism, where the inverse temperature $\beta=1/T$ is given by the size of the Wick-rotated time direction, compactified on a circle. The string spectrum then contains a winding mode that develops a tachyonic instability as one approaches the Hagedorn temperature $T \simeq T_H \simeq M_s$, present due to the exponential degeneracy of the string excitation modes. The black hole-string transition is thus expected to be understood as the condensation of this tachyonic winding mode. Recent works have consolidated the picture of a Black Hole-String transition by combining different perspectives, as well as extended it to more general settings \cite{Chen:2021dsw,Susskind:2021nqs,Brustein:2021cza,Bedroya:2022twb,Balthazar:2022szl,Balthazar:2022hno,Halder:2023kza,Urbach:2023npi,Emparan:2024mbp,Ceplak:2024dxm,Chu:2024ggi,Bedroya:2024igb,Chu:2025fko,Chu:2025kzl,Seitz:2025wpc,Urbach:2026qph, Emparan:2026wlm}.

The Black Hole-String Correspondence is, however, specific to the limit in moduli space of a weakly coupled string becoming light. A natural question is whether an analogous correspondence holds in other limits of the string moduli space, where an infinite tower of states is also expected to become light according to the Distance Conjecture \cite{Ooguri:2006in}. The Emergent String Conjecture \cite{Lee:2019wij} states that any infinite distance limit in moduli space corresponds either to a weakly-coupled critical string becoming tensionless, or to a decompactification limit, where a compact direction grows to infinite size (in the right duality frame). In decompactification limits, the relevant tower of light states is not a tower of string oscillators but a tower of $\rm{KK}$ modes. The species scale $\Lsp = \Mpld / \Nsp^{1/(d-2)}$ \cite{Dvali:2007hz, Dvali:2007wp, Dvali:2009ks}, which plays the role of UV cutoff, is then given by the higher dimensional Planck mass, whereas in weakly-coupled string limits it is given by the string scale, and the minimal black hole has entropy $\SBH \sim \Nsp$. This suggests that a \emph{Black Hole-Tower Correspondence} should exist, as a generalization of the Black Hole-String Correspondence to other infinite distance limits in moduli space, in which the microstates of the black hole are accounted for by the degrees of freedom of the corresponding tower of light modes predicted by the Distance Conjecture. Such a correspondence was proposed and studied at the level of free species in \cite{Herraez:2024kux} (see also \cite{Cribiori:2023ffn, Basile:2023blg} for related ideas, and \cite{Herraez:2025clp} for a review), where it was shown that the thermodynamic quantities of the tower, namely entropy, energy and temperature, match those of the minimal black hole at the correspondence point $T \sim \Lsp$, $S \sim \Nsp$, both for towers formed by string oscillators and for towers formed by KK-modes. 

In this paper we extend this correspondence in several ways. First, we extend the free analysis including multiple towers of different kinds at the same time, showing how the pure KK and the pure string-oscillator limits are extreme limits of the combined analysis. We compute the combined partition function of the thermal ensemble containing string oscillators and KK modes at fixed temperature, and determine the parameters that controls which tower dominates the thermodynamics. The two relevant parameters are the deviation from the Hagedorn temperature, $\epsilon \equiv (\beta-\beta_H)/\beta_H$, and the number of KK modes below the temperature, $N_T=T/\mKK$. The string-oscillator contribution dominates the entropy in the near-Hagedorn regime, $\epsilon \ll 1$, and the KK gas only takes over when $\epsilon$ reaches $\mathcal{O}(1)$, independently of $N_T$. A finer threshold when $\epsilon\sim N_T^{-2}$ determines whether the thermal winding mode is able to resolve the KK circle. For $\epsilon\ll N_T^{-2}$ it does not, and the $d$-dimensional description is recovered, with the ensemble connecting the $d$-dimensional HP/free string with the black hole. In contrast, for $ N_T^{-2}\ll \epsilon\ll 1$ the circle is resolved, and the relevant saddles are the HP/free string on the KK circle and the black string wrapping the KK cricle (which from the $d$-dimensional perspective still looks like a black hole).
Whether the near-Hagedorn regime is reached before strong gravitational effects become important, and thus whether a small black hole transitions first into the HP/free string ball (resolving the circle or not) or directly into the gas of KK modes depends on the ratio $M_s/M_{\text{Pl},D}$

Second, we extend the Black Hole-KK tower correspondence beyond the free species approximation, analogously to what the HP solution represents compared to the free-string case, by including the gravitational backreaction of the $\rm{KK}$ gas. We find that the tachyon condensation perspective for the HP solution has no analogue for a pure $\rm{KK}$ tower, which is to be expected from the fact that there is no Hagedorn transition for a decompactification limit. This can also be seen more concretely, since the scaling symmetry underlying the HP solution is absent for KK modes, and a classical saddle with non-vanishing KK mode profiles would break uniformity along the compact direction, which is in tension with the entropy of a uniform black string. Instead, we find that the tree level contribution of the $\rm{KK}$ tower to the on-shell action vanishes, as expected, and we compute its one loop energy-momentum tensor. We show that the effective action for a tower of $\rm{KK}$ modes on a curved background $\mathbb{R}^{d} \times S^1_\beta$ essentially reorganizes as a $(d+1)$-dimensional radiation gas, making the higher dimensional origin of the tower manifest.  We then focus on the $d+1=5$ case as a representative example, and analyze it in detail. We solve the resulting backreacted Einstein equations for a static spherically symmetric ansatz with $S^2 \times S^1$ symmetry and equation of state $p = \rho/4$ corresponding to five-dimensional radiation. The solution exhibits a two-branch structure parameterized by the central density $\rho_c$, with the temperature reaching a maximum at finite $\rho_c$. The thermodynamically preferred branch then matches the black string thermodynamics at the correspondence point, where $S \sim \Nsp$. In the weak-backreaction limit we show analytically that gravitational corrections to the entropy are under control for $T \ll \Lsp$, confirming the self-consistency of the free species analysis. We confront this 5d analysis with the study of the same system from the 4d theory, where the tower gives rise to an exotic equation of state, and find parametric agreement with the 5d analysis in the common regime of validity. We also argue that as a consequence of the fact that gravitational backreaction becomes important near $T\sim \Lsp$ (which coincides with $M_{\text{Pl},D}$ for the pure KK-tower), when $M_s \ll M_{\text{Pl},D}$ the $\rm{KK}$ gas transitions into a string-dominated regime before entering the strongly coupled gravitational regime, consistent with the analysis of \cite{Herraez:2024kux, Emparan:2024mbp}.

Third, we address the question of whether the two phases are continuously connected, since the matching of thermodynamic quantities at the correspondence point does not suffice to conclude that the tower ensemble and the black hole/black string can be smoothly deformed into each other. For the black hole-string transition this question was examined in \cite{Chen:2021dsw} using two complementary perspectives, which we review and extend to the black hole-tower case, using again the addition of an extra KK circle as a proof of concept. On the worldsheet, we explicitly construct linear sigma model families that describe the two Euclidean saddle topologies corresponding to the black hole and the HP/free string side, including the extra KK circle. In the heterotic string case the family interpolates smoothly, extending the result of \cite{Chen:2021dsw}, whereas in the type II case the obstruction found there remains. From the spacetime perspective, we compare the topology of both backgrounds at two different levels. On the one hand, their cobordism classes agree in every structure we check, meaning there is no topological obstruction for the transition. On the other hand, their brane charge lattices, captured by the bordism groups of each background, differ by two extra towers of charges on the black hole side, associated with branes wrapping the horizon and the cigar-like geometry. Therefore, the type II transition is perturbatively obstructed, but consistent with a non-perturbative, charge-violating process of the type predicted by the Cobordism Conjecture \cite{McNamara:2019rup}. Both angles, namely the linear sigma model description and the spacetime bordism characterization also fit in a coherent picture, where the index jump is accounted for by the appearance of extra cycles supporting the extra brane charges. 

The structure of this paper is as follows. In section \ref{s:blackholetower} we review the Black Hole-Tower Correspondence and the thermodynamics of free towers, including the combined analysis of mixed KK and string-oscillator towers, and also review the Horowitz--Polchinski solution, explaining why its KK-backreacted counterpart must be qualitatively different. In section \ref{sec:notfree.species} we derive the one loop effective action for the $\rm{KK}$ tower, solve the backreacted Einstein equations in five and four dimensions, and analyze the thermodynamics and regime of validity of the resulting self-gravitating solution. In section \ref{s:connection} we construct the linear sigma models for the black hole-tower transition in the presence of a KK circle, compute the cobordism classes and brane charge lattices of the two backgrounds, and combine the worldsheet and spacetime perspectives. We conclude and discuss some open directions in section \ref{s:conclusions}. Appendices collect technical material, including the heat-kernel computation of the one-loop effective potential in Appendix \ref{ap:heat_kernel}, the effective action of the KK tower in terms of Eisenstein series in Appendix \ref{ap:eisenstein}, details on the GSO projection and the string spectrum in the presence of the thermal circle in Appendix \ref{ap:GSO}, a careful analysis of the physics of the box in which our KK gas is confined, and why it does not affect our results, in Appendix \ref{ap:boundary}, the full-fledged resolution of the backreacted Einstein equations for 5d radiation in the presence of a compact KK circle in Appendix \ref{ap:proof}, and the bordism computations and related definitions in Appendix \ref{ap:bordism.products.spheres}.

\section{The Black Hole–Tower Correspondence} \label{s:blackholetower}

The notion of a fundamental cutoff for Effective Field Theories (EFTs) coupled to gravity, often referred to as the quantum gravity cutoff or \emph{species scale} \cite{Dvali:2007hz, Dvali:2007wp, Dvali:2009ks}, arises from the fact that a theory with a large number of light degrees of freedom ceases to be well described by a local EFT at energies well below the Planck scale. This was first noted in the one-loop contribution to Einstein-Hilbert gravity, where, in the presence of $\Nsp$ light species, the graviton kinetic term receives one-loop corrections that become comparable to its tree-level value at an energy scale
\begin{equation}\label{eq:species.scale}
    \Lsp = \frac{M_{\text{Pl},d}}{\Nsp^{1/(d-2)}}\,,
\end{equation}
where \emph{light} means the species have masses at or below the species scale $\Lsp$. The species scale can also be inferred from the structure of higher-curvature corrections to the gravitational action, since it appears as the scale suppressing a certain class of higher-curvature operators \cite{vandeHeisteeg:2022btw,Cribiori:2022nke,vandeHeisteeg:2023dlw,Castellano:2023aum,Calderon-Infante:2025ldq}.
Therefore, it sets the maximal energy scale that can be described in a local, gravitational EFT, or equivalently, the minimal length that can be probed.  In the context of black holes, this implies that the smallest possible semiclassical black hole that can be described by such gravitational EFT has a radius of order $\RBH \sim \Lsp^{-1}$. The entropy of a $d$-dimensional black hole of radius $\RBH$ scales with the area as $
\SBH \sim (\RBH M_{\text{Pl},d})^{d-2}$.
Thus, for a black hole of minimal radius, $\RBH \sim \Lsp^{-1}$, the entropy evaluates to 
\begin{equation}\label{eq:species.entropy}
\SBH \sim \left(\frac{M_{\text{Pl},d}}{\Lsp}\right)^{d-2} \sim \Nsp,
\end{equation}
which scales with the number of species. In the presence of general towers, this has been recently explored in the context of the Swampland program \cite{Cribiori:2023ffn, Basile:2023blg, Basile:2024dqq, Herraez:2024kux,Herraez:2025clp}. In particular, eq. \eqref{eq:species.entropy} motivates assigning thermodynamic quantities to general towers in analogy with black hole thermodynamics. These quantities were initially postulated on this basis in \cite{Cribiori:2023ffn} and subsequently derived from first principles in the free limit for general towers in \cite{Herraez:2024kux}. Specifically, the entropy of the tower is taken to be $S_{\text{sp}} \sim \Nsp$, corresponding to the minimum entropy that can be associated to a theory in the absence of interactions. The associated temperature is fixed by requiring that a black hole with entropy $\SBH =S_{\text{sp}}\sim \Nsp$ has the same temperature, yielding $T_{\text{sp}} \sim M_{\text{Pl},d}\, S_{\text{sp}}^{-1/(d-2)}  \sim \Lsp$.
Hence, the temperature of the tower is naturally of order the species scale. Finally, the energy (or mass) associated with a system of non-interacting species is, in Planck units, $E_{\mathrm{sp}} \sim M_{\text{BH,min}} \sim \Lsp^{3-d} \sim \Nsp \Lsp$.

In the context of string theory, this relation between black holes and species \cite{Dvali:2009ks} was already noted in the celebrated \emph{Black Hole-String Correspondence} \cite{Susskind:1993ws,Horowitz:1996nw}: if one considers a black hole and decreases the string coupling adiabatically, the black hole shrinks (in string units), eventually reaching the string scale and transitioning to a free, highly excited string. One can then explain the black hole entropy as microstates in a free string theory, where the microstates associated to a minimal black hole are identified with those of a weakly coupled free string. In subsequent work it was found that, in the thermal description of this transition, including the gravitational backreaction of the string yields an intermediate self-gravitating saddle: the Horowitz--Polchinski solution \cite{Horowitz:1996nw, Horowitz:1997jc}. This has been explored further in \cite{Chen:2021dsw}. The transition between a string in a compact background and a black-brane wrapping that background, which will be a recurrent theme of this work, was recently studied in \cite{Emparan:2024mbp}.

The \emph{Black Hole-Tower Correspondence} was formulated in \cite{Herraez:2024kux} as a generalization of the aforementioned black hole-string correspondence. The setup considers adiabatically varying the value of \emph{any} modulus controlling the ratio between the $d$-dimensional Planck scale and the species scale, i.e., $\phi\simeq (\Mpld/\Lsp)^{d-2}$ (equivalently, $\phi \simeq \Nsp$). This means any modulus that can be taken to infinite proper distance in moduli space. In addition to limits where the string coupling is made arbitrarily small, decompactification limits are then naturally included in this picture. This is consistent with the Emergent String Conjecture \cite{Lee:2019wij}, which states that all infinite-distance limits in moduli space must correspond to either decompactification limits or the limit of weakly-coupled fundamental string becoming tensionless.

Given a moduli dependent species scale near some infinite distance limit, $\Lsp(\phi)$, the black hole-tower correspondence states that there exists a point in moduli space $\phi_*$ where the transition between a black hole and a tower can occur, and that such point exists for every (neutral) black hole. In short, making the dependence between the Planck mass and species scale explicit, one can write the entropy of the minimal black hole and the tower of species as 
\begin{equation}
    \SBH \, \sim \, \dfrac{1}{\phi^{\frac{1}{d-3}}}\left(\MBH^{d-2} \,  \Lsp ^{2-d}\right)^{\frac{1}{d-3}},\qquad S_{\text{tower}}\,\sim \, \phi^{\frac{1}{p+1}}\left(\frac{M}{\Lsp}\right)^{\frac{p}{p+1}}\, ,
\end{equation}
where we are using the parameterization of the tower in which $p$ encodes the number of dimensions that are being decompactified  and one recovers the string case for $p\to \infty$. At the transition point we have 
\begin{equation}
    S\sim \phi_\ast \sim N_\ast \sim M_\ast/\Lsp\, , \qquad  T_*\sim \Lsp\, , \qquad
    M_\ast\simeq \left(\dfrac{\Lsp}{\Mpld}\right)^{3-d} {\Mpld}\,.
\end{equation}
We recover the black hole-string transition for $\phi=g_s^{-2}$, and we have a black hole-KK tower transition for $\phi=\mathcal{V}_p$, with $\mathcal{V}_p$ the volume of the $p$-dimensional internal manifold being decompactified (in higher-dimensional Planck units). When the modulus controlling the tower is the $p$-dimensional volume of an internal manifold, this transition is to a uniform black $p$-brane wrapping the $p$ decompactified directions. Such black brane solutions are known to exhibit Gregory-Laflamme instabilities \cite{Gregory:1993vy}, where the $p$-brane has lower entropy than a higher-dimensional black hole of the same mass. This has been studied recently in the present context \cite{Bedroya:2024uva, Herraez:2024kux} but the details on the dynamics of the instability are not fully understood and will not play a key role in our discussion.\footnote{See also \cite{Figueras:2026epx} for a recent numerical GR study of the Gregory-Laflamme instability.}

\subsection{Free strings and towers in the thermal circle}
\label{sec:freestringsthermalcircle}
We now review the formalism used to describe the self interactions of the string at finite temperature, and recover the results of \cite{Herraez:2024kux} for the entropy of towers of species with polynomial and exponential degeneracies, but expressed in terms of the thermal partition function.

Consider a free complex scalar field on $S^1_\beta \times \mathbb{R}^{d-1}$, where $\tau \sim \tau+\beta$ is the Wick-rotated, time-like coordinate $\tau= i\,t$. The euclidean action is given by 
\begin{equation}
  S_E[\phi] \;=\; \beta\int d^{d-1} x\,\,
\phi\,
  \big(- \nabla^2 + m^2\big)\,
  \phi .
\end{equation}
The thermal partition function is then the Euclidean path integral
\begin{equation}
  Z = \int\!\mathcal{D}\phi\, e^{-S_E[\phi]}
= \dfrac{1}{\det{\big(-\nabla^2+m^2\big)}}
\end{equation}
and 
\begin{equation}
    \log{Z}= -\log{\det{\big(-\nabla^2+m^2\big)}}
\end{equation}
Using Schwinger reparameterization
\begin{equation}
  \log\det D = -\int_\varepsilon^\infty \frac{d\ell}{\ell}\;
  \Tr\, e^{-\ell D},
\end{equation}
we can then directly compute the partition function, as the one-loop determinant in this case is a standard heat kernel result \cite{Vassilevich:2003xt}
\begin{equation}
  \log Z= \int_\varepsilon^\infty \frac{d\ell}{\ell}\;
  \frac{\,V_{d-1}}{(4\pi \ell)^{\frac{d-1}{2}}}\;
  e^{-\ell m^2},
\end{equation}
namely the one-loop determinant of a free scalar on the thermal circle. The relevant configurations will be those in a $d$-dimensional box of size $L$, and in particular in the limit $L \lesssim m^{-1}$. In this regime, the spatial momenta are quantized with spacing $\sim 2\pi/L \gtrsim 2\pi m$, so all non-zero spatial momentum states lie above the mass $m$, which sets the natural scale of the proper-time integrand $e^{-\ell m^2}$. Their contributions are then exponentially suppressed relative to the zero-momentum mode, and the partition function effectively reduces to the $d=1$ case.

\begin{equation}
  \log Z =\int_{\varepsilon}^{\infty} \frac{d\ell}{\ell}\, e^{-\ell m^2} \,=-\gamma_E - \log(\varepsilon\, m^2) + ...\,
\end{equation}
where $\gamma_E$ is the Euler–Mascheroni constant. Performing minimal subtraction regularization and
introducing a renormalization scale  $\mu$ as
\begin{equation}
     -\gamma_E - \log \varepsilon = \log \mu^2 \, ,
\end{equation}
we obtain
\begin{equation}
  \log Z
  \;=\; - \log\,\frac{m^2}{\mu^2}.
\end{equation}
The theory's cutoff is given by the temperature $T$, so it is natural to take the renormalization scale $\mu\simeq T$. This is also the characteristic scale set by the lowest Matsubara modes \cite{Kapusta:2006pm, Bellac:2011kqa}, such that in a finite temperature QFT it suppresses higher loop logarithmic corrections. For the free string near the Hagedorn temperature $T_H$ (i.e., the maximum temperature at which the free-string ensemble is well-defined) we have \cite{Horowitz:1996nw, Horowitz:1997jc} (see section (\ref{sec:horowitz.polchinski}) and appendix \ref{ap:GSO} for more details)
\begin{equation}
\label{eq:winding.mass}
    m^2=m_w^2= M_s^4 \ \dfrac{T_H^2-T^2}{T^2 T_H^2}\, ,
\end{equation}
and the partition function is given by
\begin{equation}
     \log Z
  \;=\; - \log\left(\dfrac{M_s^4}{T^4}\ \dfrac{T_H^2-T^2}{T_H^2}\right).
\end{equation}
The massless theory also includes the graviton as well as a dilaton and the radion of the thermal circle, but their interactions are assumed to be suppressed, and their tree level contribution to the entropy is nevertheless vanishing. In fact, one of the most surprising facts about the thermal winding is that it leads to a non-vanishing entropy at tree level.
At leading order near $T \to T_H$, the entropy and energy are given by
\begin{equation}\label{eq:freestringentropyenergy}
    S\, \simeq\, 2\  \dfrac{T_H^2}{T_H^2-T^2}\,,\qquad E\simeq TS\, .
\end{equation}
The proximity to the Hagedorn temperature is parameterized by dimensionless quantity
\begin{equation}\label{eq:epsilon}
    \epsilon \,\equiv\, \frac{\beta-\beta_H}{\beta_H} \,=\, \frac{T_H - T}{T} \, \ll\, 1\,.
\end{equation}
In terms of $\epsilon$, the leading near-Hagedorn behaviour of the winding mass and the free-string thermodynamics reads
\begin{equation}\label{eq:freestring.epsilon}
    m_w^2 \,\simeq\,  2\epsilon\, \frac{M_s^4}{T_H^2}\, ,\qquad
    S \,\simeq\, \frac{1}{\epsilon}\,,\qquad
    E \,\simeq\, \frac{T_H}{\epsilon}\,,
\end{equation}
making the Hagedorn divergence clear. As we will see in section \ref{sec:regimes}, the interplay between the
winding sector and the KK tower determines which of the two towers dominates the thermodynamics.

Adding an additional circle, such that the scalar sees a geometry $S^1_\beta \times S^1_\KK\times \mathbb{R}^{d-1}$, and expanding the field in Fourier modes along the additional $S^1_{\KK}$, each mode behaves as a $(d-1)$-dimensional field such that 
\begin{equation}
  \left(- \nabla_{d}^2 + m^2\right)\,=  \left(- \nabla_{d-1}^2 +p^2_{\KK}+ m^2\right).
  \end{equation}
  Since the momentum in the circle is quantized, we have a tower of $\rm{KK}$ modes with masses $p^2_{\KK}=n^2 \mKK^2$, populated up to the thermal cutoff at $N_T$, the number of species that are \emph{active} at temperature $T$.\footnote{More generally, $N_T$ denotes the number of \emph{active} species contributing to the ensemble at temperature $T$, irrespective of whether they are $\rm{KK}$ or stringy in origin. For $p$ simultaneously decompactified directions with degeneracy $N(M)\sim M^p$ this reads $N_T \sim (T/m_{\text{tower}})^p$, with the $p\to\infty$ limit smoothly recovering the string oscillator case (see \cite{Herraez:2024kux} for the general construction). At the correspondence point $T \to \Lsp$ one has $N_T \to \Nsp$, recovering the total species count associated with the species scale.} Throughout this paper we often work in regimes where only the $\rm{KK}$ tower lies below the species scale, or gives the dominant contribution. For the single compact direction considered here this gives $N_T = T/\mKK$, and $N_T$ effectively counts the $\rm{KK}$ species. The partition function describing the system inside a $(d-1)$-dimensional box of size $L\lesssim T^{-1}$ (i.e.\ the $(d-1)=0$ partition function) is then 
  \begin{equation}
  \label{eq:tower.partition}
        \log Z
  \;=\;- \sum_{n=0}^{N_T}\, \log\left(\frac{n^2 \mKK^2 +m^2}{T^2}\right)\,.
\end{equation} 
The temperature dependence of each term in \eqref{eq:tower.partition} enters both via the explicit factors of $T$ and the winding mass, $m_w^2(T)$. Writing the winding mass \eqref{eq:winding.mass} as $m_w^2 = M_s^4\left(T^{-2}-T_H^{-2}\right)$, one gets $T\,\partial_T m_w^2 = -2M_s^4/T^2$, so the near-Hagedorn (divergent) part of the entropy reads
\begin{equation}
    \label{eq:winding.tower.entropy}
    S \,\simeq \, \frac{2 M_s^4}{T^2} \ \sum_{n=0}^{N_T}\, \frac{1}{\,n^2\mKK^2 + m_w^2\,}\,.
\end{equation}
 The term inside the summation is a Lorentzian in $n$ of width $m_w/\mKK$, which means only the KK states of the thermal scalar with $n\,\mKK \lesssim m_w$ contribute significantly, while heavier states only give  a convergent, $\epsilon$-independent contribution.\footnote{The $T$ dependence of $N_T$ in the upper limit gives a contribution to the entropy of the form $N_T\log[(N_T^2 \mKK^2+m_w^2)/T^2]$, which is subdominant and can be neglected.}

Let us analyze the two key regimes. For $\mKK \gg m_w$, i.e., for $\epsilon \lesssim 1/N_T^2$, the tower does not contribute significantly to the partition function and the $n=0$ mode dominates, recovering the results in the absence of the extra circle, eq. \eqref{eq:freestringentropyenergy}. In the regime $\mKK \ll m_w$, the width of the lorentzian becomes very large and we can safely approximate the sum by an integral, which gives
\begin{equation}
\label{eq:freeKKandstringtower}
    S \,\simeq\, \pi\,\frac{m_w}{\mKK}\, \frac{T_H^2}{T_H^2-T^2} \,\simeq\, b\,\frac{N_T}{\sqrt{2\epsilon}}\,, \qquad E \,\simeq\, TS\, , 
\end{equation}
where we defined $b \,\equiv\, \pi\,M_s^2/T_H^2\,$. The result is valid up to corrections suppressed by $m_w/T$ and terms of order $N_T$ and $\log(1/\epsilon)$. The free-string divergence is thus enhanced by the number of KK copies of the winding mode that are lighter than $m_w$, namely $( m_w/\mKK)$, but
\emph{not} by $N_T$. Correspondingly, the divergence is softened from $1/\epsilon$ to $1/\sqrt{\epsilon}$.

\begin{figure}[t]
    \centering
        \centering        \includegraphics[width=0.75\textwidth]{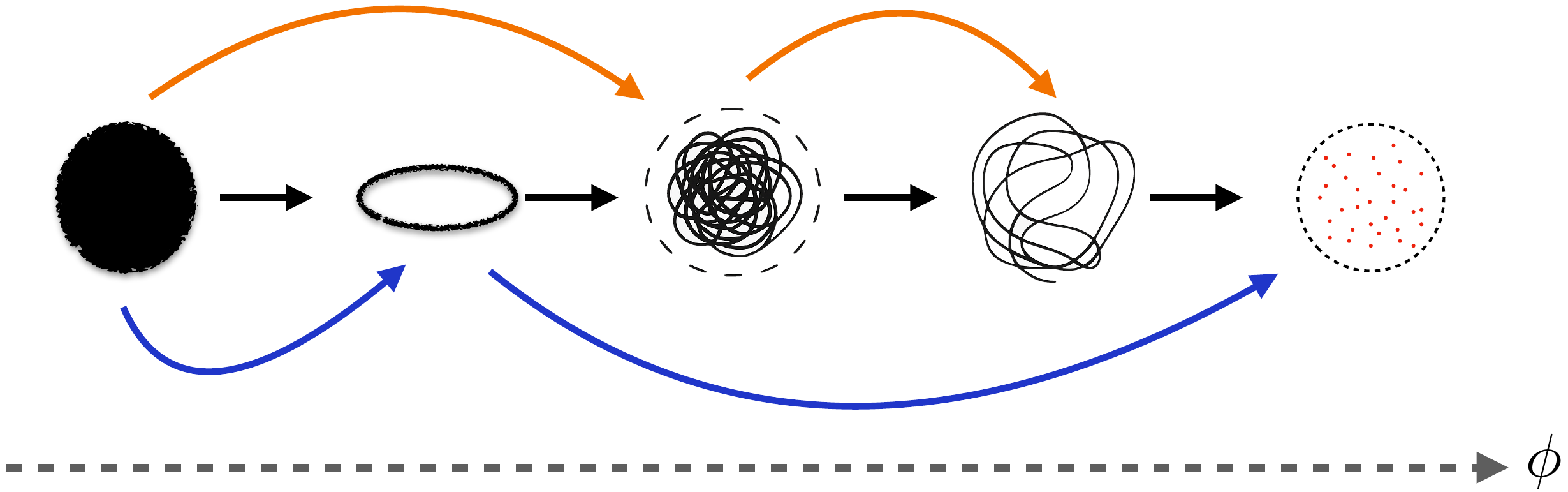}
    \caption{From left to right: black hole, black string, Horowitz--Polchinski solution, free string, and gas of species. Here $\phi$ parametrizes a direction in moduli space. Orange arrows indicate the black hole-string transition, while blue arrows indicate a black hole transitioning directly into a gas of KK modes. Black arrows represent a generic mixed limit of the black hole-tower correspondence.}
    \label{fig:bhtransitions}
\end{figure}

In the absence of light string oscillator modes (e.g.,  for $g_s\gtrsim 1$, where the higher-dimensional string scale is not parametrically below the higher-dimensional Planck scale, $M_{\text{Pl},D}$) one can in principle consider just the tower of $\rm{KK}$ modes, and neglect the winding mode. In this limit the $\rm{KK}$ gas alone can simultaneously match the energy, entropy and size of the minimal black hole \cite{Herraez:2024kux}, without the need of an intermediate string dominated regime analogous to the Horowitz--Polchinski phase. These two situations are shown schematically in Fig. \ref{fig:bhtransitions}. Then, if there is no light winding mode spectrum, we have a pure $\rm{KK}$ tower with partition function\footnote{The $n=0$ mode is the underlying massless higher-dimensional zero mode. Its order-one contribution at $T \simeq 1/L$ is subleading to the tower-dominated result $\log Z \simeq 2  N_T$ obtained below, and we omit it here to avoid formally divergent logarithms in the $d=0$ approximation.}
  \begin{equation}
        \log Z
  \;=\; \sum_{n=1}^{N_T}-\log\left(\frac{n^2 \mKK^2}{T^2}\right)\,,
\end{equation}
which at leading order gives $\log Z \simeq 2T/\mKK$ and therefore
\begin{equation}\label{eq:purekkentropyenergy}
    S\simeq 4 \dfrac{T}{\mKK}\,,\qquad E\simeq 2\dfrac{T^2}{\mKK}\,.
\end{equation}

The free string and non-interacting $\rm{KK}$ towers cases treated above are motivated by the observation, as noted in \cite{Basile:2023blg, Herraez:2024kux}, that the close relation between minimal black holes and towers of species provides bottom-up evidence for the Emergent String Conjecture \cite{Lee:2019wij}. The latter states that at infinite distance in moduli space an infinite tower of light species appears, and that this tower must consist of either $\rm{KK}$ modes or string excitation modes. From a bottom-up perspective, it was argued that only power-law spectra, as in the case of Kaluza–Klein towers, and exponential spectra, as in the case of string oscillator towers, can have a thermodynamic correspondence with minimal black holes \cite{Basile:2023blg,Basile:2024dqq, Herraez:2024kux} (see also \cite{Bedroya:2024ubj} for a complementary analysis based on unitarity of the S-matrix).

\subsection{Multiple towers and relevant physical regimes}\label{sec:regimes}

In generic string compactifications, both $\rm{KK}$ and string oscillator towers are simultaneously present. Having computed their free thermodynamics separately, we now turn to examine what determines which tower dominates the transition to a black hole.

When both towers contribute, the total entropy in the free limit combines \eqref{eq:purekkentropyenergy} and \eqref{eq:freeKKandstringtower} into
\begin{equation}\label{eq:combined.entropy}
    S \,\simeq\,
    \begin{cases}
        \; c\,N_T \,+\, \dfrac{1}{\epsilon}\,, &
        \epsilon \,\lesssim\, \dfrac{1}{N_T^2}\ll 1\,,\\
        \; N_T\left(c \,+\, \dfrac{b}{\sqrt{2\epsilon}}\right)\,, &
        \dfrac{1}{N_T^2} \,\lesssim\, \epsilon \,\ll\, 1\,,
    \end{cases}
\end{equation}
where $c$ is an order-one constant accounting for the KK replicas of the string massless sector (in the higher-dimensional theory), each contributing as in \eqref{eq:purekkentropyenergy}. In the first regime the  winding KK replicas with $n\geq1$ contribute an additional $\epsilon$-independent piece, of order $ \sum_{n\geq 1} M_s^4/(n \mKK T)^2 \sim N_T^2$, which is subleading to the $\epsilon^{-1}$ term in the regime $\epsilon \lesssim N_T^{-2}$ and we therefore do not display it explicitly.
The constant $b$ is the order-one number defined below \eqref{eq:freeKKandstringtower}, and the thresholds are understood up to $(T_H/M_s)$-dependent order-one factors. The two expressions match parametrically at $\epsilon \sim N_T^{-2}$, where both give $S \sim N_T^2\sim\epsilon^{-1}$.

The structure of \eqref{eq:combined.entropy} makes the role of the different scales transparent. Since $b$ and $c$ are both order one, throughout the near-Hagedorn regime, $\epsilon \ll 1$, the winding sector always dominates the entropy. It ceases to do so only when $\epsilon$ itself becomes order one, and thus $m_w \sim M_s$. That is, the thermal scalar is as heavy as the generic string oscillators, the near-Hagedorn effective description of section \ref{sec:freestringsthermalcircle} ceases to be valid, and the entropy reduces to that of the pure KK gas, $S \simeq c\,N_T$, as in \eqref{eq:purekkentropyenergy}. This boundary is independent of $N_T$, as it must be, since both entropies in \eqref{eq:combined.entropy} are extensive along the KK circle, so their competition is local and cannot depend on the size of the circle. In other words, whether stringy winding physics is relevant is \emph{not} controlled by any comparison between $\epsilon$ and $N_T$ (even though this is related to the number of dimensions in which the transition takes place), but only by whether temperatures parametrically close to $T_H$ are reached at all. As discussed below, this depends only on the hierarchy between $M_s$ and $M_{\text{Pl},d+p}$.

Additionally, within the near-Hagedorn description, the threshold $\epsilon \sim N_T^{-2}$ (i.e.,  $m_w \sim \mKK$) separates two qualitatively different but still winding-dominated regimes.\footnote{The $p>1$ case can be parameterized by a per-level degeneracy
$d_n=n^{p-1}$, such that $N_T=(T/\mKK)^p$ and the sum goes up to level
$n_{\rm max}=T/\mKK$. The generalization of eq. \eqref{eq:combined.entropy} for $p$ isotropic internal directions then takes the following forms in the two relevant regimes: $S\simeq N_T\left(c+\epsilon^{p/2-1}\right)$ for $N_T^{-2/p}\lesssim \epsilon \ll 1$, and $S\simeq c\,N_T+\epsilon^{-1}$ for
$\epsilon\lesssim  N_T^{-2/p}\ll 1$.} For $\epsilon \ll N_T^{-2}$ (which requires $T$ to be tuned very close to $T_H$) the winding length scale $m_w^{-1}\gg\mKK^{-1}$, and thus the thermal scalar cannot resolve the compact circle. Its KK replicas are gapped and the $d$-dimensional description is recovered. The interacting counterpart of this is the $d$-dimensional  Horowitz--Polchinski solution reviewed in section \ref{sec:horowitz.polchinski}, uniform along the KK circle. For $N_T^{-2} \ll \epsilon \ll 1$ the circle is instead resolved, $m_w^{-1} \ll \mKK^{-1}$, and the KK replicas of the winding scalar enhance the free entropy as in \eqref{eq:freeKKandstringtower}. The interacting counterparts are the Horowitz--Polchinski strings on a circle of \cite{Emparan:2024mbp,Chu:2024ggi,Chu:2025fko}, with their uniform and localized branches. A complete treatment of the thermodynamics in this intermediate regime, connecting the enhanced free-gas entropy to those saddles, is beyond the scope of this work.

In terms of the lower-dimensional string coupling, ${g_{s,d}^{-2}=g_{s,d+p}^{-2}(M_s/\mKK)^p}$, the near-Hagedorn entropy reaches $S\sim g_{s,d}^{-2}$ at the correspondence point. The $\rm{KK}$ contribution can be identified with the entropy of the compact dimensions, while the Hagedorn-divergent piece corresponds to the string propagating on this background.

The decomposition above tells us which tower dominates the thermodynamics at a given temperature. The question relevant for the nature of the black hole-tower transition is whether the system enters the near-Hagedorn regime $\epsilon \ll 1$, where the winding sector takes over, before it enters the strongly-coupled gravitational regime. At weak string coupling the mass hierarchy $\mKK \leq M_s \ll M_{\text{Pl},d+p}$ ensures that string oscillation modes become relevant well before gravity becomes strongly coupled (i.e. near $M_{\text{Pl},d+p}$), and the black hole transition is captured by a Horowitz--Polchinski solution \cite{Horowitz:1997jc}. 
Conversely, when $M_s\gtrsim M_{\text{Pl},d+p}$, the strong gravitational regime is reached before a Hagedorn temperature, and a direct black hole-KK tower transition can take place without an intermediate string dominated phase. One of the goals of this paper is to analyze this second scenario in more detail, specifically by solving the backreacted Einstein equations for a gas of $\rm{KK}$ modes in five spacetime dimensions (section~\ref{sec:self.grav.bh.tower}). To make these qualitative statements precise and place them within the general species-scale discussion, we need to examine how the species scale behaves in the presence of multiple light towers of states \cite{Castellano:2021mmx}. The idea is as follows. One starts by identifying the mass scale and degeneracy of the lightest tower and computing the species scale associated with it. If other towers are present, one checks whether their mass scale lies below this initial estimate. If not, the species scale remains the same. If they do, the number of species must be updated, and the species scale is recalculated. This process is repeated until no further towers lie below or at the resulting species scale.

Let us now focus on our case of interest. For simplicity, we consider a setup with a single Kaluza–Klein tower and a single string tower.\footnote{This can be done without loss of generality for every tower with $\mKK\leq M_s$, since multiple towers can be combined into a single ``effective'' tower, as explained in \cite{Castellano:2021mmx}.} As indicated by the procedure outlined above, we begin by computing the species scale associated with the $\rm{KK}$ tower. From equation (\ref{eq:species.scale}), we know that this corresponds to the higher-dimensional Planck scale $M_{\text{Pl},d+p}$ \cite{Castellano:2021mmx}. Whether this estimate of the species scale is correct depends on the value of the higher dimensional string coupling. The number of string excitation levels that can fit below the tower is

\begin{equation}
	N \sim \frac{M_{\text{Pl},d+p}}{M_s} = \frac{1}{g_{s,d+p}^{\frac{2}{d+p-2}}}\, .
\end{equation}

Hence, for parametrically small values of the $(d+p)$-dimensional string coupling $g_{s,d+p}$, string excitations become relevant in determining the species scale. Moreover, because of the exponential degeneracy of the string spectrum, it is expected that most of the modes contributing to the number of species $\Nsp$ will come from string excitations and, as a result, the species scale is lowered to the string scale $M_s$ \cite{Castellano:2021mmx,Marchesano:2022axe,Castellano:2022bvr,vandeHeisteeg:2022btw}.

\begin{figure}[ht]
    \centering
    \begin{minipage}{0.48\textwidth}
        \centering
        \includegraphics[width=\textwidth]{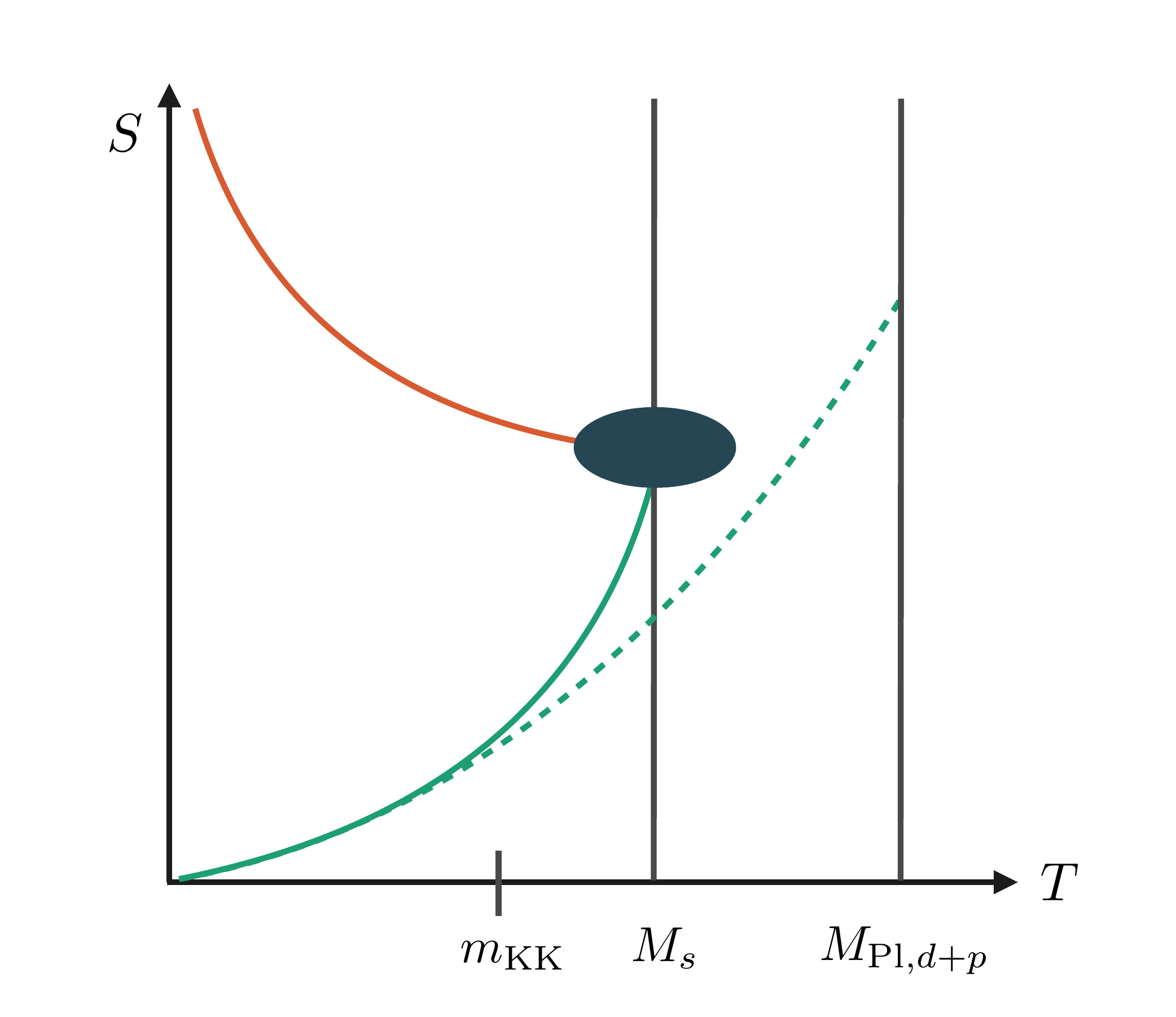}
        \subcaption{$M_s \ll M_{\text{Pl},d+p}$} \label{fig:lower}
    \end{minipage}
    \hfill
    \begin{minipage}{0.48\textwidth}
        \centering
        \includegraphics[width=\textwidth]{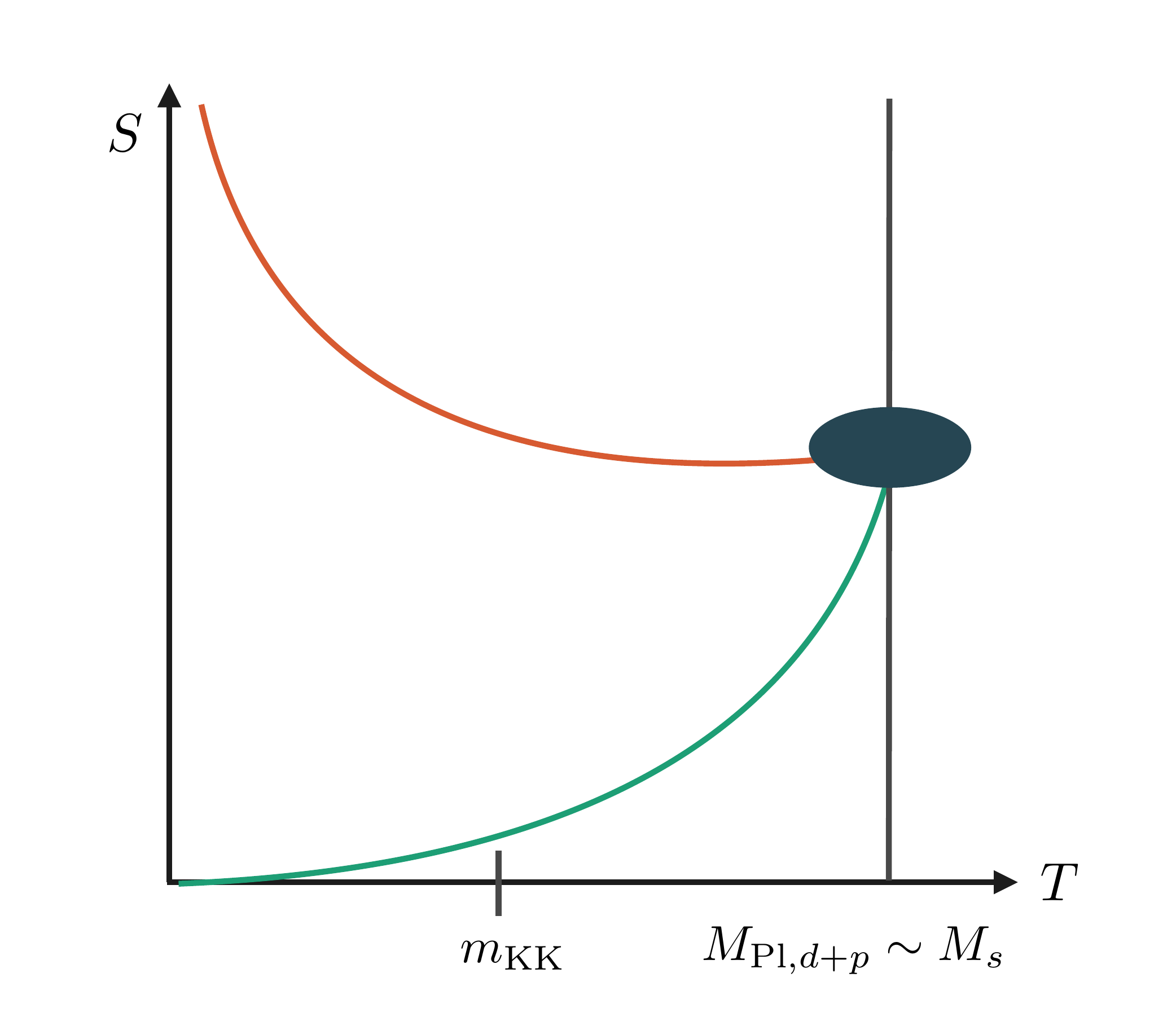}
        \subcaption{$M_s \gtrsim M_{\text{Pl},d+p}$} \label{fig:same}
    \end{minipage}
    \caption{Entropy as a function of the temperature for the $\rm{KK}$ particles in a box (green line) and the black hole (red line). (a) When the string scale is parametrically lower than the higher dimensional Planck scale, the $\rm{KK}$ box transitions into a free string, which can then become a black hole or black string. The dashed green line represents the entropy of a pure $\rm{KK}$ box, that is a box in which string excitations are not taken into account. (b) If the string scale and the higher dimensional Planck mass are parametrically the same, the gas of $\rm{KK}$ species can transition directly into a black hole.}
\end{figure}

To sum up, we distinguish two different situations depending on the hierarchy between the string and Planck scales
\begin{itemize}
	\item If $M_s \ll M_{\text{Pl},d+p}$ in the limit under consideration, no direct transition is expected between a gas of $\rm{KK}$ species and the black hole. Starting from a box with $\rm{KK}$ modes and increasing its temperature, the string scale is reached before the Planck scale (i.e., the regime $\epsilon\ll1$ above is the relevant one). At that point, string oscillator modes become relevant and dominate over the $\rm{KK}$ modes (see Fig. \ref{fig:lower}). In this limit, a $\rm{KK}$ tower–string transition takes place. Within this regime, one can further distinguish two cases, which we review in detail in section \ref{sec:horowitz.polchinski}, depending on the scale of the extra dimensions:
	\begin{itemize}
		\item[$\circ$] If $M_s \sim m_{\text{KK}}$, the internal space cannot be resolved, and a black hole-string transition in $d$ dimensions effectively takes place. This is the case originally studied in the literature \cite{Susskind:1993ws,Horowitz:1996nw}.
		\item[$\circ$] If $M_s \gg m_{\text{KK}}$, the extra dimensions become visible before the string scale is reached. As $T\to T_{H} $ the gas first probes the regime described in the second line of \eqref{eq:combined.entropy} ($\mKK\ll m_{w}$) and only for temperatures tuned very close to $T_H$ the regime described by the first line applies ($\mKK\gtrsim m_{w}$). In the former regime the string gas can either localize in the internal dimensions or remain uniform along them. These are the localized and uniform solutions reviewed in section \ref{sec:horowitz.polchinski}, namely a localized one that connects to a black hole-string transition in $(d+p)$ dimensions, and a uniform one that connects to a black $p$-brane-string transition (studied in detail in \cite{Emparan:2024mbp} for the case of a black string, that is, $p=1$), formally equivalent to a black hole-string transition in $d$ dimensions. In the latter regime, the winding length is larger than the size of the internal circle so only the uniform solution is possible.
	\end{itemize}
	\item On the other hand, if the species scale is given by the higher dimensional Planck scale (which would correspond to $M_s \gtrsim M_{\text{Pl},d+p}$), a black hole-KK tower transition in $d$ dimensions can occur, which is just the $d$-dimensional avatar of the $(d+p)$-dimensional black brane--KK tower transition (see Fig. \ref{fig:same}). As the temperature of a $\rm{KK}$ gas increases, the strongly coupled gravitational regime is reached without an intermediate Hagedorn-like ensemble, and thus a direct black hole/brane–KK tower transition can take place without an intermediate string dominated regime. This scenario will be analyzed in more detail in section \ref{sec:self.grav.bh.tower}, where for technical convenience we restrict ourselves to one single compact direction ($p=1$). The qualitative picture set out above applies to any finite $p$.
\end{itemize}
In the following, we discuss how these scenarios are realized within string theory. 

\subsection{The Horowitz--Polchinski solution}\label{sec:horowitz.polchinski}

As reviewed in the previous sections, black holes at weak string coupling ---identified in section \ref{sec:regimes} as the regime $M_s \ll M_{\text{Pl},d+p}$--- are expected to transition into a highly excited near-Hagedorn string. A better understanding of this transition can be gained by examining the intermediate regime, where self-gravitational effects become significant. Horowitz and Polchinski \cite{Horowitz:1996nw,Horowitz:1997jc} famously found a classical saddle to the (euclidean) equations of motion that describes this phenomenon. Their solution is known as the Horowitz--Polchinski (HP) string. Intuitively, the HP string can be understood as a free string that effectively ``curls up'' due to its self-gravitation.

Near-Hagedorn strings are most naturally studied in the imaginary-time formalism. Finite temperature quantum field theory is obtained by Wick-rotating to Euclidean time and compactifying it on a circle of circumference $\beta=1/T$. In string theory, this compactification gives rise to string winding modes along the thermal circle. As the temperature reaches the Hagedorn scale $T=T_H$ the lowest lying winding mode becomes tachyonic, signaling a phase transition. This is the so-called \emph{thermal scalar}. For instance, in the type II string theories \cite{Polchinski:1998rr}, the mass spectrum is
\begin{equation}\label{eq:thermal.circle.spec}
    M^2 = \frac{2}{\alpha'}(N + \tilde{N} - 1) + \frac{4\pi^2n^2}{\beta^2} + \frac{w^2 \beta^2}{4\pi^2\alpha'^2},
\end{equation}
subject to the level-matching condition $N - \tilde{N} = nw$ and the GSO projection \cite{Gliozzi:1976qd}. Usually the GSO projection excludes $N=\tilde{N}=0$, but for states winding around the thermal circle the projection is reversed for odd $w$ \cite{Atick:1988si} (see appendix \ref{ap:GSO}). As a result the choice $w=1$ allows for $N=\tilde{N}=0$, yielding a scalar degree of freedom $\chi$ with mass
\begin{equation}\label{eq:winding}
    m^2_{\chi}(\beta) = \frac{\beta^2 - \beta_H^2}{4\pi^2\alpha'^2},
\end{equation}
where $\beta_H = 2\pi \sqrt{2\alpha'}$ is the inverse Hagedorn temperature. This thermal scalar appears in all critical string theories, though the value of $\beta_H$ depends on the specific model.

As the temperature approaches the Hagedorn scale, the thermal scalar becomes almost massless and enters the low energy dynamics \cite{Horowitz:1997jc,Chen:2021dsw,Bedroya:2024igb,Emparan:2024mbp,Balthazar:2022hno,Balthazar:2022szl,Urbach:2022xzw,Urbach:2023npi,Matsuo:2022kvx}. The effective action capturing this effect is given by
\begin{equation}\label{eq:exact.HP}
	I_{\text{HP},d} = \frac{\beta}{16\pi G_{N}}\int d^{d-1}x\sqrt{g}\,e^{-2\phi_{d-1}}\Big[-\mathcal{R}-4(\nabla \phi_{d-1})^2+(\nabla \varphi)^2+(\nabla \chi)^2+m^2(\varphi)\chi^2\Big],
\end{equation}
where $\varphi$ parameterizes fluctuations of the thermal circle, $\phi_{d-1} = \phi_d-\varphi/2$ is the lower-dimensional dilaton, and $\chi$ is the thermal scalar.

Expanding near the Hagedorn point with $\chi,\varphi\sim \epsilon\ll 1$, with $\epsilon$ the near-Hagedorn parameter defined in \eqref{eq:epsilon}, one finds
\begin{equation}\label{eq:winding.expanded}
	m^2(\varphi)=m_{w}^2+\frac{\kappa}{\alpha'}\varphi + \mathcal{O}(\varphi^2), \qquad 
	m_{w}^2\simeq \frac{\kappa(\beta-\beta_H)}{\alpha'\beta_H},
\end{equation}
where $\kappa=\alpha'\partial_{\varphi}m^2(\varphi)$ is an order-one constant.
Setting the dilaton and metric to constants, which is consistent because their fluctuations scale as higher powers of $\epsilon$, the action reduces to
\begin{equation}\label{eq:originalHP}
	I_d = \frac{\beta}{16\pi G_N}\int d^{d-1}x\Big[(\nabla \varphi)^2+(\nabla \chi)^2+\Big(m_w^2+\frac{\kappa}{\alpha'}\varphi\Big)\chi^2\Big].
\end{equation}
This truncated theory enjoys a scaling symmetry \cite{Emparan:2024mbp,Balthazar:2022hno,Balthazar:2022szl},
\begin{equation}\label{eq:scale.inv}
	(x,\varphi,\chi,m_{w}^2)\;\longrightarrow\;(\lambda^{-1/2}x,\, \lambda \varphi,\, \lambda \chi,\, \lambda m_{w}^2).
\end{equation}
The symmetry fixes the scaling of the fields with $\epsilon$, justifying the truncation. For example, the dilaton equation of motion implies $\phi_d\sim \epsilon^2$, and the same holds for metric fluctuations.

The equations of motion are
\begin{equation}
	\nabla^2\chi - \Big(m_w^2 + \frac{\kappa}{\alpha'}\varphi\Big) \chi = 0, 
	\qquad
	2\nabla^2\varphi - \frac{\kappa}{\alpha'}\chi^2 = 0.
\end{equation}

Rescaling the variables removes explicit dependence on $m_w$ and $\alpha'$, showing that the rescaled fields are order-one and $\chi,\varphi\sim \epsilon\sim\alpha'm_w^2\ll1$, consistent with expectations. The characteristic size of the solution is
\begin{equation}
    \ell \sim \frac{1}{m_{w}} \sim \sqrt{\frac{\beta_H}{\beta-\beta_H}},
\end{equation}
so the string ball expands as $T\to T_H$.

Evaluating the action on-shell, the entropy is 
\begin{equation}\label{eq:entropy.HP}
	S = \frac{\kappa}{\alpha'} \frac{\beta_H}{16\pi G_N} \int d^{d-1} x |\chi|^2
	= \frac{2\alpha'(d-3)\Omega_{d-2}}{\kappa} \frac{\beta_H}{16\pi G_N} \eta^{\frac{d-5}{2}} m_w^{5-d}\, ,
\end{equation}
where $\eta$ is the lowest lying discrete eigenvalue of the rescaled non-local equation for $\chi$, fixed by the normalization $\int d^d\hat x\,|\hat\chi(\hat x)|^2 = 1$ (see e.g.\ \cite{Chen:2021dsw,Bedroya:2024igb} for details).
Corrections to the entropy can be found by integrating the Clausius relation $dM = TdS$ \cite{Chen:2021dsw}. Comparing the classical action to quantum fluctuations gives the regime of validity,
\begin{equation}\label{eq:validity.HP}
     g_s^{\frac{4}{7-d}} \, \lesssim \alpha' m_{w}^2 \, \ll \, 1,
\end{equation}
or equivalently, in terms of the mass, using $M\simeq S/\beta_H$,
\begin{equation}\label{eq:validity.masses.HP}
     g_s^{\frac{4}{d-7}} \, \lesssim \, \sqrt{\alpha'} M \, \ll \, \frac{1}{g_s^2}.
\end{equation}
Normalizable solutions exist only for $d=4,5,6$ \cite{Chen:2021dsw,Chu:2024ggi,Bedroya:2024igb,Chu:2025fko,Chu:2025kzl}. The interpretation is sharpest in $d=4$, where the black hole transitions into a self-gravitating string ball that evaporates and expands until it becomes a free string. In $d=5$, the window degenerates to $M\sim M_s g_s^{-2}$ and self-interactions become negligible at the transition scale. In $d=6$ the window inverts and the mass grows as $T_H$ is approached, the HP ball coexists with black holes and free strings but is entropically subdominant. In these higher dimensions the transition has been argued to produce \emph{atypical} string states \cite{Horowitz:1996nw,Horowitz:1997jc}, though this point remains debated \cite{Damour:1999aw,Damour:1999iq}.

Due to the HP string's success in describing the gravitational self-interactions in the BH-string transition, a natural question is whether a similar strategy can be applied when studying the more general black hole-tower correspondence. We revisit the two regimes identified in section (\ref{sec:regimes}) ($M_s \ll M_{\text{Pl},d+p}$ and $M_s \gtrsim M_{\text{Pl},d+p}$) and assess whether an HP-like saddle is available in each. Throughout, we will assume that in addition to the thermal circle, there is a (space-like) compact direction. As explained in section \ref{sec:regimes}, depending on the size of this compact direction, we expect a black hole to correspond to either a highly excited string or a gas of $\rm{KK}$ species.

\subsubsection*{Black hole-string correspondence $M_s \ll M_{\text{Pl},d+1}$}

This is the situation depicted in \figref{fig:lower} and it corresponds to a weak string coupling regime. The $(d+1)$-dimensional low energy effective action will then be described by the massless string modes, with one of the space directions compactified along a circle of radius $R_{\KK}$. In that case, the self-gravitating solution is given by a HP string with one compactified direction. This situation was already explored in \cite{Emparan:2024mbp,Chu:2024ggi,Chu:2025fko}. Two families of solutions were found\footnote{In \cite{Emparan:2024mbp}, solutions with an extra compact dimension are termed (uniform or non-uniform) HP strings, and they refer to the original Horowitz-Polchinski solution as the HP ball.}, those that are uniform along the compact $\rm{KK}$ circle and those that are not. The uniform solutions were shown to be thermodynamically favored compared to their non-uniform counterparts. This uniform HP string would then transition into a uniform black string.\footnote{ For $m_{\text{KK}}\ll m_w$ 
there is an approximate scaling symmetry $(x,\varphi,\chi_n,m_{w}^2)\;\longrightarrow\;(\lambda^{-1/2}x,\, \lambda \varphi,\, \lambda \chi_n,\, \lambda m_{w}^2),$ which becomes exact as $ m_{\text{KK}}/m_w\to 0$. In this limit one can then compute the entropy in the same way as HP, and one finds that it corresponds to that of the HP solution uniformly smeared along the compact direction, a Horowitz-Polchinski string \cite{Emparan:2024mbp}.}

\subsubsection*{Black hole-KK tower correspondence $M_s \gtrsim M_{\text{Pl},d+p}$}

This corresponds to the situation shown in \figref{fig:same}. From the higher dimensional point of view strings are strongly coupled, and the low energy dynamics cannot be described by the string effective action. Instead, large wavelength excitations are described by some EFT coupled to Einstein gravity. As a first approximation, we can write the following low energy effective action
\begin{equation}
\begin{aligned}
S_{\text{tower}} = \frac{\beta}{16\pi G_N} \int d^{d-1}x \sqrt{g}\, \Big( -\mathcal{R}_{d-1}+ (\partial\varphi_{\text{KK}})^2
+ \sum_{n}^{N_T} \big(|\partial\chi_n|^2 + m_n(\varphi_{\text{KK}})^2 |\chi_n|^2\big) \Big)\,,
\end{aligned}
\end{equation}
with $m_n(\varphi)^2=n^2 m_{\text{KK}}^2 e^{-2\varphi_{\text{KK}}}$, where $\varphi_{\text{KK}}$ parameterises the size of the $\rm{KK}$ circle. The $\chi_n$ represent the $\rm{KK}$ modes of a field in the higher dimensional theory. Expanding $e^{-2\varphi_{\text{KK}}}\simeq1-2\varphi_{\text{KK}}$ and keeping $\chi,\varphi_{\text{KK}}\ll1$ gives
\begin{align}
    \nabla^2\chi_n &=m_n^2(1-2\varphi_{\text{KK}})\chi_n, \\
    \nabla^2\varphi_{\text{KK}} &=-\sum_n m_n^2\chi_n^2.
\end{align}
In contrast to the HP case, there is no scaling symmetry: $\varphi_{\text{KK}}$ cannot be rescaled to make the system invariant. The scaling argument underlying the HP solution is therefore absent and higher-point interactions cannot be neglected. This is expected from the fact that in the HP case the winding mass, $m_w$, and the cubic coupling, $\kappa/\alpha^\prime$, were independent, whereas in this setup both the relevant mass and the (cubic) couplings are proportional to $m_n$, so one cannot scale them independently to find an overall rescaling that can leave the equations of motion invariant.

This could have been expected from the fact that a classical solution with non-vanishing $\rm{KK}$ profiles would necessarily break uniformity along the compact circle, leading to an entropy inconsistent with that of a uniform black string. Indeed, a gas of $\rm{KK}$ modes reproduces the entropy of a uniform black string, not that of a localized black hole. For this reason the HP formalism, which relies on scaling symmetry and the dominance of a single winding mode, cannot be directly generalized to describe a self-gravitating $\rm{KK}$ tower. In the next section we will explore an alternative approach that preserves uniformity along the compact directions.

\section{The Black Hole–Tower correspondence with gravitational backreaction }\label{sec:notfree.species}

In section \ref{s:blackholetower} we computed the thermodynamic quantities of a tower of free $\rm{KK}$ species from microstate counting. We will now re-derive these results using the imaginary-time formalism. This method will then allow us to compute the gravitational backreaction that a box of $\rm{KK}$ species induces.

The (classical) action for a tower of free $\rm{KK}$ particles in $d$ spacetime dimensions is given by 
\begin{equation}\label{eq:action.free.tower}
    I_{\KK} = \int d^d x \, \sum_{n=1}^{N_T} \left[|\partial \chi_n|^2 + n^2 m_{\KK}^2 |\chi_n|^2 \right ]
\end{equation}
where $N_T$ is the number of species and the $\chi_n$ are complex scalars (following the same normalization as in section \ref{sec:freestringsthermalcircle}). Furthermore, the species are restricted to move inside a box of size $L$. 

Similarly to the case of the free string, the action vanishes on-shell. Consequently, the dominant contributions to the partition function will come from the one-loop effective potential. This can be done by computing the heat kernel of the operator $D_ n =- \nabla^2 + n^2 m_{\KK}^2$ on a flat $\mathbb{R}^{d-1}\times S^1_\beta$ background. The effective potential in flat space reads \cite{Vassilevich:2003xt,Aparici:2025kjj} (see also Appendix \ref{ap:heat_kernel} for a short review)
\begin{align}
    I_{\KK} = - \int d^{d}x \sqrt{g_{d}
    } \, \sum_{n=1}^{\infty} \, \frac{\pi^{\frac{d-1}{2}}}{\beta^{d}} \int_0^\infty \frac{ds}{s^{\frac{d-1}{2}+1}} \,e^{-n^2 \frac{m_{\KK}^2 \beta^2}{4\pi^2} s}\,\theta_3(e^{-s})  \, .
\end{align}

For $N_T \gg 1$, the sum over $\rm{KK}$ levels can be approximated by a Gaussian integral, $\sum_{n=1}^{N_T} e^{-n^2 m_{\KK}^2\beta^2 s/4\pi^2} \simeq \int_0^\infty dn\, e^{-n^2 m_{\KK}^2 \beta^2 s/4\pi^2}$, which is performed explicitly to give
\begin{equation}
\sum_{n=1}^{N_T}
e^{- n^2 m_{\KK}^2\beta^2 s/4\pi^2}
\;\simeq\;
\frac{\sqrt{\pi}}{2}
\left(\frac{4\pi^2}{m_{\KK}^2\beta^2 s}\right)^{1/2}
\text{Erf}\!\left(\frac{\sqrt{s}}{2\pi}\right) .
\end{equation}
After regularization the leading behaviour comes from large $s$, where $\mathrm{Erf}(\sqrt{s}/(2\pi)) \to 1$, and the action takes the form at leading order 
\begin{align}
    I_{\KK} = - \int d^{d}x \sqrt{g_{d}
    }  \, \dfrac{1}{m_{\KK}\beta} \frac{\pi^{\frac{d-1}{2}}}{\beta^{d}} \int_0^\infty \frac{ds}{s^{\frac{d}{2}+1}}\pi^{3/2}\,\left(\theta_3(e^{-s}) -\dfrac{\sqrt{\pi}}{\sqrt{s}}\right) \, ,
\end{align}
where we also have subtracted a reference flat space heat kernel in order to cancel the UV divergence at $s\to0^{+}.$\footnote{This is equivalent to subtracting the $T\to 0$ thermal vacuum, as discussed in appendix \ref{ap:eisenstein}.} Both the final dependence on $\beta$ and the power of the Schwinger integral point towards the higher dimensional origin of the tower of states. In fact, if we recognize $\int\sqrt{g_{yy}} \,dy=2\pi R_{\rm KK}=2\pi\mKK^{-1}$, this is exactly the one loop potential for a single massless scalar in $d+1$ dimensions, including all order 1 factors. The relevant contribution to the effective action is therefore given by
\begin{equation}\label{eq:eff.action.species}
    I_{\KK} = -\frac{1}{d} |\sigma_{d+1}| \dfrac{2\pi}{m_{\KK}} \int d^{d} x \sqrt{g_{d}}\, \beta ^{-(d+1)}
\end{equation}

where $|\sigma_{d+1}|$ is related to the Stefan--Boltzmann in $(d+1)$ spacetime dimensions,  given by\footnote{Here $\sigma_{d+1}$ is the energy density of a single massless 
scalar, $\rho=\sigma_{d+1}T^{d+1}$, rather than the usual 
Stefan--Boltzmann constant, defined using the outgoing energy flux of a photon gas. The two are related by $\sigma_{\rm SB}=\frac{\Omega_d}{2\pi\Omega_{d-1}}(d-1)\,\sigma_{d+1}$, where $(d-1)$ comes from the number of polarizations of a photon in $d+1$ dimensions and the unit sphere ratio gives the density to flux projection.} 
\begin{equation}
\label{eq:sigmad}
    |\sigma_{d+1}| = \frac{d}{2}\pi^{\frac{d}{2}}\int_0^\infty \frac{ds}{s^{\frac{d}{2}+1}} \,\left (\theta_3(e^{-s})-\sqrt{\frac{\pi}{s}}\right ) = \frac{2d}{\Omega_d}\zeta(d+1)\,,
\end{equation}
where we have used $\Omega_d = 2\pi^{\frac{d+1}{2}}\Gamma\!\left(\frac{d+1}{2}\right)^{-1}$  and the definition of $\theta_3$, eq. (\ref{eq:theta3}). 

Since the one-loop energy of the $\rm{KK}$ tower naturally re-organizes as a higher-dimensional radiation gas, it is instructive to treat the $\rm{KK}$ circle on equal footing with the thermal circle from the start. In Appendix~\ref{ap:eisenstein} we carry out this computation by writing the Casimir energy of a massless scalar on $S^1_\beta \times S^1_{\KK}$ via the real-analytic Eisenstein series, which makes the modular $\beta \leftrightarrow R$ exchange manifest. After subtracting the $\beta \to \infty$ thermal vacuum \cite{Kapusta:2006pm}, the two relevant asymptotic limits are a $(d+1)$-dimensional radiation gas (for $\beta/R \to 0$, i.e.\ $T \gg m_{\KK}$, with contribution $I_1$) and a $d$-dimensional one with no support along the $\rm{KK}$ direction (for $\beta/R \to \infty$, contribution $I_2$). The ratio of these two pieces yields the identity
\begin{equation}\label{eq:I1.I2.ratio}
    \dfrac{I_1}{I_2} \simeq\dfrac{R}{\beta} = N_T\,,
\end{equation}
showing that the tower contribution to the effective action is parametrically $N_T$ times the lower-dimensional single-particle one. This result is one of the key motivations for the subsequent analysis.

With the effective action at hand, we can now compute the $(d+1)$-dimensional energy momentum tensor of the species in the limit $T/m_{\KK}\gg 1$. We work in a convention where the metric components are dimensionless
\begin{equation}\label{eq:metric.euclidean}
    ds^2 = g_{00}\, d\tau^2 + g_{ij} \,dx^i dx^j\,,
\end{equation}
and the time coordinate $\tau$ is periodic in a circle of length $\beta=1/T$. Making the local dependence explicit in (\ref{eq:eff.action.species}) one has
\begin{equation}
    I_{\KK} = -\frac{1}{d}|\sigma_{d+1}| T^{d+1} \int d^{d+1} x \sqrt{g_{d-1}} \left (\sqrt{g_{00}}\right )^{-d}\sqrt{g_{\KK}}
\end{equation}

The energy momentum tensor generated by this action is given by
\begin{equation}\label{eq:energy.momentum.species}
    T_{00} = \frac{2}{\sqrt{g}}\frac{\delta I}{\delta g^{00}} = -|\sigma_{d+1}| T^{d+1}g_{00}^{-\frac{d-1}{2}}\,, \quad T_{ij}=  \frac{2}{\sqrt{g}}\frac{\delta I}{\delta g^{ij}} =\frac{1}{d}|\sigma_{d+1}| T^{d+1} g_{00}^{-\frac{d+1}{2}} g_{ij} \,.
\end{equation}

In the rest frame of the metric (\ref{eq:metric.euclidean}), the four-velocity reads $u^{\mu} = (i/\sqrt{g_{00}},\vec{0})$, meaning that $T^{0}_{0}=-\rho$, $T^i_i=\rho/d$, where $\rho = |\sigma_{d+1}| T^{d+1} g_{00}^{-\frac{d+1}{2}}=|\sigma_{d+1}|T_{\text{loc}}^{d+1}$, with $T_{\text{loc}}\equiv T /\sqrt{g_{00}}$ the local temperature. Note that this energy-momentum tensor recovers the energy of the free species in a flat background after performing the integral over spacetime, indeed ${E \sim  L^{d-1}\,R\,\rho\sim N_T L^{d-1} T^{d}}$.

\subsection{Gravitational backreaction of a circle} \label{sec:self.grav.bh.tower}

Before turning to the detailed computation, let us remark a key difference between the BH--KK tower correspondence and the BH--string case already at the free-tower level. For the BH--string transition reviewed in section~\ref{sec:horowitz.polchinski}, the free string matches the minimal black hole in mass and entropy at the correspondence point, but their \emph{typical} sizes differ by $\sim \sqrt{N}$, where $N$ is the string excitation number. Resolving this size mismatch is precisely the motivation behind the Horowitz--Polchinski self-interacting solution. For the BH--KK tower transition \cite{Herraez:2024kux}, by contrast, the free analysis produces a match in mass, entropy \emph{and} characteristic size $L\simeq \Lsp^{-1}$ simultaneously with those of the minimal black hole. There is no analogous size mismatch to resolve. This is a strong indication that no self-interaction of the tower itself is needed, only the gravitational backreaction of the (otherwise free) $\rm{KK}$ gas on the geometry should matter. The remainder of this section is devoted to confirming this picture explicitly.  

We are thus interested in a setup with background $\mathbb{R}^{d-1}\times S^1_\beta$, and a configuration with a box of $\rm{KK}$ species at thermal equilibrium at temperature $T=1/\beta$. The classical Euclidean action vanishes \cite{Ortin:2015hya}, so the energy momentum tensor has a first contribution at one loop, given by \eqref{eq:energy.momentum.species}, and our goal is to compute how Einstein equations backreact due to its presence. In a perturbative expansion, which becomes an arbitrarily good approximation as we approach the infinite distance point where the $\rm{KK}$ tower becomes light, this gives the leading contribution to the energy-momentum tensor, with higher point interactions being subleading. 

For concreteness, from now on we focus on matching the $4d$ case with background $\mathbb{R}^{3}\times S^1_\beta$ and $N_T$ KK species, and the $5d$ case with background $\mathbb{R}^{3}\times S^1_\beta \times S^1_{\rm KK}$.  As we have argued above, the resulting energy-momentum tensor for such $4d$ theory  is intrinsically $d+1=5$ dimensional. We begin by solving the full Einstein equations in $5d$ with a circular spatial dimension and then make contact with the $4d$ description of having a $4d$ radiation gas with $N_T$ species. 

We start with the Euclidean action\footnote{In our setup this needs to be supplemented by boundary Gibbons Hawking terms and the action of the shell itself, which will be kept implicit throughout.}
\begin{equation}
    I = -\frac{1} {16\pi G_5}\int d^5x \sqrt{g_5}\, R + I_{\KK}\, ,
\end{equation}
which includes the Einstein Hilbert term and the one loop effective potential $I_{\KK}$.  Varying with respect to the metric yields Einstein's equations
\begin{equation}
    G_{\mu\nu} = 8\pi G_5 \, T_{\mu\nu}\,.
\end{equation}
We consider static euclidean gravitational configurations, with $S^2\times S^1$ symmetry and metric given by
\begin{equation}
ds^{2} = e^{2\psi(r)} d\tau^{2} + e^{2\Lambda(r)} dr^{2} + r^{2} d\theta^{2} + r^{2}\sin^{2}\theta d\varphi^{2} + e^{2\gamma(r)} dy^{2},
\end{equation}
where $y\sim y+2\pi R_{\text{KK}}$ is the coordinate along the $S^1_{\rm KK}$. It is worth mentioning that a black brane solution has trivial $\gamma(r)$, since a vacuum solution with $S^2\times S^1$ symmetry  and a nontrivial radion profile at tree level necessarily leads to singular horizons \cite{Ortin:2015hya}.
We assume an energy momentum tensor corresponding to radiation in 5 dimensions
\begin{equation}
T^{M}_{\,\,\,\,N} = \rho(r)\,\mathrm{diag}\left(-1,\frac{1}{4},\frac{1}{4},\frac{1}{4},\frac{1}{4}\right),
\end{equation}
such that its equation of state is $P=\rho/4$.

The independent Einstein equations are
\begin{equation}
\frac{e^{-2 \Lambda}}{r^2}-\frac{1}{r^2} + e^{-2 \Lambda} \gamma'' - e^{-2 \Lambda} \gamma' \Lambda' + e^{-2 \Lambda} \gamma'^2 + \frac{2 e^{-2 \Lambda} \gamma'}{r} - \frac{2 e^{-2 \Lambda} \Lambda'}{r} = -8\pi G_5 \rho,
\label{eq:ein15d}
\end{equation}
\begin{equation}
\frac{e^{-2 \Lambda}}{r^2}-\frac{1}{r^2} + e^{-2 \Lambda} \gamma' \psi' + \frac{2 e^{-2 \Lambda} \gamma'}{r} + \frac{2 e^{-2 \Lambda} \psi'}{r} = 2\pi G_5 \rho,
\label{eq:ein25d}
\end{equation}
\begin{equation}
\frac{e^{-2 \Lambda}}{r^2}-\frac{1}{r^2} - e^{-2 \Lambda} \Lambda' \psi' - \frac{2 e^{-2 \Lambda} \Lambda'}{r} + e^{-2 \Lambda} \psi'' + e^{-2 \Lambda} \psi'^2 + \frac{2 e^{-2 \Lambda} \psi'}{r} = 2\pi G_5 \rho.
\label{eq:ein35d}
\end{equation}
The remaining Einstein equations are redundant, as follows from the Bianchi identity together with stress tensor conservation.

The conservation equation
\begin{equation}
\nabla_M T^{M}_{\,\,\,\,N}=0\,,
\end{equation}
gives
\begin{equation}
\rho' = -5\rho\,\psi'\,,
\label{eq:5d.conservation}
\end{equation}
which integrates to
\begin{equation}
\rho(r) = \rho_0 e^{-5\psi(r)}\,.
\end{equation}

The warping factors satisfy nontrivial algebraic relations, as shown in Appendix \ref{ap:proof},
\begin{equation}\label{eq:relation}
\gamma' = -\psi'/4\,,
\end{equation}
and
\begin{equation}
\Lambda' = -\gamma' + r \frac{\psi'^2 + \psi''}{2 + r \psi'}\,.
\end{equation}
The system then reduces to a single differential equation for $\psi$, with $\gamma$, $\Lambda$, and $\rho$ determined algebraically.
Additionally, the equations admit a scaling symmetry
\begin{equation}
r \to \frac{r}{L}\,, \qquad \rho \to L^2 \rho\,,
\end{equation}
while $\psi$, $\Lambda$, and $\gamma$ remain invariant as functions of the rescaled argument. Therefore, if $(\psi_1,\Lambda_1,\gamma_1,\rho_1)$ is a solution on a unit length box,\footnote{The box can be modeled by an infinitely thin spherical shell that accounts for the pressure jump at the boundary, such that the total energy-momentum tensor is conserved. The pressure of the shell can be computed by means of the Israel junction conditions \cite{Israel:1966rt}, once Einstein's equations are solved inside and outside the box, as detailed in Appendix \ref{ap:boundary}, however the tension of the shell is in principle a free parameter. We choose it such that the metric is continuous in every direction, but different choices of tension contribute at most $\mathcal{O}(1)$ factors, and do not modify the parametric behaviour.} a solution with size $L$ is given by 
\begin{equation}
\psi(r) = \psi_1\!\left(\frac{r}{L}\right), \quad \Lambda(r) = \Lambda_1\!\left(\frac{r}{L}\right), \quad \gamma(r) = \gamma_1\!\left(\frac{r}{L}\right),
\end{equation}
\begin{equation}
\rho(r) = \frac{1}{L^2}\rho_1\!\left(\frac{r}{L}\right).
\end{equation}

It is convenient to introduce dimensionless variables $\tilde{r}=r/L\in[0,1]$,
\begin{equation}
\tilde \rho(\tilde{r}) = 8\pi G_5 L^2 \rho(L\tilde{r}), \quad \tilde \psi(\tilde{r}) = \psi(L\tilde{r}), \quad \tilde \Lambda(\tilde{r}) = \Lambda(L\tilde{r}), \quad \tilde \gamma(\tilde{r}) = \gamma(L\tilde{r})\, .
\end{equation} 
In these variables, the Einstein equations \eqref{eq:ein15d}-\eqref{eq:ein35d} keep the same functional form with $8\pi G_5 \rho \to \tilde\rho$. From now on we set $L=1$ and \emph{drop the tildes for simplicity},  except for the central density $\tilde{\rho}_c$ (which labels the solutions), restoring $L$ via the scaling symmetry at the end. Regularity at the origin is imposed by expanding
\begin{equation}
\psi \simeq \psi_0 + \psi_1 r + \psi_2 r^2+\ldots, \quad \Lambda = \Lambda_0 + \Lambda_1 r + \Lambda_2 r^2+\ldots, \quad \gamma = \gamma_0 + \gamma_1 r + \gamma_2 r^2+\ldots,
\end{equation}
and solving the Einstein equations order by order. This then implies
\begin{equation}
\Lambda_0 = 0, \quad \Lambda_1 = 0, \quad \gamma_1 = 0, \quad \psi_1 = 0.
\end{equation}
The central density $\tilde{\rho}_c=\tilde{\rho}(0)$, is a free parameter, while $\psi(0)$ and $\gamma(0)$ are fixed by matching the metric at the boundary at $r=L$ with the exterior black string solution.

Near the origin, for $r \ll 1/\sqrt{\tilde{\rho}_c}$, we find
\begin{equation}\label{eq:small.r}
\psi(r) \simeq \frac{\tilde{\rho}_c}{6} r^2, \qquad \tilde{\rho}(r) \simeq \tilde{\rho}_c - \frac{5}{6}\tilde{\rho}_c^2 r^2.
\end{equation}
We can additionally expand the Einstein equations at large radius, for $r \gg 1/\sqrt{\tilde{\rho}_c}$, to obtain
\begin{equation}
\psi(r) \simeq \frac{2}{5}\log r, \qquad \tilde{\rho}(r) \simeq \frac{4}{11}\frac{1}{r^2},
\end{equation}
with
\begin{equation}
\Lambda(r) \simeq \frac{1}{2}\log\frac{143}{100}.
\end{equation}

In our setup the radiation is localized in a region, so that at the interface at $r=L$ we match the metric to that of the vacuum solution
\begin{equation}
ds^2 = \left(1-\frac{2G_5 M}{(2\pi R_{\mathrm{KK}}) r}\right) d\tau^2 + \left(1-\frac{2 G_5 M}{(2\pi R_{\mathrm{KK}}) r}\right)^{-1} dr^2 + r^2 d\Omega_2^2 + dy^2,
\end{equation}
where we write the metric in terms of the total mass, not the mass density of the string. We can then obtain the total ADM mass as
\begin{equation}
M = \left(1-e^{-2\Lambda(L)}\right)\frac{L (2\pi R_{\mathrm{KK}})}{2 G_5} = \left(1-e^{-2\Lambda(L)}\right)\frac{L}{2 G_4}\, 
\end{equation}
where we have defined $G_4=G_5/(2\pi R_{\mathrm{KK}})$. Continuity of $g_{tt}$ and $g_{yy}$ then fixes
\begin{equation}
\label{eq:psiLambdaequality}
\psi(L) = -\Lambda(L), \qquad \gamma(L) = 0.
\end{equation}
The temperature at the box is fixed by the density through the
Stefan--Boltzmann law,
\begin{equation}
T_{\rm loc}(L) = \left(\frac{\rho(L)}{\sigma_{5}}\right)^{\frac{1}{5}},
\end{equation}
with $\sigma_{5}=\dfrac{3\zeta(5)}{\pi^2}$ the five dimensional Stefan--Boltzmann constant defined in \eqref{eq:sigmad}. This is then related to the temperature measured at infinity by
\begin{equation}
    T= T_{\rm loc}(L)e^{\psi(L)}=e^{-\Lambda(L)}\left(\frac{\rho(L)}{\sigma_{5}}\right)^{\frac{1}{5}},
\end{equation}
where the last equality uses  the matching condition \eqref{eq:psiLambdaequality}. The entropy can then be computed directly through the Clausius equation using this temperature and the ADM mass
\begin{equation}
S = \int \frac{dM}{T}.
\end{equation}
This gives the following asymptotic behaviour\footnote{This is easily generalized to $d+p$ total dimensions:
\begin{align}\label{eq:weakp}
M &\sim L^{d-1} \RKK^{\,p}\, \rho_c, \qquad T \sim \rho_c^{\frac{1}{d+p}}, \qquad \qquad \quad S \sim L^{d-1} \RKK^{\,p}\rho_c^{\frac{d+p-1}{d+p}}, \qquad \  \text{for}\ L\ll 1/\sqrt{G_{d+p}\,\rho_c}\, ; \\ 
M &\sim \frac{L^{d-3} \RKK^p}{G_{d+p}}, \qquad \quad T \sim \frac{1}{L^{\frac{2}{d+p}} G_{d+p}^\frac{1}{d+p}}, \qquad S \sim \frac{L^{\frac{(d+p)(d-3)+2}{d+p}}\RKK^{\,p}}{G_{d+p}^{\frac{d+p-1}{d+p}}},  \qquad \text{for}\ L\gg 1/\sqrt{G_{d+p}\, \rho_c}\, .
\end{align}} 
\begin{align}\label{eq:weak}
M &\sim L^3 \RKK \rho_c, \qquad T \sim \rho_c^{\frac{1}{5}}, \qquad \quad \ \  S \sim L^3 \RKK\rho_c^\frac{4}{5}, \qquad \text{for} \ L\ll 1/\sqrt{G_5\rho_c}\, ;\\
M &\sim \frac{L \RKK}{G_5}, \qquad \quad  T \sim \frac{1}{L^{\frac{2}{5}} G_5^\frac{1}{5}}, \qquad S \sim\dfrac{ L^{\frac{7}{5}}\RKK}{G_5^{\frac{4}{5}}}\, , \qquad \ \  \text{for}\ L\gg 1/\sqrt{G_5\rho_c}.
\end{align}

Far from these asymptotics, we solve the system numerically by fixing $L=1$ and scanning over $\tilde{\rho}_c$. The resulting profiles for the metric functions and thermodynamic quantities allow us to extract $T$, $M$, and $S$ as functions of $\tilde{\rho}_c$.
\begin{figure}
    \centering
    \begin{tabular}{c c}
       \includegraphics[width=0.49\linewidth]{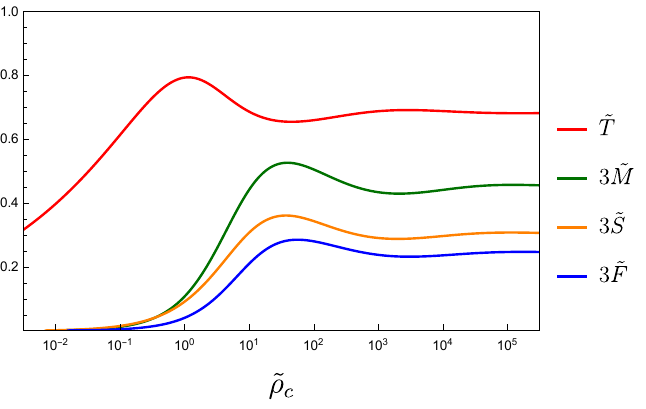}   &  \includegraphics[width=0.49\linewidth]{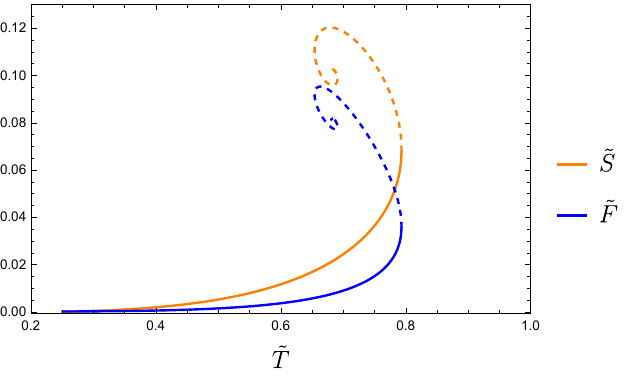} \\
    \end{tabular}

    \caption{\textit{Left:} Dimensionless Temperature $\tilde T$, Mass $\tilde M$, entropy $\tilde S$ and free energy $\tilde F$ vs the central density $\tilde \rho_c.$ The mass here was defined by matching the $g_{rr}$ coordinate to the exterior Schwarzschild $M=(1-e^{-2\Lambda(L)})\frac{\pi L\, \RKK}{G_5}$, the temperature was defined as $T=(\rho(L)/\sigma_5)^{\frac{1}{5}}e^{-\Lambda(L)}$ and the entropy as $dS=dM/T$. A factor of 3 has been added for better visibility. \textit{Right}: Free energy and entropy as a function of temperature. We have two branches, corresponding to small (solid) and large (dashed) $\rho_c$. The large density branch is actually more entropic after the maximum temperature, but it has larger free energy. This suggests that the large density branch is the less stable of the two.}
    \label{fig:tms1}
\end{figure}
The results are shown in Fig.~\ref{fig:tms1}. Here we see that the temperature reaches a maximum at lower $\tilde{\rho}_c$ than the mass and entropy, implying a change of sign of the specific heat. It becomes negative after the maximum temperature, signaling a thermodynamic instability, in analogy with the behaviour of self gravitating radiation in a box \cite{Sorkin:1981wd} and in AdS \cite{Page:1985em}. The free energy also supports this, as despite the large density branch being the more entropic of the two, it has larger free energy, and is therefore thermodynamically disfavored. The solution being unstable signals that the static ansatz is no longer well justified, and one should do a time-dependent computation. This is in fact the case for computing collapse in numerical relativity \cite{Choptuik:1992jv, Gundlach:2007gc}, which cannot be observed in a static ansatz. In this case, given the symmetry of the system, $S^2\times S^1$ with the $S^1$ wrapping the internal direction, the endpoint of collapse will be the black string.

It is convenient to introduce reduced variables for the temperature, mass, and entropy. These quantities are defined with respect to the $L=1$ solution and can subsequently be rescaled to obtain their values for arbitrary $L$. We define
\begin{equation}\label{eq:rescaling.T}
    \tilde T = (8\pi G_5)^{\frac{1}{5}}\sigma_{5}^{\frac{1}{5}}L^{\frac{2}{5}}T = \sigma_{5}^{\frac{1}{5}} (L\Lsp)^{-\frac{3}{5}} L T
\end{equation}
\begin{equation}\label{eq:rescaling.M}
    \tilde M = \frac{8\pi G_5 M}{L R_{\KK}} =(L\Lsp)^{-3} \frac{L}{R_{\KK}} L M
\end{equation}
\begin{equation}\label{eq:rescaling.S}
    \tilde S = \left(\dfrac{8\pi G_5}{L^3}\right)^{\frac{4}{5}}\sigma_5^{-\frac{1}{5}}\frac{L}{R_\KK} S = \sigma_5^{-\frac{1}{5}}(\Lsp L)^{-\frac{12}{5}} \frac{L}{R_\KK} S \,,
\end{equation}
where we used that the species scale is $\Lsp = M_{\text{Pl},5} = (8\pi G_5)^{-\frac{1}{3}}$. These also obey
\begin{equation}
    \tilde S = \int \frac{d \tilde M}{\tilde T}.
\end{equation}
The scaling then informs us of the behaviour of the solution at the correspondence point, where $\tilde T, \tilde M, \tilde S$ are of order 1. In this regime, the mass is given by
\begin{equation}
    M\simeq \dfrac{R_{\KK}}{G_5}L \simeq  L M_{\mathrm{Pl,4}}^{2}\,,
\end{equation}
which means $L$ is (parametrically) the Schwarzschild radius of the configuration. The instability occurs at $\tilde{T} \sim \tilde{M} \sim \tilde{S} \sim 1$ independently of $L$, and at this point \emph{compactness} parameter  $G_4 M/L$ (c.f. eq. \eqref{eq:compactness}) is of order one. Along the trajectory $T\sim 1/L$ that be impose below, this parameter coincides with the one controlling the heat kernel expansion, so the perturbative expansion is reliable precisely on the branch where the radiation gas is stable, and breaks down at the onset of the instability in the correspondence point. The connection to the correspondence point requires additionally imposing the trajectory $L \sim T^{-1}$, along which higher-order corrections remain under control \cite{Herraez:2024kux}. In the frozen species limit $L^{-1}\to T\to \Lsp$ we have that the temperature, mass and entropy match those of the minimal black hole around the point where the quantities are maximized. Since $\tilde T, \tilde M, \tilde S\simeq 1$ we have
\begin{equation}
    S\simeq \dfrac{\RKK}{L}=\dfrac{\Lsp}{\mKK}=\Nsp .
\end{equation}
A more complete treatment of the near-collapse regime would require two additional ingredients not addressed here: first, recomputing the equation of state self-consistently at each point in the backreacted geometry, rather than assuming the flat-space one-loop result. For the low-density branch this is justified, as the metric near the center of the configuration is approximately flat and the curvature remains small inside the box, but during collapse large curvatures are necessarily involved. Second, as mentioned already, relaxing the static ansatz entirely to follow the time-dependent collapse dynamics. Both of these are left for future work, but neither affects the thermodynamic diagnosis of the instability, which is already visible within the static solution on the low-density branch. The instability of the self-gravitating radiation gas and the matching of thermodynamic quantities at the species scale therefore together constitute the gravitationally backreacted version of the Black Hole-Tower Correspondence: the free species analysis of section \ref{s:blackholetower} is confirmed and extended to include gravitational interactions, with the black string solution emerging as the preferred phase at the correspondence point.

\subsection{Weak-backreaction limit and regime of validity}\label{sec:weak.backreaction}
Equation \eqref{eq:weak} implies that in the weak backreaction limit (i.e. $G_5 \rho_c L^2 \ll 1$) the mass distribution can be very well approximated by a constant density ball. Furthermore, the warping satisfies $\gamma' = -8\pi G_5\rho_c r/12 + \mathcal{O}(G_5\rho_c L^2)$ (see equations \eqref{eq:small.r} and \eqref{eq:relation}). Hence, in the weakly backreacted regime the time component of Einstein's equations yields
\begin{equation}
    e^{2\Lambda} \approx \frac{1}{1- 2\pi G_5 \rho_c r^2}\, .
\end{equation}
We can use this result to obtain an approximate expression for the entropy of the self-gravitating system. First, recall that this thermodynamic quantity is derived directly from the on-shell evaluation of the action. Second, since the Ricci scalar vanishes for the specific geometries under consideration (the energy-momentum tensor for radiation is traceless), the classical bulk action contribution drops out, leaving the one-loop effective action as the sole non-vanishing contribution to the partition function, which evaluates to
\begin{align}
    \log \mathcal{Z} = - I_{\KK} &= \frac{\sigma_{5}}{4} T^{5} \int d^{5} x \sqrt{g_{d-1}} (\sqrt{g_{00}})^{-4} \sqrt{g_{\KK}} \\
    &=\frac{\sigma_5}{4}\, T^4\, 2\pi \RKK \int_0^L 4\pi r^2\, dr\;
    e^{\Lambda+\gamma-4\psi} \nonumber\\
    &= \frac{2\pi^2}{3}\, \sigma_{5}\, N_T\, L^3 T^3\left(1+\mathcal{O}(G_5\rho_c L^2)\right)\, ,
\end{align}
where in the last step we used that, once the matching conditions at the boundary are
imposed, $\Lambda$, $\gamma$ and $\psi$ are all of order $G_5\rho_c L^2$ throughout
the box. This gives an entropy
\begin{equation}
S = \frac{\partial}{\partial T} (T\log\mathcal{Z})=\frac{10\pi^2}{3}\sigma_{5}N_T L^3 T^3 +\mathcal{O}(G_5\rho_c L^2)\,,
\end{equation}
whose first term coincides with the entropy of the free tower in the same spherical box obtained from taking the flat-space limit of eq. \eqref{eq:eff.action.species}, namely $S_{\rm free}=\tfrac{5}{4}\,\sigma_5\, T^4 \cdot \tfrac{4\pi}{3}L^3 \cdot 2\pi\RKK$. Parametrically, at $T\sim 1/L$, it also reproduces the box counting of \eqref{eq:purekkentropyenergy}. Self-gravitational corrections to this expression are controlled by the parameter $G_5\rho_c  L^2 \ll 1$. Importantly, this implies that the asymptotic dependence of the entropy on the number of species does \emph{not} change when gravitational interactions are taken into account, at least in a weakly coupled regime. It follows that at temperatures below the species scale, the gravitational backreaction of the gas of $\rm{KK}$ particles in a box is negligible. Consequently, if the string scale is parametrically lower than the species scale, the $\rm{KK}$ gas should first transition into a Horowitz--Polchinski string regime. It would then evolve into a black string, in the number of dimensions where it exists \cite{Emparan:2024mbp}. The finding that the gravitational backreaction is weak is key for this conclusion, as otherwise the configuration could not transition into a Horowitz--Polchinski regime.

This observation also clarifies a technical assumption underlying the HP solution of section~\ref{sec:horowitz.polchinski}. There, the metric is set to a flat background, with the dilaton and metric fluctuations argued to be of order $\epsilon^2$ using the scaling symmetry of the truncated theory. The weak-backreaction result above provides the dynamical justification for this choice. Namely, along the trajectory $T \sim 1/L$ that approaches the correspondence point from below, the parameter $G_5 \rho_c  L^2$ stays parametrically small all the way up to the 5d Planck scale, so the spacetime curvature inside the box is correspondingly small when $L\sim M_s^{-1}$, and the HP flat-metric ansatz is self-consistent throughout the entire $M_s \ll M_{\text{Pl},5}$ regime where HP applies.

\subsubsection*{Higher-order corrections: approaching the correspondence point}
An important point in this discussion is to determine the regime in which the approximations at hand remain valid, specifically when higher order terms in the heat kernel expansion of the one loop potential become significant. In general, the full one loop effective action can be written as
\begin{equation}
    I_{\KK} = \sum_{k\geq0} I^{(k)}_{\KK} \,,
\end{equation}
where
\begin{equation}
    I_{\KK}^{(k)} = -\pi^{\frac{d-1}{2}} T^{d-2k}\int d^dx \, \sqrt{g_{d-1}} \sqrt{g_{00}}^{2k-d+1}  a_{2k}(x) \sum_n^{N_T}\int_0^{\infty} \frac{ds}{s^{\frac{d-1}{2}+1-k}} e^{-\frac{n^2 m_{\KK}^2 \beta_{\rm loc}^2}{4\pi^2}s} \theta_3(e^{-s})\,, 
\end{equation}
where $\beta_{\rm{loc}} = \beta\sqrt{g_{00}}$ is the local inverse temperature, which controls the Boltzmann suppression of the KK levels. Summing over the tower then yields a factor $T_{\rm{loc}}/\mKK = N_T/\sqrt{g_{00}}$, whose redshift factor accounts for the
different power of $g_{00}$ in \eqref{eq:generic} below.

In appendix \ref{ap:heat.kernel}, we show that all terms in the heat kernel expansion can be written approximately as
\begin{equation}\label{eq:generic}
    I_{\KK}^{(k)} = -c_{d,k} N_T T^{d-2k} \int d^dx \, \sqrt{g_{d-1}} \sqrt{g_{00}}^{2k-d}  a_{2k}(x)\,,
\end{equation}
for some constant, order-one coefficients $c_{d,k}$, defined also in \ref{ap:heat.kernel}. In other words, for each term in the heat kernel expansion, species with masses of the same order or below the temperature can be effectively treated as massless. This approximation is valid as long as the number of species is parametrically large, namely $N_T \gg 1$. Using this, we can estimate the scaling of higher order terms in the heat kernel expansion. In particular,\footnote{Here $R^k$ denotes an invariant combination of $k$ Riemann tensors}
\begin{equation}
    a_{2k}(x) \sim R^{k} \sim \left ( \frac{G_d M}{L^{d-1}} \right )^{k} \sim \left ( G_d N_T T^{d}\right )^k\,.
\end{equation}

Using this, we find that
\begin{equation}
    I_{\KK}^{(k)} = c_{d,k} N_T T^{d-2k} \int d^dx \, \sqrt{g_{d-1}} \sqrt{g_{00}}^{2k-d}  a_{2k}(x) \sim N_T L^d T^{d-2k}\left ( G N_T T^{d}\right )^k \,.
\end{equation}
Then, we obtain
\begin{equation}\label{eq:hk.ratio}
    \frac{I_{\KK}^{(k)}}{I_{\KK}^{(0)}} \sim \left (G_d N_T T^{d-2} \right)^k \sim \left (N_T \left (\frac{T}{M_{\text{Pl},d}}\right )^{d-2} \right)^k\,.
\end{equation}
For temperatures $T \ll \Lsp$, the term in brackets is parametrically smaller than one, confirming the self-consistency of the solution in this regime. At the correspondence point $T \to \Lsp$, $N_T \to \Nsp$, and the ratio reduces to $(\Lsp/M_{\text{Pl},d})^{d-2} \Nsp = 1$. The perturbative expansion then breaks down near the correspondence point.

The structure of the analysis is in fact governed by two independent dimensionless parameters. The first is the \emph{compactness} of the matter distribution,
\begin{equation}\label{eq:compactness}
    \mathcal{C} \equiv \frac{G_d M}{L^{d-3}} \sim G_d N_T T^{d} L^2\,,
\end{equation}
which controls whether the radiation gas remains stable or undergoes gravitational collapse. The second is the EFT \emph{validity} parameter,
\begin{equation}\label{eq:eft.validity}
    \mathcal{V} \equiv G_d N_T T^{d-2} \sim N_T \left(\frac{T}{M_{\text{Pl},d}}\right)^{d-2}\,,
\end{equation}
which is just the $k=1$ term in the heat-kernel ratio \eqref{eq:hk.ratio} and controls the relative size of higher-order curvature corrections. These two are independent in general ($\mathcal{C}$ depends on $L$, $\mathcal{V}$ does not) but along the trajectory $T \sim 1/L$ that approaches the correspondence point they are identified, $\mathcal{C}\big|_{T=1/L} \simeq \mathcal{V}$, and both reach the common value
\begin{equation}\label{eq:critical.value}
    \mathcal{C}\big|_*\simeq \mathcal{V}\big|_*\simeq 1
\end{equation}
at $T \to \Lsp$, $L \to \Lsp^{-1}$. This is the precise version of the statement that the EFT validity and the onset of gravitational collapse coincide at the correspondence point, consistent with the free-tower analysis up to corrections of order $\Nsp^{-1/(d-2)}$.

\subsection{Gravitational backreaction of a gas of $\rm{KK}$ species}

Lastly, we complement the $5d$ analysis with a $4d$ perspective on the same problem. For $m_{\KK} \gg T$, only the $\rm{KK}$ zero mode is populated, the stress tensor reduces to $T^{\,\,M}_N = \rho\,\mathrm{diag}(-1,\tfrac{1}{3},\tfrac{1}{3},\tfrac{1}{3},0)$, the $G_{44}$ Einstein equation becomes redundant, and the system collapses to a standard $4d$ TOV equation \cite{Tolman:1939jz} with radiation equation of state $w=1/3$. This is a useful check on the 5d formalism. More interesting is the limit $m_{\KK} \ll T$, where the entire tower contributes. Plugging the 5d metric ansatz into \eqref{eq:ein15d} and using the algebraic relations \eqref{eq:relation} and \eqref{eq:small.r}, the terms that appear in the $4d$ eom yield a TOV-like sector
\begin{equation}
\frac{e^{-2 \Lambda}}{r^2}-\frac{1}{r^2}- \frac{2 e^{-2 \Lambda} \Lambda'}{r} = -8\pi G_4 \rho_4\,,\qquad
\frac{e^{-2 \Lambda}}{r^2}-\frac{1}{r^2} + \frac{2 e^{-2 \Lambda} \psi'}{r} = 8\pi G_4 P_4\,,
\end{equation}
with effective four-dimensional density and pressure
\begin{align}\label{eq:5to4d.EOS}
    \rho_4 = \dfrac{3}{4}(2\pi \RKK)\rho = \tilde\sigma_4 N_T T^4 \,,\qquad P_4 = \dfrac{5}{9}\,\rho_4\,,
\end{align}
where $\tilde\sigma_4=\frac{3\pi}{2}\sigma_5$, with $\sigma_5$ defined in eq.\eqref{eq:sigmad}, and recall $\rho$ is the 5d energy density. The unexpected feature is the \emph{exotic} equation of state $w_4 = 5/9$, in contrast with the naive expectation $w_4 = 1/3$ for $N_T$ copies of $4d$ massless radiation. The reason for this is the warping of the KK circle, i.e., $\gamma^\prime\neq 0$. The $\gamma$-terms in eq. \eqref{eq:ein15d} capture the pressure of the gas along the compact direction, and feed it back into the effective $4d$ energy-momentum tensor changing the pressure-to-density ratio from the standard value. Let us remark that this identification of $w_4=5/9$ only holds at leading order in the weak-backreaction or near-origin expansion, $r \ll 1/\sqrt{G_4 \, \rho_4(0)}$. Beyond this order the system cannot be described by a perfect fluid with constant $w_4$, as can be seen from the conservation laws. The exact 4d fluid with constant $w_4=5/9$ would yield $\rho_4 \sim e^{-14\psi/5}$, whereas the exact 5d conservation law gives $\rho\sim e^{-5\psi}$ (cf. \eqref{eq:5d.conservation}). These agree only at leading order, where the density is effectively constant. Consistently, the equation of state departs from $w_4=5/9$ on the large-density branch, as noted below. Within its regime of validity, $w_4=5/9$ is a smoking gun for the higher dimensional origin of the gas, which a naive 4d treatment of $N_T$ species misses. However, as we show in the following, the parametric dependence of the key thermodynamic quantities on $N_T$ coincides with the one captured by naive 4d analysis with $N_T$ species. In the rest of this subsection we work with a general equation of state $P_4 = w \rho_4$, keeping in mind that the case of interest is $w=5/9$.

Since the ADM mass at the boundary depends only on $\Lambda(L)$,
\begin{equation}
    M\simeq \dfrac{1}{2 G_4}L(1-e^{-2\Lambda(L)})\, ,
\end{equation}
the equation of state $w$ enters only through the definition of the temperature. That is, the exact value of $w$ affects overall $\mathcal{O}(1)$ constants only, not the parametric dependence on $L$, $N_T$, or $T$. We therefore expect the $5d$ thermodynamic results to extend to $4d$ up to $\mathcal{O}(1)$ factors, as confirmed with the numerical analysis below. In the near-origin regime $r \ll 1/\sqrt{G_4 \, \rho_4(0)}$, the surviving metric components are
\begin{equation}
    \psi = \psi_0 + \dfrac{3w+1}{12} 8\pi G_4\rho_{4}(0) \ r^2\, , \qquad  \Lambda = \dfrac{1}{6} 8\pi G_4\rho_{4}(0)\ r^2 \,.
\end{equation}

In the opposite branch, for $r\gg 1/\sqrt{G_4 \, \rho_4(0)}$, the system is no longer described as an ideal fluid, and the 4d conservation equation is broken, signaling that the purely $4d$ interpretation breaks down. This is consistent with the $5d$ diagnosis that the small-density branch is the thermodynamically preferred one.

To examine the $4d$ interpretation in the regime where it is valid, we put the configuration inside a spherically symmetric box of size $L$, with ansatz 
\begin{equation}
    ds^2 = e^{2\Phi}d\tau^2 + e^{2\Lambda} dr^2 + r^2d\Omega^2_{2}\,,
\end{equation} 
The time component of Einstein's equations gives \cite{Misner1973}
\begin{equation}\label{eq:mass.inside}
    e^{2\Lambda} = \frac{1}{1-\frac{2 G_4 m(r)}{r}}\,,\qquad m(r) = \int 4\pi r^{2} \rho_{4}(r)\, dr\,,
\end{equation}
with $m(r)$ the total mass-energy inside radius $r$ \cite{Misner1973}. Combined with covariant conservation of the energy-momentum tensor, $\nabla_\mu T^{\mu\nu} = 0$, the $rr$ component yields the TOV equations \cite{Oppenheimer:1939ne,Tolman:1939jz}
\begin{align}
\label{eq:TOVequations1}
    \frac{dm}{dr} &= 4\pi r^{2}\rho_{4}\,,\\
    \frac{d\rho_{4}}{dr}&=-\frac{1+w}{w}\,G_4\, \rho_{4}\, \frac{m+4\pi w r^3\rho_{4}}{r^2\left( 1-\frac{2G_4 m}{r} \right) }\,,
    \label{eq:TOVequations2}
\end{align}
where regularity at the origin fixes $m(0)=0$ and the central density $\rho_{4}(0)=\rho_c$ labels the family of solutions. Finding the backreacted metric is equivalent to integrating this system.

Numerically integrating \eqref{eq:TOVequations1}-\eqref{eq:TOVequations2} from the centre $(m(0)=0,\,\rho_{4}(0)=\rho_c)$ outward and matching at $r=L$ to an exterior Schwarzschild geometry with mass $m(L)$ via an infinitely thin shell that compensates the pressure jump (as in the $5d$ case), the local-temperature continuity condition at the boundary reads
\begin{equation}
    T_{\text{loc}}(L) = \left (\frac{\rho_{4}(L)}{\tilde\sigma_4 N_T(L)}\right)^{\frac{1}{4}} = T \left (1-\frac{2 G_4 m(L)}{L}\right )^{-\frac{1}{2}}\,,
\end{equation}
where the number of active species is itself set by the local temperature, $N_T(L) = T_{\text{loc}}(L)/\mKK$ (the KK mass is not redshifted at the wall, since the 5d matching fixed $\gamma(L)=0$). Substituting, the first equality becomes $\rho_{4}(L) = (\tilde\sigma_4/\mKK)\,T_{\text{loc}}(L)^5$, which is consistent with the 5d scaling $\rho = \sigma_5\, T_{\text{loc}}^{5}$ after using \eqref{eq:5to4d.EOS}. The temperature is then given by
\begin{equation}
    T = \left ( \frac{\rho_{4}(L) \mKK}{\tilde\sigma_4}\right)^{\frac{1}{5}}\left (1-\frac{2 G_4 m(L)}{L}\right )^{\frac{1}{2}}\,.
\end{equation}

As in the $5d$ case, the TOV equations enjoy a scaling symmetry \cite{Vaganov:2007at} that allows us to introduce reduced quantities
\begin{align}
    \tilde \rho_{4}(r) &= 8\pi G_4 L^2 \rho_{4} \left (Lr\right )\,, \\
    \tilde m(r) &= \frac{8\pi G_4 m\left (Lr \right)\,}{L}\, ,
\end{align}
for $r\in [0,1]$. 

The rescalings of temperature, mass and entropy from the 5d solution \eqref{eq:rescaling.T}-\eqref{eq:rescaling.S} apply here with $\sigma_5 \to \tilde\sigma_4$. The numerical integration reveals the same structure as in $5d$, we refer to the existing literature \cite{Sorkin:1981wd, Page:1985em, Vaganov:2007at, Chavanis:2007kn, Hammersley:2007ahw, Li:2008xw, Li:2009jq} for plots similar to those in Figure \ref{fig:tms1}. As shown there, the rescaled temperature and mass increase with $\rho_c$, reach a peak, and then enter a regime of damped oscillations, in close analogy with self-gravitating radiation in AdS space. The $4d$ description is strictly valid only at small central densities, $G_4 \rho_c L^2 \ll 1$. Within this regime, the maxima of the rescaled quantities $\tilde T,\tilde M,\tilde S \simeq 1$ fix the correspondence point, where in the frozen limit $L\to 1/T$ the temperature reaches $T\simeq \Lsp$ and the entropy reaches
\begin{equation}
    S\simeq \Nsp \,.
\end{equation}
This is the parametric $4d$ counterpart of the $5d$ result above. The fact that the parametric dependence on $N_T$ is unchanged, despite the non-trivial $w=5/9$, confirms that the exact value of $w$ only affects $O(1)$ prefactors. Therefore, the $5d$ and $4d$ analyses are parametrically equivalent in their common regime of validity. In the weak-backreaction limit, the entropy is given by the number of active $\rm{KK}$ species $N_T$ up to corrections controlled by $G_4 \rho_c L^2 \ll 1$.

\section{The Connection between the Self-Gravitating Towers and Black Holes}
\label{s:connection}
In sections \ref{s:blackholetower} and \ref{sec:notfree.species} we have established the following thermodynamic statement. The free and subsequently self-gravitating KK tower, followed along the trajectory $T\sim 1/L$, reaches the correspondence point with the same mass, temperature, size and entropy as the minimal black hole (or black string), with $S\simeq \Nsp$. However, the matching of thermodynamic variables at a point does not guarantee that both phases be continuously connected. That, is, it does not answer whether there exists a one-parameter family of configurations that interpolates between them, or if instead some invariant distinguishes them and forces the transition to proceed through a genuine jump. This is the main question behind \cite{Chen:2021dsw} for the black hole-string transition, and in this section we extend their two complementary approaches to the black hole-tower case.

The two solutions differ in the topology of their associated Euclidean saddles. In the Euclidean black hole the thermal circle caps off at the horizon, yielding a cigar-like geometry with topology $B^2\times S^{d-2}$, whereas in the Horowitz--Polchinski and the free string (or tower) it does not shrink, yielding a cylinder $\mathbb{R}^{d-1}\times S^1$. The first probe to address the question is the worldsheet. A linear sigma model is a 2d supersymmetric field theory that in the IR flows to a non-linear sigma model on its vacuum manifold. If a single family of linear sigma models interpolates between the two topologies as its parameters are varied, it provides an explicit continuous path of worldsheet theories that connects both phases. On the contrary, if a deformation invariant, such as a protected index, jumps between the two phases, no such continuous path exists within that family of theories, signaling an obstruction. 

The second probe to address the question is the target space topology itself, independently of any worldsheet construction. In this case, two key questions diagnose the transition. First, whether the two manifolds are cobordant, such that there exists a $(d+1)$-dimensional manifold that interpolates between the two. This is a kinematic pre-requisite for the existence of a finite-action, topology-changing process. Second, one can ask whether the two backgrounds support the same lattices of conserved brane charges, which are captured by their bordism groups and refine the K-theory comparison in \cite{Chen:2021dsw}. The answers to these two questions need not be the same, and indeed they turn out to be different for our backgrounds. The brane-charge lattices for both backgrounds differ, which may constitute a perturbative obstruction to the transition, whereas both belong to the same bordism class in all the structures we checked, meaning there is no topological obstruction to a potentially charge-violating, non-perturbative process, connecting the two. This last process is of the kind predicted by the Cobordism Conjecture \cite{McNamara:2019rup}. 

In section \ref{sec:lsm.review} we review the linear sigma model construction of \cite{Chen:2021dsw}, and then construct the corresponding models for the black hole-tower transition in section \ref{sec:lsm.tower}. Then, we compute cobordism classes and brane-charge lattices of the two phases in \ref{sec:bordism}. We finally combine the worldsheet and spacetime perspectives in section \ref{sec:lsmandbordism}.

\subsection{Review of sigma models for Horowitz-Polchinski/Black Hole transition}
\label{sec:lsm.review}
In \cite{Chen:2021dsw} it was shown that a linear sigma model can smoothly interpolate between the black hole and Horowitz-Polchinski solutions in heterotic string theories, but not in Type II. In the following, we briefly review their construction.

\subsubsection*{Linear sigma model for the Heterotic String}

The worldsheet theory of a heterotic string has $(0,1)$ supersymmetry. This can be realized in a superspace with bosonic coordinates $x^-$,$x^+$ and a single fermionic coordinate $\theta = \theta^+$. The model considered in \cite{Chen:2021dsw}, includes $n$ scalar superfields $\Phi_i(x,\theta)=\phi_i(x)+i\theta\psi_{+,i}(x)$, and $m$ Fermi superfields $\Lambda_\alpha(x,\theta)=\lambda_{-,\alpha}+\theta F_{\alpha}$, where $\phi_i$ are real scalars, $\psi_{+,i}$ and $\lambda_{-,\alpha}$ are fermion fields of the indicated chirality, and $F_\alpha$ are auxiliary fields. To generate a potential, one can further include a term of the form
\begin{equation}\label{eq:full.potential}
    \int d^2 x\, d\theta \sum_{\alpha=1}^m \Lambda_\alpha W_\alpha(\Phi_i),
\end{equation}
for some functions $W_{\alpha}$. After integrating over $\theta$ one finds 
\begin{equation}
    \int d^2x \sum_\alpha F_\alpha W_\alpha(\phi_i)\,+\, i\sum_{\alpha,i} \frac{\partial W_\alpha}{\partial\phi_i}\lambda_{-,\alpha}\psi_{+,i}.
\end{equation}
After integrating out the auxiliary field $F$, the first term gives rise to a potential of the form
\begin{equation}\label{eq:potential}
    V(\phi_i) = \frac{1}{2}\sum_\alpha W_\alpha(\phi_i)^2.
\end{equation}
At low energies, the model flows to the vacuum manifold
\begin{equation}\label{eq:general.locus}
    W_1(\phi)=\dots=W_m(\phi_i)=0,
\end{equation}
and, if $n>m$, it will lead to a non-linear sigma model with a target space of dimension $n-m$.

Since both the black hole and HP string are $d$-dimensional, one can use a linear sigma model with $n=d+1$ scalar superfields and $m=1$ Fermi superfields. The superfields are organized into a $(d-1)$-plet $\vec{\mathcal{Y}}$ and a two component vector $\vec{\mathcal{X}}$, so that their bottom components read
\begin{equation}
    \mathbf{Y}=(Y_1,\dots, Y_{d-1}), \quad \mathbf{X} = (X_1,X_2).
\end{equation}
The function $W$ is chosen to be \cite{Chen:2021dsw}
\begin{equation}
    W(\mathbf{X},\mathbf{Y})\ = \mu\left((\mathbf{Y}^2+a)(\mathbf{X}^2-b)+c \right ),
\end{equation}
where $a$, $b$ and $c$ are dimensionless positive constants and $\mu$ is a constant with units of mass. The target space is then given by 
\begin{equation}\label{eq:locus}
    (\mathbf{Y}^2+a)(\mathbf{X}^2-b)+c=0,
\end{equation}
which can be rewritten as
\begin{equation}\label{eq:locus2}
    \mathbf{X}^2=b-\frac{c}{\mathbf{Y}^2+a}.
\end{equation}
First, note that as $|\mathbf{Y}|$ increases, $|\mathbf{X}|$ tends to the constant value $b^{1/2}$, while for decreasing $|\mathbf{Y}|$, $|\mathbf{X}|$ becomes smaller. Whether this describes a Euclidean Black Hole or the HP string depends on whether $\mathbf{X}$ is allowed to reach $0$ or not. Intuitively, this corresponds to whether the geometry caps-off, forming a cigar-like geometry, or remains open, giving rise to a cylindrical one.

Since $\mathbf{X}^2$ is non-negative, $|\mathbf{Y}|$ cannot take any value in general. Concretely, if $c>ab$, one has $\mathbf{Y}^2\geq c/b-a$. In this situation, one finds that $\mathbf{X}$ is restricted to take values in a disk of radius $b^{1/2}$ and, once the value of $\mathbf{X}$ is set, $\mathbf{Y}$ is restrained to a $({d-2})$-sphere. Hence, the topology is given by $B^2\times S^{d-2}$, which corresponds to the euclidean black hole solution, with $B^2$ the $2$-ball.  On the other hand, for $c<ab$, $\mathbf{Y}$ can take any value and, once $\mathbf{Y}$ is fixed, the possible values of $\mathbf{X}$ form a circle $S^1$, as can be seen in equation (\ref{eq:locus}). This indicates that the topology is given by $\mathbb{R}^{d-1}\times S^1$, corresponding to the HP or free-string solutions. Lastly, it is worth considering what happens when $c=ab$. In this case, the target space (\ref{eq:locus}), has a conical singularity at $\mathbf{X}=\mathbf{Y}=0$. A way to see this is by expanding the potential around this point. One finds then,
\begin{equation}
    a\mathbf{X}^2-b\mathbf{Y}^2=0
\end{equation}
which is the equation of a cone. 

Some intuition can be gained about what this sigma model represents by plotting equation (\ref{eq:locus}) at specific values of $a,b,c$.  Indeed, one finds that the locus takes the shape of a cigar for $c > ab$, that of a cylinder for $c<ab$, and a conical singularity arises at the critical point $c=ab$.
\subsubsection*{Linear sigma model for Type II strings}

In the case of type II strings, the worldsheet theory has $(1,1)$ supersymmetry, which requires a superspace with two bosonic coordinates $x^+,x^-$ and two fermionic ones $\theta^+,\theta^-$. In this case, it is enough to consider scalar superfields, which take the form $\Phi(x,\theta)=\phi(x)+i\theta^+\psi_+(x)+i\theta^-\psi_-+i\theta^+\theta^-F$, where $\phi$ is a real scalar field, $\psi_{\pm}$ are fermionic fields of the indicated chirality, and $F$ is an auxiliary field. In addition, there is also a $\mathbb{Z}_2$ chiral $R$-symmetry $\tau$ required for the GSO projection \cite{Seiberg:1986by}, which acts on the fermionic coordinates as $\theta^{\pm}\rightarrow\pm\,\theta^{\pm}$. As a consequence, any potential of the form (\ref{eq:full.potential}), has to be invariant under this symmetry. Since the measure $d^2\theta$ is odd under it, the superpotential $W$ is also required to be $\tau$-odd. For these purposes, it is useful to consider fields that are even or odd under the $\tau$ symmetry. Specifically, we consider $n$ even superfields $\Phi_i$, which transform as
\begin{equation}
    \Phi_i(x,\theta^+,\theta^-) \rightarrow  \Phi_i(x,\theta^+,-\theta^-)\,,
\end{equation}
and $m$ odd superfields $\mathcal{P}_\alpha$ that transform as
\begin{equation}
    \mathcal{P}_\alpha(x,\theta^+,\theta^-) \rightarrow - \mathcal{P}_\alpha(x,\theta^+,-\theta^-)\,.
\end{equation}
Then, one can write the following $\tau$-odd potential
\begin{equation}
    W = \sum_\alpha \mathcal{P}_\alpha W_{\alpha}(\Phi_i)\,.
\end{equation}
Now, we integrate over the fermionic coordinates, keeping only the pure bosonic terms, one obtains
\begin{equation}
    i\int d^2x \,\left(\sum_\alpha G_\alpha W_\alpha(\phi_k) + \sum_{\alpha,i} F_iP_\alpha\partial_iW_\alpha(\phi_k)\right)\,.
\end{equation}
Here, $P_\alpha$, $\phi_i$ are the bottom components of $\mathcal{P}_\alpha$ and $\Phi_i$ and, $G_\alpha$ and $F_i$ are their respective auxiliary fields. Integrating out $G_\alpha$ and $F_i$, one finds the following potential
\begin{equation}
    V(\phi_i)=\frac{1}{2}\sum_\alpha W_\alpha(\phi_k)^2+\frac{1}{2}\sum_i \left ( \sum_{\alpha} P_\alpha \partial_i W_\alpha(\phi_k)\right )^2\,.
\end{equation}
In order to have $V(\phi_i)=0$, it is required not only that equation (\ref{eq:general.locus}) is satisfied, but also that $P_\alpha \partial_i W_\alpha(\phi_k)=0$. In general, this last equality only holds when $P_\alpha = 0$, except for when the matrix whose entries are given by $\partial_i W_\alpha(\phi_k)$ does not have full rank. In other words, when the Jacobian of the map $f : \mathbb{R}^n \rightarrow \mathbb{R}^{m}$, defined by
\begin{equation}
    f(\phi_k)=(W_1(\phi_k),\dots,W_m(\phi_k)),
\end{equation}
does not have full rank when evaluated at a given point in $f^{-1}(0)$. When this happens, $0$ is not a regular value of $f$, and a singularity in the target space can appear. At such points, a new direction in the vacuum manifold parameterized by the $P_\alpha$ opens up, and the $R$-symmetry is spontaneously broken.

In the case of the black hole-string transition, one can consider a model with $n=d+1$ scalar even superfields, called
$\vec{\mathcal{Y}},\vec{\mathcal{X}}$ in the same way as for the heterotic string and a $\tau$-odd superfield $\mathcal{P}$. Their bottom components are $\mathbf{X}$, $\mathbf{Y}$ and $P$, respectively. The superpotential is chosen to be the same as in the (0,1) case\footnote{The choice of potential here is arbitrary, so some statements are model-dependent; however the general lessons are expected to hold regardless.}
\begin{equation}
    W = \mu \, \mathcal{P}\left((\vec{\mathcal{X}}^2-b)(\vec{\mathcal{Y}}^2+a)+c\right).
\end{equation}
One finds the constraint,
\begin{equation}\label{eq:locus3}
    (\mathbf{Y}^2+a)(\mathbf{X}^2-b)+c=0,
\end{equation}
which is also present in the case of the heterotic string, together with the extra constraint
\begin{equation}
    P{\mathbf{X}}({\mathbf{Y}}^2+a)=P{\mathbf{Y}}({\mathbf{X}}^2-b)=0.
\end{equation}

For $P=0$ one gets the same geometric branch, cigar or cylinder depending on the couplings, found in the heterotic counterpart. However, a new branch appears for $P\neq 0$. In this latter case, the other fields can only take the value $\mathbf{X}=\mathbf{Y}=0$, and the locus eq. \eqref{eq:locus3} also imposes $c=ab$. Classically, a new branch of vacua parameterized by $P$ opens up precisely at the transition point, which is where the target space develops a conical singularity, and the $R$-symmetry is spontaneously broken along this branch. Note that this is a classically flat direction with no clear geometric interpretation away from the point $P=0$.
This classical picture, although suggestive, is incomplete, as emphasized in \cite{Chen:2021dsw}. Since the branch $P\neq0$ spontaneously breaks the $R$-symmetry, an effective superpotential $W_{\text{eff}} (P)$ is generated after the massive fields $\vec{\mathcal{X}}$ and $\vec{\mathcal{Y}}$ are integrated out, and the classical flat direction gets lifted to a finite set of massive supersymmetric vacua located at exponentially large values of $|P|$. Counting these massive supersymmetric vacua at fixed couplings, one finds two of them in the HP background ($c<ab$) and either zero or four in the black hole background ($c>ab$), so that the number jumps by two across the transition, signaling an obstruction. This obstruction can be understood as follows. The Witten index of a $(1,1)$ sigma model on a compact target space\footnote{Note that we are considering the corresponding compact versions of the BH and HP/free string saddles, as explained in more detail around eq. \eqref{eq:compact.topologies}.} equals its Euler characteristic, which in our case takes the \emph{different} values ${\chi(S^2\times S^{d-2})=2\,  (1+(-1)^d)}$, and ${\chi(S^{d})=1+(-1)^d}$, and an index cannot change under continuous deformations of the theory unless states run in/out to/from infinity in field space. The transfer of the massive vacua in the $P\neq 0$ branch through $|P|\to \infty$ at the transition point ($c=ab$) is precisely the  crossing through a non-compact theory, allowing for the index jump. For odd $d$ both Euler characteristics vanish, but the same conclusion follow from a refined index, namely the Lefschetz number of a reflection symmetry \cite{Chen:2021dsw}. Given a symmetry $T$ that commutes with the supercharges, one can define this index in the same way as the Euler characteristic, weighting the supersymmetric ground states by the action of $T$.  Taking $T: \mathcal Y_1 \to -\mathcal Y_1$ this gives ${L_T(S^2\times S^{d-2})=2\,  (1-(-1)^d)}$, and ${L_T(S^{d})=1-(-1)^d}$, so one can reach the same conclusion for odd $d$. Therefore, there is no smooth family of $(1,1)$ linear sigma models interpolating between the two phases. On the contrary, the heterotic family finds no obstruction, since no flat direction opens at the transition point (since there is no $R$-symmetry that forces the inclusion of a $\mathcal{P}$ field), and there are no invariants that change as one crosses from one phase to the other.

The distinction between the two string theory embeddings can be summarized in a way that anticipates the rest of this section. In the heterotic case, the protected invariants of the $(0,1)$ worldsheet theory are believed to be cobordism invariants of the cobordism class of the target space \cite{Chen:2021dsw}. As explained in section \ref{sec:bordism}, the two saddle topologies are cobordant, so no protected invariant can change, consistent with the smooth interpolation displayed by the linear sigma model. In type II, on the other hand, the protected index is the Euler characteristic of the target space. The latter counts the non-trivial cycles weighted by the parity of their dimensionality. As we will see in section \ref{sec:bordism}, each such cycle also carries its own tower of wrapped brane charges, so the jumping index appears to track the brane charge content of the background. At this point this is simply a suggestive coincidence, which we will make precise and exploit in section \ref{sec:lsmandbordism}. The relation between this number and the D-brane charges supported by the target space is precisely what \cite{Chen:2021dsw} explain using the framework of K-theory, and as we reviewed the number jumps by two units across the transition point, as also detected by the linear sigma model via the transfer of the two massive vacua of the $P\neq 0$ branch through infinity as one necessarily crosses the non-compact theory. This obstruction is thus related to the charges that the background can support, rather than to the cobordism class of the background itself. Section \ref{sec:bordism} makes this distinction precise, showing how the cobordism classes of the two backgrounds agree in all the structures checked, thus leaving room for a non-perturbative, charge-violating process of the kind the Cobordism Conjecture proposes.

\subsection{Linear sigma models for the BH Tower correspondence}
\label{sec:lsm.tower}
The Black Hole–Tower correspondence motivated the  construction of a linear sigma model for the transition in the formalism of the heterotic string. We choose a model with $n=d+3$ scalar superfields, where $d$ is the number of non-compact directions and $m=2$ fermi superfields, so that the target space has $n-m=d+1$ dimensions. We organize the superfields in a $(d-1)$-dimensional vector $\vec{\mathcal{Y}}=(\mathcal{Y}_1,\dots,\mathcal{Y}_{d-1})$ and two doublets: $\vec{\mathcal{X}}=(\mathcal{X}_1,\mathcal{X}_2)$ related to the thermal circle, $\vec{\mathcal{Z}}=(\mathcal{Z}_1,\mathcal{Z}_2)$, representing the $\rm{KK}$ circle. We call the fermi superfields $\Lambda_1$ and $\Lambda_2$, so that the potential term in the action reads
\begin{equation}\label{eq:potential_tower}
    \int d^2 x d\theta \, \left (\Lambda_1 W_1(\vec{\mathcal{X}},\vec{\mathcal{Y}},\vec{\mathcal{Z}}) +\Lambda_2 W_2(\vec{\mathcal{X}},\vec{\mathcal{Y}},\vec{\mathcal{Z}})\right).
\end{equation}
We choose the functions $W_1$ and $W_2$ to be
\begin{equation}
    W_1 = \mu_1\left((\vec{\mathcal{X}}^2 -b)(\vec{\mathcal{Y}}^2+a)+c\right),
\end{equation}
and
\begin{equation}
    W_2 = \mu_2 \left( \vec{\mathcal{Z}}^2-e\right ).
\end{equation}
where $\mu_1$ and $\mu_2$ are constants with dimensions of energy, and $a,b,c,e$ are dimensionless positive constants. We call the bottom components of the scalar superfields $\textbf{Y}=(Y_1,\dots,Y_{d-1})$, $\textbf{X}=(X_1,X_2)$ and $\textbf{Z}=(Z_1,Z_2)$. Integrating equation (\ref{eq:potential_tower}) in the fermionic variables, we obtain a potential of the form (\ref{eq:potential}), which gives rise to a non linear sigma model with a target space defined by
\begin{equation}\label{eq:locus4}
    (\mathbf{Y}^2+a)(\mathbf{X}^2-b)+c=0,
\end{equation}
and
\begin{equation}
    \textbf{Z}^2-e=0.
\end{equation}
The first equation is the same as that for the regular black hole-string transition. Hence, we deduce that for $c>ab$, this parametrizes a $(d+1)$-dimensional Euclidean black string, topologically $B^2\times S^{d-2}\times S^1_{\KK}$. On the other hand, the complementary regime, $c<ab$, describes a $\mathbb{R}^{d-1}\times S_\beta^1 \times S^1_{\KK}$ topology, where $\beta$ and $\KK$ refer to the thermal and $\KK$ circles. Depending on the size of the circles, this parametrizes either the HP solution with a tower, or the tower backreacting on itself. When $|\mathbf{X}|>|\mathbf{Z}|$ (equivalently $\mKK>T$) the HP solution dominates, while for $|\mathbf{X}|<|\mathbf{Z}|$ (or also $\mKK<T$),  this is the tower backreacting on itself. This family of worldsheet theories thus recovers the phase structure presented in \ref{sec:regimes} and \ref{sec:notfree.species}. The black string saddle corresponds to the minimal black hole limit, the regime $\mKK\gg T$ describes the $d$-dimensional Horowitz-Polchinski/free string saddle that does not resolve the extra circle, and the regime $\mKK\ll T$ includes the self-gravitating tower phase (when embedded in  weakly coupled string theory setup, captured by such a worldsheet theory, which requires $\mKK \ll M_s\lesssim M_{\mathrm{Pl,}d+1}$). Since the heterotic linear sigma model interpolates smoothly between the two saddles, the smoothness found in \cite{Chen:2021dsw} for the black hole-string transition carries over to the black hole-tower transition.

In the case of Type II the same type of obstruction remains when adding an extra compact circle. The latter enters as a spectator, identical in the two saddles. Since $\vec{\mathcal{Z}}$ is a spectator field, we can decouple it from the rest and treat it as a separate $(1,1)$ SCFT. Its supersymmetric ground states are $H^*(S^1)$, of dimension $b_0 + b_1 = 2$, with two massless states of opposite fermion parity \cite{Witten:1982df, Witten:1982im}. This doubles the number of vacua on both the geometric and $P\neq0$ branches so that one has either 0 or 8 on the black hole side, and 4 on the string/HP side. Although strictly speaking these are not massive supersymmetric vacua, as they have a massless sector, they are still supersymmetric ground states. Since the Euler characteristic vanishes for a $S^1$ we can consider the index associated to $T_Z:\mathcal Z_1\to -\mathcal Z_1$ such that $L_{T_{Z}}(S^2\times S^{d-2}\times S^1)=2(1+(-1)^d)(1-(-1))=4(1+(-1)^d)$ and $L_{T_{Z}}(S^d\times S^1)=(1+(-1)^d)(1-(-1))=2(1+(-1)^d)$ and again for even $d$ the jump in the index accounts for the jump in the number of supersymmetric ground states, and for odd $d$ one can consider an additional reflection on $\mathcal{Y}$.  Note moreover that each $W_{1,2}$ is multiplied by its own $R$-symmetry-odd superfield, $\mathcal{P}_{1,2}$, but the presence of an extra $\mathcal{P}$ does not open a new non-geometric branch, since $\vec{\mathcal{Z}}$ never vanishes at the transition locus. On the other hand, the fields that describe the thermal cigar-like or cylinder topologies are the same as in the model reviewed in section \ref{sec:lsm.review}, and thus share the same extra branch, the same massive vacua drifting to infinity at the transition locus, and the same index jump as one crosses via the non-compact theory in that sector. One may hope to solve this problem by allowing the superpotential to depend non-trivially on the $\textbf{Z}$ coordinates\footnote{One can in principle include a coupling between the KK circle and the rest of the sigma model, e.g. $W_1 = \mu_1\left((\vec{\mathcal{X}}^2 + \vec{\mathcal{Z}}^2-e-b)(\vec{\mathcal{Y}}^2+a)+c\right)$, that will render the exact same geometric and $P$ branches. Also note that the spin structure on the KK circle depends on the exact details of the ambient space. As a curve on $\mathbb{R}^2$ it inherits anti-periodic boundary conditions, but as e.g. a Hopf fiber in $\mathbb{C}^2$ they can be either periodic or anti-periodic \cite{Milnor1963}.}, but such a realization remains elusive. In the following we analyze the transition from the complementary target space point of view, analyzing the topological data that distinguish the two backgrounds and how that diagnostics fits with the worldsheet perspective presented so far.

\subsection{Cobordism classes and brane charges of the two backgrounds}
\label{sec:bordism}
As seen in the previous section, the topological obstruction present in $(1,1)$ sigma models when trying to deform a free string into a black hole persists in the presence of additional compact dimensions. In \cite{Chen:2021dsw}, it was also argued from a spacetime perspective that a topological obstruction to deforming the Horowitz--Polchinski/free string solution into a black hole in Type II theories could be traced to the different K-theories on each side. 
To compare both Euclidean backgrounds, which have different number of compact directions, the geometries are capped off at infinity, following \cite{Chen:2021dsw}
\begin{equation}
\label{eq:compact.topologies}
    \mathbb{R}^{d-1}\times S^1\to S^d\qquad     B^2\times S^{d-2}\to S^2 \times S^{d-2}.
\end{equation}
The different K-theories on each side would then correspond to different brane content, and therefore different topological charges on each manifold \cite{Minasian:1997mm, Witten:1998cd}. With more recent insights into topology change and the classification of branes in string theory, we would like to revisit this. 

Two distinct questions arise when comparing the topology of these two backgrounds. First, whether both manifolds are cobordant, so that a topology-changing process interpolating between them is kinematically allowed. Second, which conserved brane charges are supported by each background, and whether they coincide. Both questions are logically independent, and the key result of this subsection is that the classes agree in every structure we examine, while the charge lattices differ.

Two closed $n$-dimensional manifolds $M,\, N$ are said to be cobordant if their disjoint union corresponds to the boundary of some $(n+1)$-dimensional manifold $W$,  such that $\partial W=M\sqcup \bar N$, with the bar indicating reversed orientation. The two being cobordant is a necessary condition for a finite-action process connecting the two manifolds. In fact, it implies the existence of some kinematically allowed process between the two, even in the presence of a topological obstruction, although a dynamical realization of this process is not guaranteed \cite{Buratti:2021fiv}. 
One may however hope that in a consistent theory of quantum gravity such kinematically allowed processes admit a dynamical realization, possibly non-perturbative. In a Euclidean path integral formulation the transition amplitude between two manifolds is given by \cite{Gibbons:1976ue}
\begin{equation}
    Z(M\to N)\simeq \sum_{W}\int \mathcal{D} g e^{-I_E[W,g]},
\end{equation}
with $W$ the bordism between $M$ and $N$. This can be made sense of by considering a saddle point approximation, such that the amplitude is interpreted as a sum over gravitational instantons mediating the transition \cite{Hawking:1978pog}. Additionally, in the case the two manifolds have trivial bordism classes, these can be smoothly deformed to the trivial configuration (bordism to nothing). The Cobordism Conjecture \cite{McNamara:2019rup} states that in any consistent theory of quantum gravity all backgrounds must be trivial in bordism. As such, two such backgrounds could then be dynamically deformed to nothing and back. This effectively connects the two as they would be separated by finite energy domain walls.

The cobordism classification of branes also refines the K-theoretic one, capturing topological charges, including non-supersymmetric branes, beyond the K-theory spectrum \cite{McNamara:2019rup, Blumenhagen:2021nmi}. Since, as we show below, the brane charge lattices of the two backgrounds differ, any connecting process must absorb the charges that are supported only on one side. In type II, this constitutes a perturbative obstruction. The defects predicted by the Cobordism Conjectures are precisely objects on which the corresponding branes can end, so processes involving them do not conserve such charges. The mismatch between both sides thus provides the defect data that a non-perturbative transition must involve, and the obstruction of \cite{Chen:2021dsw} is expected to be removed at this level.

\subsubsection*{Cobordism classes of the two backgrounds}
Let us then consider the bordism classes of $S^d$ and $S^2\times S^{d-2}$ by computing the relevant bordism invariants on each side of the transition. 
The invariants that are relevant for the structures considered here are the Stiefel--Whitney and Pontrjagin classes, as well as Chern classes of the determinant line bundle in the $\Spin^c$ case. Since spheres have stably trivial tangent bundles \cite{MilnorStasheff1974}, all their Stiefel--Whitney and Pontrjagin classes are trivial:
\begin{equation}
    w_0(S^n)=1\, ,\quad w_i (S^n)=0\, , \quad p_0(S^n)=1\, ,\quad p_i (S^n)=0\, , \qquad \text{for $i>0$}
\end{equation}
Given a product manifold, the Whitney sum formula gives \cite{MilnorStasheff1974}
\begin{equation}
    w_k (X\times Y)=\sum_{k=i+j} w_i(X)\smile w_j(Y)\, ,
\end{equation}
with the classes on the right understood as pulled back to $X$ and $Y$.
This implies that for $S^2\times S^{d-2}$ all $w_i$ classes are trivial, just like for $S^d$. Therefore, the BH and HP/free string solutions are \emph{unoriented} cobordant. To show that they also fall in the same \emph{oriented} cobordism class, one can check that the Pontrjagin classes, $p_i$, also match on both sides. These similarly obey a Whitney sum formula \cite{MilnorStasheff1974}.\footnote{For the Pontrjagin classes the Whitney sum formula holds modulo elements of order two \cite{MilnorStasheff1974}. It is exact in our case, since the integral cohomology of $S^2\times S^{d-2}$ is torsion-free.}
\begin{equation}
   p_k (X\times Y)=\sum_{k=i+j} p_i(X)\smile p_j(Y),
\end{equation}
implying they are trivial on products of spheres, and thus the solutions are cobordant and, in fact, null-bordant due to their trivial characteristic classes. Alternatively one can show null-bordance directly, since the two backgrounds bound explicitly, namely $S^d=\partial B^{d+1}$ and $S^2\times S^{d-2}=\partial (B^3\times S^{d-2})$. These fillings also allow one to extend the statement to the more refined structures below.

The natural next step is to keep adding structure. To that end, we now focus on \emph{spin} bordism, whose non-trivial groups $\Omega_n^{\rm Spin}$  for $n\leq 10$ appear for $n=0,1,2,4, 8, 9, 10$ \cite{Blumenhagen:2022bvh}. For manifolds of dimension $n=0,4,8$, the relevant cobordism invariants are built from the $\hat{A}$ genus and the $L$-genus
\begin{equation}
    \hat{A}(TM)= 1-\dfrac{1}{24}p_1+\dfrac{1}{5760}(7p_1^2-4p_2)+\ldots\qquad  L(TM)= 1+\dfrac{1}{3}p_1-\dfrac{1}{45}(p_1^2-7p_2)+\ldots
\end{equation}
which are formal polynomials built from Pontrjagin classes. Then, since spheres have trivial Pontrjagin classes, for all $3\leq d\leq8$ the BH and HP/free string are \emph{spin} cobordant. 
In fact, one can prove the backgrounds are spin cobordant without resorting to the invariants. As mentioned above, the two backgrounds bound explicitly, $S^d=\partial B^{d+1}$ and $S^2\times S^{d-2}=\partial (B^3\times S^{d-2})$, and since all the manifolds involved are simply connected, they admit a unique spin structure. Hence, the fillings directly show the triviality of both cobordism classes for every $d\geq4$. This extends the discussion beyond $d\leq 8$, where it is known that the bordism groups contain torsion factors that are not detected by the invariants above \cite{McNamara:2019rup, Debray:2023yrs}.\footnote{Note that this agreement of bordism class between both backgrounds is not automatic. For instance, in $d=4$, the K3 surface is a spin manifold that bounds no spin manifold. It generates $\Omega_4^{\rm Spin}\cong\mathbb{Z}$, as detected by its $\hat A$-genus, $\hat A(\mathrm{K3})=2$. A transition between $S^4$ and K3 would therefore be obstructed already at the level of bordism classes, so the agreement found here for the black hole and string saddles is a nontrivial check.}

For completeness, we also check that the two sides are Spin$^c$ cobordant, i.e., they allow for a spin structure with fermions charged under a $U(1)$ symmetry. Note that all spin manifolds are Spin$^c$, as can be seen by choosing a trivial $U(1)$ gauge bundle. The converse, on the other hand, is not true in general. The bordism group $\Omega^{\text{Spin}^c}_{n}$ is deeply related to the classification of branes in Type II string theory \cite{Blumenhagen:2021nmi}. The Spin$^c$ bordism invariants are built from Pontrjagin classes together with the Chern classes of the determinant line bundle \cite{Blumenhagen:2022bvh}. Equipping both sides with the trivial determinant line bundle, all such invariants vanish, so the black hole and free string are Spin$^c$ cobordant for any $3\leq d\leq 8$. Once again, the explicit fillings $B^{d+1}$ and $B^3\times S^{d-2}$ also guarantee triviality of the Spin$^c$ bordism class for every $d\geq 4$. As we mentioned, any spin manifold admits a canonical Spin$^c$ structure (choosing the trivial $U(1)$ gauge bundle). The fillings are spin, and thus carry the same canonical Spin$^c$ structure, which restricts to that of their corresponding boundaries, meaning both backgrounds belong to the trivial Spin$^c$ bordism class.

The structures considered so far are the ones relevant for type II strings. For the heterotic string, however, one can include a further refinement in the tangential structure, namely a (twisted)\emph{string} structure. The latter requires a trivialization of the class $\lambda=p_1/2$ of the target space manifold, twisted by the gauge bundle (which is trivial in our cases) \cite{Killingback:1986rd, Witten:1987cg,Tachikawa:2021mby}. The two backgrounds admit such a \emph{string} structure, since they are stably parallelizable (as mentioned above), and $\lambda$ is precisely one of the stable characteristic classes that vanishes in that case. Furthermore, for $d\geq 4$ we also expect these structures to extend over the explicit fillings described above, in analogy with the spin structures, so that the triviality of the bordism classes of both backgrounds also extends to the heterotic string case.

One may ask whether this conclusion also survives after the duality group of the theory is taken into account. In Type IIB string theory, for instance, the relevant tangential structure is not simply Spin. Since the duality group acts on fermions it must be refined to a mixed Spin--U-duality tangential structure, with the associated bordism classes detected by $\eta$-invariants. The corresponding bordism groups have been worked out for $d=10$ and $d=9$ Type IIB \cite{Debray:2023yrs, Braeger:2025kra}, but the duality groups become increasingly complicated as the dimension is lowered, so computing these for lower dimensions is quite challenging. In principle, one can equip both the $S^d$ and $S^2\times S^{d-2}$ geometries with the trivial duality bundle, such that the difference in their bordism classes reduces to those of the underlying Spin bordism. Whether they can instead be distinguished by a nontrivial duality bundle is a more involved question that we leave open.

Let us close with a clarification on the physical, non-compact saddles. Bordism classes are defined for compact manifolds, which is the reason why the analysis above was carried out for the capped backgrounds \eqref{eq:compact.topologies}. Nevertheless, and as noted in \cite{Chen:2021dsw}, promoting the coupling $c$ to an extra coordinate, the locus ${(\mathbf{Y}^2+a)(\mathbf{X}^2-b)+c=0}$ in $\mathbb{R}^{d+2}$ is a smooth $(d+1)$-dimensional manifold diffeomorphic to $\mathbb{R}^{d+1}$. Its slices for constant $c$ interpolate between the cylinder and the cigar-like geometry with fixed asymptotics, and since its tangent bundle is trivial it admits the Spin, Spin$^c$ and string structures discussed above.

\subsubsection*{Brane charges for the two backgrounds}
Although as we have seen $S^d$ is cobordant to $S^{2}\times S^{d-2}$, the two manifolds will generally carry different topological charges. These are captured by the bordism groups of each background. For a fixed tangential structure $\xi$, the group $\Omega_{n}^\xi(X)$ classifies closed $n$-dimensional manifolds with $\xi$-structure equipped with a map into $X$, modulo manifolds that extend to a compact $(n+1)$-dimensional one with boundary mapped into $X$. Physically, the pair describes a brane worldvolume, or the cycle it wraps, placed in the background $X$, and its class is the conserved topological charge of the configuration \cite{McNamara:2019rup, Blumenhagen:2021nmi}. The collection of such groups gives the lattice of brane charges that the background $X$ can support, with the bordism groups of the point (which we denote $\Omega_{n}^\xi(\text{pt})\equiv\Omega_{n}^\xi$ to avoid cluttering the notation) simply describing branes mapped to a point, and thus present in any background. Recall that the $\Spin$ and $\Spin^c$ bordism groups of the point are, up to $n=10$ (with $n \mathbb{Z}_i$ denoting the direct sum of $n$ copies of $\mathbb{Z}_i$) \cite{Blumenhagen:2022bvh}, 
\begin{center}
\begin{tabular}{ |c|c c c c c c c c c c c| } 
\hline
n & 0 & 1 & 2 & 3 & 4 & 5 & 6 & 7 & 8 & 9 &10 \\
\hline
$\Omega^{\text{Spin}}_{n}$& $\mathbb{Z}$ & $\mathbb{Z}_2$  & $\mathbb{Z}_2$  & 0  & $\mathbb{Z}$  & 0  & 0 & 0 & $2\,\mathbb{Z}$  & $2\,\mathbb{Z}_2$& $3\,\mathbb{Z}_2$   \\ 
$\Omega^{\text{Spin}^c}_{n}$& $\mathbb{Z}$ & $0$  & $\mathbb{Z}$  & 0  & $2\,\mathbb{Z}$  & 0  & $2\,\mathbb{Z}$ & 0 & $4\,\mathbb{Z}$  & 0&  $4\,\mathbb{Z}\oplus \mathbb{Z}_2$ \\ 
\hline
\end{tabular}\,.
\end{center}
The bordism groups of $S^k$ are also known (see e.g. \cite{Blumenhagen:2022bvh})
\begin{equation}
    \Omega^{\xi}_n(S^k)=\Omega_{n}^{\xi}\oplus \Omega_{n-k}^{\xi},
\end{equation}
with $\xi$ the choice of tangential structure, here either Spin or Spin$^c$.

The generalization to a product of spheres analogously decomposes over the cells of the product. We compute this in appendix \ref{ap:bordism.products.spheres}, using the suspension and splitting of the product. This yields the following result for the product of a manifold with a sphere
\begin{equation}
\label{eq:bordism.product}
    \Omega_{n}^{\xi}(X\times S^k) =  \Omega_{n}^{\xi}(X)\oplus \Omega_{n-k}^{\xi}(X)\, .
\end{equation}

Focusing on $X=S^2$ and $k=d-2$ we get
\begin{equation}
\label{eq:lattice.bh}
     \Omega_{n}^{\xi}(S^2\times S^{d-2})= \Omega_{n}^{\xi}\oplus \Omega_{n-2}^{\xi}\oplus \Omega_{n+2-d}^{\xi}\oplus \Omega_{n-d}^{\xi},
\end{equation}
and comparing with
\begin{equation}
\label{eq:lattice.HP}
     \Omega_{n}^{\xi}(S^d)= \Omega_{n}^{\xi}\oplus \Omega_{n-d}^{\xi},
\end{equation}
we see that the manifolds have different bordism groups. 

 This comparison has a transparent physical interpretation. The backgrounds share the point part, $\Omega_n^\xi$, which describes unwrapped branes, and the $\Omega_{n-d}^\xi$, which describes branes wrapping the whole space. The black hole background supports two extra towers of charges with no counterpart in the Horowitz--Polchinski/free string side, corresponding to branes wrapping the two extra cycles of black hole topology, namely the $S^2$ (corresponding to the capped cigar-like manifold) and the $S^{d-2}$ (corresponding to the horizon). For the latter case, the charge sits in the $\Omega_0^\xi=\mathbb{Z}$ part of $\Omega_{d-2}^{\xi}(S^2\times S^{d-2})$, with the integer counting the wrapping number, whereas in the Horowitz--Polchinski/free string side the corresponding summand of $\Omega_{d-2}^{\xi}(S^d)$ would be associated to the part $\Omega_{-2}^\xi=0$, so there is no slot for it on this side of the transition. This is the bordism counterpart of the observation in \cite{Chen:2021dsw} that the black hole admits a brane wrapping its horizon with no analogous brane wrapping on the Horowitz--Polchinski/free string side.  Indeed, exploiting the relation between Spin$^c$ (Spin) cobordism and $\rm K$ ($\rm KO$) theory this also tells us that different branes can be wrapped on each manifold. This holds true both in $\mathcal N=2$ Type II string theory, where branes are classified by $\rm K$-theory and in the $\mathcal N=1$ Type I where instead we use $\rm KO$-theory as the branes are classified by real vector bundles. The comparison between this charge-level mismatch, the class-level agreement established above, and the worldsheet analysis of the previous subsections is the subject of section \ref{sec:lsmandbordism}. 

The whole analysis also extends to the black hole-tower case, where both backgrounds have the extra spectator $S_{\KK}^1$. At the level of the charge lattice, \eqref{eq:bordism.product} shows that multiplying by $S^1$ simply adds a direct summand $\Omega_{n-1}^{\xi}(X)$ to each side, built from the lattice of $X$, so the mismatch remains. In fact, we can see the total number of charges on each side doubles with the addition of $S_{\KK}^1$. The bordism groups are now given by
\begin{equation}
\label{eq:lattice.bh2}
     \Omega_{n}^{\xi}(S^2\times S^{d-2}\times S^1)= \Omega_{n}^{\xi}\oplus \Omega_{n-2}^{\xi}\oplus \Omega_{n+2-d}^{\xi}\oplus \Omega_{n-d}^{\xi}\oplus\Omega_{n-1}^{\xi}\oplus \Omega_{n-3}^{\xi}\oplus \Omega_{n+1-d}^{\xi}\oplus \Omega_{n-1-d}^{\xi},
\end{equation}
and
\begin{equation}
\label{eq:lattice.HP2}
     \Omega_{n}^{\xi}(S^d\times S^1)= \Omega_{n}^{\xi}\oplus \Omega_{n-d}^{\xi}\oplus \Omega_{n-1}^{\xi}\oplus \Omega_{n-d-1}^{\xi}.
\end{equation}
Focusing again on  $n=d-2$ we see that on the black hole side there are two additional towers of charges, $\Omega_{d-3}^{\xi}\oplus \Omega_{d-5}^{\xi}$, while on the HP/free string side there is one, $\Omega_{d-3}^{\xi}$. 
This matches the fact that the non-geometric vacua also doubled when adding a spectator circle to the (1,1) sigma model. At the level of bordism classes, the product manifolds are no longer simply connected, and their spin (or string) structures require a choice of fermion periodicity along the $S_{\KK}^1$. The choice is, however, the same in both backgrounds, since the transition does not modify the spectator circle, and for both choices the two backgrounds continue to be explicitly null-bordant, as can be seen by using the previous fillings multiplied by $S_{\KK}^1$. These cap the thermal direction but not the spectator circle, so the antiperiodic spin structure is not exclusively selected.  In fact, we can consider any closed manifold $Y$ with some tangential structure, and if $X$ is a boundary $\partial W=X$ preserving that same tangential structure, then $\partial(W\times Y)=X\times Y$, and $X\times Y$ is also null-bordant.

\subsection{The obstruction from both perspectives: Indices vs. brane charges}
\label{sec:lsmandbordism}
The worldsheet analysis of sections \ref{sec:lsm.review} and \ref{sec:lsm.tower} and the spacetime investigation of section \ref{sec:bordism} are two probes of the same physics. The type II obstruction can be understood as a chain of equalities. The jump in the number of massive vacua in the $P\neq 0$ branch across the transition equals the jump in the Euler characteristic of the target space topologies. The latter precisely corresponds to the (weighted) count of the extra cycles of $S^2\times S^{d-2}$ relative to $S^d$, and each of them supports precisely the extra tower of wrapped brane charges that appear in \eqref{eq:lattice.bh} relative to \eqref{eq:lattice.HP}. The first equality is a worldsheet bookkeeping device, while the second is elementary topology and the third is the spacetime charge interpretation. For odd $d$ the change in $\chi$, as well as the (weighted) count of extra cycles, vanish, but the extra towers of charges persist. The worldsheet bookkeeping in this case is the Lefschetz refinement of the index used in \cite{Chen:2021dsw}. In fact, the lattice was formulated in the language of K-theory in \cite{Chen:2021dsw}, with $\chi=\text{rk} K^0 -\text{rk} K^1$ and both ranks protected separately. The cobordism classification of section \ref{sec:bordism} refines that charge lattice \cite{McNamara:2019rup,Blumenhagen:2021nmi}.

The identification can be pushed a bit further. The ground states of a $(1,1)$ linear sigma model on a compact target space are its harmonic forms, and the extra cohomology of $S^2\times S^{d-2}$ with respect to $S^d$ corresponds precisely to the capped cigar-like sphere, $S^2$, and the horizon, $S^{d-2}$. The two massive vacua that leave the theory as the transition point is crossed (leaving through infinity in the non-compact theory living precisely at the transition point) re-appear precisely in the black hole side as the extra supersymmetric ground states supported by the two cycles that provide the charges for the lattice mismatch. All the ingredients for this identification appear in \cite{Chen:2021dsw}, but the unified interpretation is what we want to emphasize here.

Comparing the type II and heterotic models in this light shows that the distinction between them can be understood as a consequence of the different things the two quantized worldsheet theories are computing. The $(1,1)$ theory quantizes the supercharges into the de Rham operator, whose ground states are harmonic forms, and with protected index given by the Euler characteristic, which has an interpretation in terms of the cycles of the target space and therefore in terms of their charge lattice. One should mention that this is the relevant invariant for the black hole string transition, but it can become degenerate, for instance for odd $d$ or in the presence of a spectator circle, in which case the reflection-refined indices introduced in section \ref{sec:lsm.tower} become the relevant ones. On the other hand, the $(0,1)$ worldsheet theory quantizes the supercharge into the Dirac operator, whose ground states are harmonic spinors, and with protected index given by the $\hat{A}$-genus, which is a cobordism invariant of the class. The former disagrees between our two target space topologies, producing the (perturbative) obstruction in type II, whereas the latter vanishes on both target space topologies, as they belong to the trivial cobordism class, so no protected quantity jumps, and hence produces no obstruction in the heterotic case. Incidentally, the only part of the Euler characteristic that is a bordism invariant is $\chi$ modulo two, and the type II index jumps precisely by two, making it consistent with the fact that both backgrounds are in the same cobordism class but still the obstruction appears. 

To sum up, the mismatch of brane charge lattices is a perturbative obstruction in the type II side, since one background supports conserved wrapped brane charges without a counterpart in the other background. In contrast, the fact that both backgrounds belong to the same cobordism class (the trivial one) means that there is no topological obstruction for an interpolating geometry. The Cobordism Conjecture \cite{McNamara:2019rup} claims that there are no exact global topological charges in Quantum Gravity, and the defects that trivialize the bordism charges are precisely the ones expected to provide the non-perturbative ingredients that a charge-violating process requires.\footnote{Note that a conserved charge measured at infinity need not be sourced by a brane. As noted in \cite{Chen:2021dsw}, the D0-brane charge of the type IIA Horowitz--Polchinski solution is realized on the black hole side by an $F_2$ RR-flux proportional to the normalizable harmonic form on the cigar, without any brane source.} As shown at the end of section \ref{sec:bordism}, both statements remain unchanged in the presence of the spectator KK circle. The transition between the black string and Horowitz--Polchinski/free string-like solutions wrapping the extra circle stays perturbatively obstructed in type II, but not topologically, whereas in the heterotic string there is no obstruction for it to be realized already at the level of the perturbative worldsheet.

\section{Conclusions and Outlook}
\label{s:conclusions}
In this work we have extended the \emph{Black Hole-Tower correspondence} in three directions. We have analyzed the combined thermodynamics of string-oscillator and KK towers, we have included the gravitational backreaction of the KK tower in the analysis (in analogy with the role of the Horowitz-Polchinski solution in the black hole-string correspondence), and we have studied whether the tower and black hole phases can be continuously connected. A detailed summary of these results can be found in section \ref{s:Intro}.

The main conclusion of the thermodynamic analysis is that the Black Hole-Tower correspondence, and particularly its realization in the presence of KK-towers, is robust against gravitational backreaction. The one-loop effective action of the KK tower reorganizes into a higher dimensional radiation gas, and the resulting self-gravitating solution still reaches the correspondence point with $S\sim\Nsp$. This extends the matching of the free species case in \cite{Herraez:2024kux} to the self-interacting regime. This fills the gap for pure decompactification limits that the Horowitz--Polchinski solution \cite{Horowitz:1997jc} filled for emergent string limits, with a key qualitative difference that we have highlighted, namely that no classical condensate analogue of the winding tachyon appears, and the interpolating configuration is instead a quantum gas of KK modes whose non-vanishing energy-momentum tensor appears at one-loop. The two-branch structure of the solutions and the instability of the high-density branch point towards the black string wrapping the KK circle as the natural endpoint of this collapse, and the weak-backreaction analysis shows that gravitational corrections remain controlled up to the correspondence region, with the free analysis being recovered in the relevant limit. Together with the analysis of the combined towers at the free level, which identifies whether a shrinking black hole first encounters the string-oscillator-dominated or the KK-dominated regime in terms of the ratio $M_s/M_{\mathrm{Pl,}D}$, this provides a significant step towards a consistent picture of minimal black holes in decompactification limits, as well as their possible transitions.

Analysing whether the two phases can be continuously connected provided an interesting contrast, parallel to the one observed in \cite{Chen:2021dsw}. In the heterotic embedding of the transition, the explicit linear sigma model families smoothly interpolate between the two saddles, also in the presence of the extra KK circle, whereas in type II the transition is obstructed. Their cobordism classes agree in all the structures we checked (including the string structure relevant for the heterotic theory), so there is no obstruction for a topology-changing transition. In contrast, the brane charges that the two backgrounds can support are different, precisely by the towers of charges associated with the extra cycles in the black hole backgrounds, as we show explicitly by computing the bordism groups associated to both topologies. The worldsheet and spacetime perspectives fit into a coherent picture, as the jump in the index that obstructs the smooth interpolation for the type II sigma models is precisely accounted for by the extra cycles that correspond to the extra brane charges. The obstruction is thus perturbative, and the agreement of the cobordism classes leaves room for a non-perturbative, charge-violating process as the ones predicted by the Cobordism Conjecture \cite{McNamara:2019rup}, which in this case would account for the transition between competing saddles. 

These results also open several concrete directions. From the gravitational side, a completely self-consistent resolution of Einstein's equations as the correspondence point is approached would require recomputing the equation of state at each step in the numerical iteration, and also following the time-dependent solutions describing the collapse, for which our analysis provides initial data and the expected endpoint. The generalization to compact manifolds of higher dimensionality is also a natural extension, replacing the single KK circle by more general geometries, which may exhibit some qualitatively different features according to the combined tower analysis in the free limit.

From the linear sigma model perspective, the type II perturbative obstruction motivates the search for other families in which the extra KK circle enters the superpotential in a more general way, potentially avoiding the index-related obstructions, although such constructions remain elusive. Complementarily, the bordism analysis can be extended by including the duality structures for type II theories, for which the bordism groups are currently known only in high dimensions \cite{Debray:2023yrs,Braeger:2025kra}. It would be informative to also follow the states directly through the transition, in analogy to the transport of boundary states through phase transitions of gauged linear sigma models \cite{Herbst:2008jq}, applied to the boundary states that support the brane charges. Finally, identifying the concrete defects that would trivialize these bordism charges, and could therefore mediate the non-perturbative transition, would hint towards the concrete dynamical process that could take place in the perturbatively obstructed settings.

\section*{Acknowledgments}
We thank Ivano Basile, Roberto Emparan, Dieter L\"ust, Juan Maldacena, Miguel Montero, Mikel Sanchez-Garitaonandia, Ashoke Sen, Marija Tomasevic, and Irene Valenzuela for insightful discussions and comments on this draft. The work by MA is supported by the fellowship LCF/BQ/DFR25/12000054 from “La Caixa” Foundation (ID 100010434) and through the grants CEX2025-001574-S and PID2024-156043NB-I00, funded by MCIN/AEI/10.13039/\\501100011033, and ERDF, EU. We would also like to thank the Simons Center for Geometry and Physics for its hospitality during the Simons Summer Physics Workshop 2026, where part of this work was carried out.

\newpage
\appendix

\section{The heat kernel expansion}\label{ap:heat_kernel}

The heat kernel $K(x,y\,|\,s)$ associated to a kinetic differential operator $D$ acting on the Hilbert space of square-integrable functions (with respect to the $x$ coordinate) is a function that solves
\begin{equation}\label{eq:heat_kernel_def}
    \left(\partial_s + D\right) K(x,y\,|\,s) = 0
\end{equation}
subject to the distributional initial condition $K(x,y\,|\,0)= \delta(x,y)$. Labeling an orthonormal spectral basis $\{f_i(x)\}_i$ of $D$ with a (possibly continuous) index $i$ and their respective eigenvalues with $\lambda_i$, one can write the following formal expression for the heat kernel
\begin{equation}\label{eq:heat_kernel_gen}
    K(x,y\,|\,s) = \sum_i \, f_i(x)^* \, f_i(y) \, e^{- \lambda_i s} \, .
\end{equation}
In addition, the Green's function of the associated differential operator can be expressed in terms of the heat kernel as
\begin{equation}\label{eq:heat_kernel_propagator}
    G(x,y) = -\int_0^{\infty} ds K(x,y \,|\,s) = \sum_i \frac{f_i(x)^{*}f_i(y)}{\lambda_i}
\end{equation}

As an example, consider the kinetic operator associated to a real scalar field of mass $m$ in $d$-dimensional flat space $D = -\nabla^2+m^2$. A (distributionally) normalized set of eigenfunctions for this operator is given by $\{e^{-ip \cdot x}\}$, with eigenvalues $\lambda_p = p^2+m^2$. Then, equation (\ref{eq:heat_kernel_gen}) yields
\begin{equation}\label{eq:heat.kernel.flat}
    K_{\mathbb{R}^d}(x,y\,|\,s) = \int \frac{d^dp}{(2\pi)^d}\, e^{-ip\cdot(x-y)} e^{-(p^2 +m^2)s} = \frac{1}{(4\pi s)^{\frac{d}{2}}}\exp\left({-\frac{(x-y)^2}{4s}-m^2 s}\right).
\end{equation}
One can also compute the heat kernel for a $q$-torus. In that case, the differential operator is the same, with the difference that we choose periodic boundary conditions along the torus cycles. The eigenfunctions form a discrete set $\{e^{-i\frac{n\cdot x}{R}}\}$ with eigenvalues $\lambda_{n}=\frac{n^2}{R^2} + m^2$, $n \in \mathbb{Z}^q$. The heat kernel then reads
\begin{equation}\label{eq:heat.kernel.torus}
     K_{T^q}(x, y \, | \, s) = \sum_{\vec{n} \in \mathbb{Z}^q} \frac{e^{-(\frac{\vec{n}^2}{R^2}+m^2)s + i \frac{\vec{n} \cdot (\vec{x} - \vec{y})}{R}}}{(2\pi R)^q} \, .
\end{equation}

Heat kernel techniques are particularly powerful in the computation of one-loop quantum corrections. Concretely, the one-loop effective action can be written purely in terms of the heat kernel as
\begin{equation}\label{eq:1loop.heat.kernel}
    I_{\text{1-loop}} = -\frac{1}{2} \, \int_0^\infty \frac{ds}{s} \int d^d x \sqrt{g}\, K(x,x\,|\,s)\,.
\end{equation}

Oftentimes it is not possible to obtain a closed form expression for the heat kernel. This can happen, for example, when considering curved spaces. In those cases, it may be possible to expand the heat kernel in an asymptotic expansion. In a general curved space one can write
\begin{equation}
    K(x,y\,|\,s) = \left (\frac{1}{4\pi s}\right )^{\frac{d}{2}}e^{-\frac{\sigma(x,y)}{4s}} \, \Omega (x,y\,|\,s)
\end{equation}
where $\sigma(x,y)$ is the square of the geodesic distance between the points $x,y$ in the chosen geometry. The function $\Omega(x,y\,|\,s)$ is often expanded in an asymptotic series as $s \rightarrow 0^{+}$
\begin{equation}
    \Omega(x,y\,|\,s) \sim 1 + \sum_{k=1}^{\infty} s^k a_{2k}(x,y) \,.
\end{equation}
This procedure is known as the \emph{heat kernel expansion} \cite{Vassilevich:2003xt}. The coefficients $a_{2k}(x,y)$ are functions of curvature invariants and can be computed iteratively using the Schwinger-de Witt procedure. For example, the first two heat kernel coefficients are given by (after evaluating at the coincidence limit)
\begin{equation}
    a_2(x,x) = \frac{1}{6} R\,,
\end{equation}
\begin{equation}
    a_4(x,x) = \frac{1}{360} \left(12\nabla^2 R + 5 R^2 - 2 R_{\mu\nu}R^{\mu\nu}+2R_{\mu\nu\rho\sigma}R^{\mu\nu\rho\sigma}\right)\,.
\end{equation}

Consider the heat kernel on a warped product geometry of the form $\mathbb{R}^d \times S^1$, with metric
\begin{equation}
    ds^2 = g_{ij}(x)\,dx^i dx^j + e^{2\Phi(x)}\,d\tau^2 \,,
\end{equation}
where $x \in \mathbb{R}^d$ and $\tau \sim \tau + \beta$ parametrizes the thermal circle. Typically, the geodesic distance $\sigma(x,y)$ vanishes when evaluated in the coincidence limit. However, if the background topology is non-trivial, additional terms can arise. In the present case, these correspond to the presence of geodesics with non-zero length that connect a point to itself by winding around the circle. Since the warping does not depend on $\tau$, the length of these paths evaluates to $n^2 \beta^2 e^{2\Phi(x)}$, where $n$ is the winding number. When evaluating the heat kernel expansion in the coincidence limit, one has to sum over these paths, obtaining
\begin{equation}
    K(x,x\,|\,s) = \left (\frac{1}{4\pi s}\right )^{\frac{d+1}{2}}\left ( \sum_{n\in \mathbb{Z}}e^{-\frac{n^2\beta^2e^{2\Phi(x)}}{4s}}\right ) \, \Omega (x,x\,|\,s)\,.
\end{equation}
Then, using the third Jacobi theta function, 
\begin{equation}\label{eq:theta3}
    \theta_3(e^{-t})= \sum_{n} e^{-n^2 t} = \left (\frac{\pi}{t}\right )^{\frac{1}{2}} \sum_{n\in \mathbb{Z}}e^{-\frac{\pi^2 n^2}{t}} \,,
\end{equation}
the heat kernel in this curved geometry can then be rewritten as
\begin{equation}
    K(x,x\,|\,s) = \left (\frac{1}{4\pi s}\right )^{\frac{d}{2}} \frac{1}{\beta e^{\Phi(x)}}\theta_3(e^{-\frac{4\pi^2}{e^{2\Phi(x)}\beta^2}s}) \, \Omega (x,x\,|\,s) \,.
\end{equation}

\subsection{Higher order terms in the heat kernel expansion}\label{ap:heat.kernel}

As stated in section \ref{sec:weak.backreaction}, every term in the heat kernel expansion for the self-gravitating $\rm{KK}$ tower can be well approximated by \cref{eq:generic}, provided that the number of species $N_T$ is sufficiently large. In general, higher order terms in the expansion take the form
\begin{equation}
    I_{\KK} = \sum_{k\geq0} I^{(k)}_{\KK} \,,
\end{equation}
where
\begin{equation}
    I_{\KK}^{(k)} = -\pi^{\frac{d-1}{2}} T^{d-2k}\int d^dx \, \sqrt{g_{d-1}} \sqrt{g_{00}}^{2k-d+1}  a_{2k}(x) \sum_n^{N_T}\int_0^{\infty} \frac{ds}{s^{\frac{d-1}{2}+1-k}} e^{-\frac{n^{2} m_{\KK}^2 \beta_{\rm loc}^2}{4\pi^2}s} \theta_3(e^{-s})\,. 
\end{equation}

Schwinger integrals for terms with $2k < d+1$ are ultraviolet divergent. As in the case of the free $\rm{KK}$ action, this divergence can be regularized by subtracting a reference heat kernel. This procedure effectively adds an (infinite) $T$-independent constant, which can be absorbed into the renormalization of terms already present in the Einstein–Hilbert action. For these terms, we approximate the sum over $\rm{KK}$ modes by an integral, just as in \ref{sec:notfree.species}. Then, for $2k < d+1$, we define the coefficients
\begin{equation}
    c_{d,k} = \pi^{\frac{d+2}{2}} \int_0^\infty \frac{ds}{s^{\frac{d}{2}+1-k}}\text{Erf}\left(\frac{\sqrt{s}}{2\pi}\right)\,\left(\theta_3(e^{-s}) -\sqrt{\frac{\pi}{s}}\right) \, .
\end{equation}

On the other hand, terms with $2k \geq d+1$ have to be treated differently. Approximating the $\rm{KK}$ sum by an integral leads to an IR divergence. Instead, we perform the integral exactly
\begin{align}
    \sum_{n=1}^{N_T}\int_0^{\infty} ds\,s^{k-\frac{d-1}{2}-1}\, e^{-\frac{n^2 m^2 \beta^2}{4\pi^2}s} \theta_3(e^{-s}) = \sum_{l\in\mathbb{Z}}\sum_{n=1}^{N_T}\frac{\Gamma(k-\frac{d-1}{2})}{\left({\frac{1}{4\pi^2}\left(\frac{n}{N_T}\right )^{\frac{2}{p}}} + l^2\right )^{k-\frac{d-1}{2}}}\,,
\end{align}
and replace the sum by an integral afterwards
\begin{equation}
    \int_1^{N_T} dx\, \frac{1}{\left(\frac{1}{4\pi^2}\left(\frac{n}{N_T}\right )^{2} + l^2\right )^{k-\frac{d-1}{2}}} 
\approx (2\pi)^{2k-d+1}N_T
\begin{cases}
\displaystyle \,\,\,\frac{1}{l^2}\frac{1}{(4\pi^2l^2+1)^{\frac{2k-d-3}{2}}} a_{2k-d+1,l}\, & l \neq 0 \\[1.5ex]
\displaystyle  \frac{1}{d-2k}\,, & l = 0
\end{cases}\,,
\end{equation}
where $a_{n,p,l}={}_2F_1\left(1, \frac{3 - n}{2}, \frac{3}{2}, -\frac{1}{4 l^2 \pi^2}\right)$ can be written in terms of the hypergeometric function. In view of this, for $2k>d+1$ we define
\begin{equation}
    c_{d,k} = (2\pi)^{2k-d+1}\pi^\frac{d-1}{2} \Gamma\left(k - \frac{d-1}{2}\right)\left (\frac{1}{d-2k}+ 2\sum_{l>0} \frac{1}{l^2}\frac{1}{(4\pi^2l^2+1)^{\frac{2k-d-3}{2}}} a_{2k-d+1,l}\right)\,.
\end{equation}
so that higher order terms in the heat kernel expansion can also be recast in the form (\ref{eq:generic}). 

Finally, the logarithmic divergent correction $d = 2k$ requires special treatment, as any of the schemes used to approximate the Schwinger integral used so far lead to a divergent result. Instead, we compute the Schwinger integral as follows
\begin{equation}
    \sum_{n=1}^{N_T}\int_0^{\infty} \frac{ds}{s^{\frac{1}{2}}}\,\, e^{-\frac{n^2 m^2 \beta^2}{4\pi^2}s} \theta_3(e^{-s}) = 4\sqrt{\pi} \sum_{n=1}^{N_T}\sum_{l>0} K_0 \left (l\frac{n}{N_T}\right) \approx 4\sqrt{\pi} \sum_{l>0} \int_0^{N_T} dx \,K_0 \left (l\frac{x}{N_T}\right )\,,
\end{equation}
here, $K_n$ is the $n$th modified Bessel function of the second kind.
The result of this integral can be written in terms of Bessel and Struve functions
\begin{equation*}
    \int_0^{N_T} dx \,K_0 \left (l\frac{x}{N_T}\right ) = \frac{1}{2} \pi  N_T ({L}_{-1}(l) K_0(l)+{L}_0(l) K_1(l))\,,
\end{equation*}
with $L_n$ the Struve functions. Defining, 
\begin{equation}
    c_{d,\frac{d+1}{2}} = 2{\pi^{\frac{d+2}{2}}} \sum_{l>0}({L}_{-1}(l) K_0(l)+{L}_0(l) K_1(l))\,,
\end{equation}
we find that the logarithmically divergent term can also be written in the form (\ref{eq:generic}).

\section{Effective action of the $\rm{KK}$ tower via the Eisenstein series}\label{ap:eisenstein}

In section~\ref{sec:notfree.species} we summarised the result that, treating the $\rm{KK}$ circle on equal footing with the thermal circle, the one-loop energy of an infinite tower of $\rm{KK}$ modes re-arranges itself as a higher-dimensional radiation gas, with the dimensional-reduction structural identity \eqref{eq:I1.I2.ratio}. We give here the detailed derivation.

To this purpose we compute the Casimir energy of a massless scalar on the torus $\mathbb{R}^{d-1}\times S^1_\beta\times S^1_{\KK}$. Starting from (absorbing some $\pi$ factors)
\begin{equation}
    V = \sum_{n,m\neq 0} \frac{1}{(n^2 R^2 + m^2 \beta^2)^{\frac{d+1}{2}}}
\end{equation}
and using the real analytic Eisenstein series
\begin{equation}
    E_{k} (\tau)\ \mathrm{Im} (\tau)^{-k} = \sum_{n,m\neq 0}\frac{1}{|n\tau + m|^{2k}}\,,
\end{equation}
we can write the potential as
\begin{equation}
    V = \left (\frac{1}{R\beta} \right )^\frac{d+1}{2} E_{\frac{d+1}{2}}\left (i\frac{\beta}{R}\right ).
\end{equation}
An important detail here is that, due to the modular invariant nature of the torus partition function, this expression is invariant under $R\leftrightarrow\beta$, which corresponds to a modular $S$ transformation. From string theory we know that in the limit $\mKK\gg T$ there is a second tower of states becoming light, but these are not captured in the EFT limit we are going to take. Here we need to subtract the thermal vacuum \cite{Kapusta:2006pm} ($\beta\to\infty$), thereby selecting a preferred direction, that being time, and breaking modular invariance. The regularized potential is
\begin{equation}
  V=V_{reg}=V-V\Big\vert_{\beta\to\infty}=\left (\frac{1}{R\beta} \right )^\frac{d+1}{2} \left[E_{\frac{d+1}{2}}\left (i\frac{\beta}{R}\right )-\left(\dfrac{\beta}{R}\right)^{\frac{d+1}{2}}\right],
\end{equation}
which we will just call $V$ by a mild abuse of notation. The one-loop effective action is then
\begin{equation}
    I_{\KK} = -C \int d^{d+1}x\sqrt{g_{d-1}}\sqrt{g_{00}}\sqrt{g_\KK}\left (\frac{1}{R\beta\sqrt{g_{00}}\sqrt{g_{\KK}}} \right )^\frac{d+1}{2} \left[E_{\frac{d+1}{2}}\left (i\frac{\beta\sqrt{g_{00}}}{R\sqrt{g_{\KK}}}\right )-\left(\dfrac{\beta\sqrt{g_{00}}}{R\sqrt{g_{\KK}}}\right)^{\frac{d+1}{2}}\right],
\end{equation}
where we have restored the local dependence in $\beta$ and $R$. Before proceeding we recall
\begin{equation}
E_k(it) = t^{-k} + \sqrt{\pi}\,\frac{\Gamma\!\left(k-\tfrac12\right)}{\Gamma(k)}\frac{\zeta(2k-1)}{\zeta(2k)}\,t^{k-1} + \mathcal{O}(e^{-2\pi/t}), \qquad t\to 0\,,
\end{equation}
\begin{equation}
E_k(it) = t^{k} + \sqrt{\pi}\, \frac{\Gamma\!\left(k-\tfrac12\right)}{\Gamma(k)} \frac{\zeta(2k-1)}{\zeta(2k)}\,t^{1-k} + \mathcal{O}(e^{-2\pi t}), \qquad t\to\infty\,,
\end{equation}
such that in the limit $\beta/R\to0$ we obtain
\begin{equation}
    I_{1} \simeq -C \int d^{d+1}x\sqrt{g_{d-1}}\sqrt{g_\KK}\sqrt{g_{00}}^{-(d)}\left (\frac{1}{\beta} \right )^{d+1},
\end{equation}
corresponding to a radiation gas in $d+1$ dimensions. In the opposite limit $\beta/R\to\infty$ we obtain, after subtracting the thermal vacuum,
\begin{equation}
    I_{2} \simeq -C\sqrt{\pi}\,\frac{\Gamma\!\left(\frac{d}{2}\right)}{\Gamma(\frac{d+1}{2})}\frac{\zeta(d)}{\zeta(d+1)} \int d^{d+1}x\sqrt{g_{d-1}}\sqrt{g_{00}}^{1-d}\left (\frac{1}{\beta} \right )^{d}\dfrac{1}{R}\,,
\end{equation}
corresponding to a radiation gas in $d$ dimensions. Notice that this has no dependence on $g_{\KK}$, so its energy momentum tensor has no support along the $\rm{KK}$ direction. We can compute
\begin{equation}
    \dfrac{I_1}{I_2}\simeq \frac{\int d^{d+1}x\sqrt{g_{d-1}}\sqrt{g_\KK}\sqrt{g_{00}}^{-d}}{\int d^{d+1}x\sqrt{g_{d-1}}\sqrt{g_{00}}^{1-d}}\dfrac{R}{\beta}\,.
\end{equation}
Then, as long as the metric is regular, the integrals only contribute numerical prefactors. This holds trivially in the limit of weak gravitational backreaction, but more generally for the entire static solution. Parametrically we then recover
\begin{equation}
        \dfrac{I_1}{I_2} \simeq\dfrac{R}{\beta} = N_T\,,
\end{equation}
which is the result quoted as \eqref{eq:I1.I2.ratio} in the main text.

\section{The GSO projection and the thermal circle}\label{ap:GSO}

As noted in section \ref{sec:horowitz.polchinski}, the GSO projection \cite{Gliozzi:1976qd} needs to be revisited in the imaginary‐time formalism. This ultimately stems from the non‐trivial boundary conditions of fermionic fields along the Euclidean time circle, which cause the GSO projection to flip sign for strings winding around the thermal circle. This effect was analyzed in detail in \cite{Atick:1988si}. In this appendix we provide the derivation underlying this phenomenon.

The additional signs appearing in the GSO projection can be computed by considering the genus-one path integral. Recall that a genus-one worldsheet can be parametrized by two periodic coordinates, $\sigma_1$ and $\sigma_2$, which can be chosen to have period one. In our convention, $\sigma_1$ is the ``spatial coordinate'' and $\sigma_2$ is the ``time coordinate.'' The difference with respect to the zero temperature case arises because, in addition to the sum over spin structures, one must also sum over winding modes along the circle, namely those for which $X^0$ is periodic up to a multiple of the circle length $\beta$, i.e.
\begin{equation}
    X^0=x^0+n\sigma_1+m\sigma_2+\dots
\end{equation}
These modes introduce new negative signs in the GSO projection that must be accounted for.

At genus one there are four spin structures, depending on whether $\sigma_1$ or $\sigma_2$ are periodic or antiperiodic. These are $(+,+)$, $(+,-)$, $(-,+)$ and $(-,-)$, where the first sign refers to $\sigma_1$ and the second to $\sigma_2$. Then, one sees that spacetime fermions (from the Ramond sector) correspond to the structures $(+,+)$ and $(+,-)$, and spacetime bosons (from the Neveu-Schwarz sector) correspond to $(-,+)$ and $(-,-)$. 

In order to compute the extra sign, one starts by considering a string with $n=0$, $m\neq0$. This string wraps $m$ times along the thermal circle. Because of this, fermions will get an extra $-1$ factor for each string winding. In other words, the spin structures $(+,+)$ and $(+,-)$ are to be corrected by a phase $(-1)^{m}$, whereas $(-,-)$ and $(-,+)$ stay the same. The next step is to generalize this to arbitrary $n$, which can be done by exploiting modular invariance. The key is to write down a modular invariant expression that reproduces the $n=0$ behaviour. Let $L$ represent any of the four spin structures. The additional sign weighting each of the spin structures is
\begin{equation}\label{eq:sign}
    U(L;\epsilon)=\phi(L)\cdot\phi(L\otimes\epsilon),
\end{equation}
where $\phi(L)$ is the parity of $L$ \footnote{Recall that $\phi(L)=-1$ for $(+,+)$ and $\phi(L)=+1$ for the other three.}, and $\epsilon$ represents the spin structure that would be given by $((-)^n,(-)^m)$. Modular invariance follows from the fact that the parity does not change under modular transformations. One can check that (\ref{eq:sign}) also gives the right behaviour when $n=0$ since, in that case, for the spacetime bosons one has $\phi(L)=\phi(L\otimes\epsilon)=+1$, and for the spacetime fermions $\phi(L\otimes\epsilon)=(-1)^m\,\phi(L)$.

We now turn to the spectrum of the string. For Type II, it takes the form
\begin{equation}
    M^2 = \frac{2}{\alpha'}(N + \tilde{N} - 1) + \frac{4\pi^2\tilde{m}^2}{\beta^2} + \frac{n^2 \beta^2}{4\pi^2\alpha'^2},
\end{equation}
subject to $N-\tilde{N}=\tilde{m}n$ and the GSO projection. Note that $\tilde{m}$ is not the same as $m$. The first represents momentum excitations along the thermal circle. The second is linked to the winding number in the worldsheet time coordinate $\sigma_2$. 

Usually, the GSO projection forbids us from taking $N=\tilde{N}=0$. This is because the spin structures $(-,-)$ and $(-,+)$ are weighted with different signs. However, in the case that $\tilde{m}=0$, $n=\pm 1$, and arbitrary $m$, one obtains
\begin{equation}
U(L, \epsilon) = 
\begin{cases} 
(-1)^m, & \text{if } L = (-, -), \\ 
(-1)^{m+1}, & \text{if } L = (-, +). 
\end{cases}
\end{equation}
Crucially, these are corrected by opposite signs. Combined with the initial sign difference between them, both $(-,-)$ and $(-,+)$ carry now the same weight. This inverts the GSO projection and allows for the choice $N=\tilde{N}=0$. Hence, a new scalar field with mass
\begin{equation}
    m^2 = \frac{\beta^2-8\pi^2\alpha'}{4\pi^2\alpha'^2},
\end{equation}
appears in the spectrum. Calling $\beta_H = 2\pi\sqrt{2\alpha'}$ and including the radion field, one obtains equation (\ref{eq:winding}).

\section{Induced energy-momentum tensor on the boundary}\label{ap:boundary}
We consider a $(d+1)$-dimensional static spacetime with symmetry
$S^{d-2}\times S^1$ and metric
\begin{equation}
ds^2
=
-e^{2\psi(r)}dt^2
+e^{2\Lambda(r)}dr^2
+r^2 d\Omega_{d-2}^2
+e^{2\gamma(r)}d\chi^2,
\qquad \chi \sim \chi + 2\pi R_{\mathrm{KK}} .
\end{equation}
The shell sits at $r=L$. Outside the shell, the geometry is the uniform
black string
\begin{equation}
ds^2_{\text{out}}
=
-f(r)dt^2+\frac{dr^2}{f(r)}+r^2 d\Omega_{d-2}^2+d\chi^2,
\qquad
f(r)=1-\frac{\mu}{r^{d-3}} .
\end{equation}
The mass density is normalized as
\begin{equation}
M=\frac{(d-2)\Omega_{d-2}(2\pi R_{\mathrm{KK}})}{16\pi G_{d+1}}\,\mu ,
\label{eq:appd:totalmass}
\end{equation}
with $M$ the Komar mass \cite{Komar:1963svp}
\begin{equation}
M=M_{\text{bulk}}+M_{\text{shell}} .
\end{equation}
For a static source, the bulk contribution is
\begin{equation}
M_{\text{bulk}}
=
\frac{(d-2)\Omega_{d-2}(2\pi R_{\mathrm{KK}})}{(d-1)(d-3)}
\int_0^L dr\, e^{\psi+\Lambda+\gamma} r^{d-2}
\left[
(d-2)(-T^t{}_t)+T^r{}_r+(d-2)T^\theta{}_\theta+T^\chi{}_\chi
\right] .
\end{equation}
For radiation
$p=\rho/d$, demanding the solution is regular at the origin we have
\begin{equation}
M_{\text{bulk}}
=
\frac{(d-2)\Omega_{d-2}(2\pi R_{\mathrm{KK}})}{8\pi G_{d+1}
(d-3)}
\,e^{\psi(L)+\gamma(L)-\Lambda(L_-)}L^{d-2}\psi'(L_-) ,
\end{equation}
where $- (+)$ means the quantity is evaluated inside (outside) the box.

In order to compute the contribution from the shell we must compute its stress energy tensor via the Israel junction conditions
\begin{equation}
S^a{}_b
=
-\frac{1}{8\pi G_{d+1}}
\left(
[K^a{}_b]-\delta^a{}_b[K]
\right) ,
\end{equation}
where
\begin{equation}
[X]=X_{\text{out}}-X_{\text{in}} .
\end{equation}
We require the tangential components of the metric to be continuous along the interface, but in principle the $g_{rr}$ does not need be.

At $r=L$, the nonzero mixed extrinsic curvatures are
\begin{equation}
K^t{}_t=e^{-\Lambda}\psi',
\qquad
K^\theta{}_\theta=\frac{e^{-\Lambda}}{L},
\qquad
K^\chi{}_\chi=e^{-\Lambda}\gamma' .
\end{equation}
The shell contribution to the Komar mass is
\begin{equation}
M_{\text{shell}}
=
\frac{(d-2)\Omega_{d-2}(2\pi R_{\mathrm{KK}})}{(d-1)(d-3)}
\,e^{\psi(L)+\gamma(L)}L^{d-2}
\left[
(d-2)(-S^t{}_t)+(d-2)S^\theta{}_\theta+S^\chi{}_\chi
\right] .
\end{equation}
Using the Israel junction conditions, this simplifies to
\begin{equation}
M_{\text{shell}}
=
\frac{(d-2)\Omega_{d-2}(2\pi R_{\mathrm{KK}})}{8\pi G_{d+1}(d-3)}
\,e^{\psi(L)+\gamma(L)}L^{d-2}
\left[
(e^{-\Lambda(L_+)}\psi'(L_+))-(e^{-\Lambda(L_-)}\psi'(L_-))
\right] .
\end{equation}
Adding bulk and shell then reproduces \eqref{eq:appd:totalmass}. Matching the solution outside the box to the black string, still leaves
a one-parameter family of possible shells, parameterized by the surface energy density $\sigma = - S^t{}_t$ which then fixes the shell stress and the total mass.

In $d=4$, the Israel junction conditions give
\begin{equation}
\sigma
=
\frac{1}{8\pi G}
\left[
\frac{2}{L}\bigl(e^{-\Lambda(L)}-e^{\psi(L)}\bigr)
+e^{-\Lambda(L)}\gamma'(L)
\right].
\end{equation}
Solving for $e^{\psi(L)}$,
\begin{equation}
e^{\psi(L)}
=
e^{-\Lambda(L)}
\left(
1+\frac{L}{2}\gamma'(L)
\right)
-4\pi G L\,\sigma .
\end{equation}

Since the exterior metric is
\begin{equation}
f(r)=1-\frac{\mu}{r},
\end{equation}
the matching condition at $r=L$ implies
\begin{equation}
\mu
=
L\left[1-e^{2\psi(L)}\right],
\end{equation}
and therefore
\begin{equation}
M
=
\frac{(2\pi R_{\mathrm{KK}})}{2G}\,\mu .
\end{equation}

The bulk contribution is
\begin{equation}
M_{\rm bulk}
=
\frac{(2\pi R_{\mathrm{KK}})}{G}\,
e^{\psi(L)-\Lambda(L)+\gamma(L)}L^{2}\psi'(L),
\end{equation}
or, equivalently,
\begin{equation}
M_{\rm shell}=M-M_{\rm bulk}.
\end{equation}

Then, once the boundary data is known, distinct configurations differ by a single parameter, that being the shell density $\sigma$, the mass of the shell is given by

\begin{align}
M_{\mathrm{shell}}
&=
\frac{(2\pi R_{\mathrm{KK}})L}{2G}
\Bigg[
1-
\left(
e^{-\Lambda(L)}
\left(1+\frac{L}{2}\gamma'(L)\right)
-4\pi G\,L\,\sigma
\right)^2
\Bigg]
\nonumber\\[4pt]
&\qquad
-
\frac{(2\pi R_{\mathrm{KK}})L^{2}}{G}\,
e^{-\Lambda(L)}\psi'(L)
\left(
e^{-\Lambda(L)}
\left(1+\frac{L}{2}\gamma'(L)\right)
-4\pi G\,L\,\sigma
\right).
\end{align}

The price that we have to pay for insisting on static solutions that are asymptotically Minkowski is this interface. Matching with Schwarzschild-like solutions outside fixes the shell pressure, but the energy density $\sigma$ can be chosen freely. We present some physically motivated choices but the exact choice of $\sigma$ is irrelevant for the parametric behaviour of our thermodynamic quantities. 

A \textit{tensionless shell} $\sigma=0$ gives
\begin{equation}
M
=
\frac{(2\pi R_{\mathrm{KK}})L}{2G}
\left[
1-e^{-2\Lambda(L)}
\left(
1+\frac{L}{2}\gamma'(L)
\right)^2
\right].
\end{equation}

A \textit{massless shell} $M_{\rm shell}=0$ implies that the total mass is carried entirely by the bulk contribution $ M=M_{\rm bulk}$. Here the tension satisfies
\begin{equation}
\sigma
=
\frac{1}{4\pi G\,L}
\left[
e^{-\Lambda(L)}
\left(1+\frac{L}{2}\gamma'(L)\right)
+
L\,e^{-\Lambda(L)}\psi'(L)
-
\sqrt{
1+L^{2}e^{-2\Lambda(L)}\psi'(L)^2
}
\right].
\end{equation}

Another choice is to impose a radially \textit{continuous shell} with $g_{rr}^{\rm -}(L)=g_{rr}^{\rm +}(L)$. Since $g_{rr}^{\rm -}(L)=e^{2\Lambda(L)},\, g_{rr}^{\rm +}(L)=\frac{1}{f(L)},$ and continuity of $g_{tt}$ gives $e^{2\psi(L)}=f(L)$.
Substituting into the Israel junction condition gives
\begin{equation}
\sigma_{\rm cont}
=
\frac{1}{8\pi G}\,e^{-\Lambda(L)}\gamma'(L).
\end{equation}
As we argue below all three choices, or generally any choice of shell with $|\sigma|\lesssim \rho_c L$, will not modify the parametric behaviour of the solution. For concreteness in the main text we take the metric to be continuous everywhere.
To compare different choices of sigma, let us consider our setup with
\begin{equation}
\psi'(r)=\frac{\kappa}{3}r,
\qquad
\Lambda(r)=\frac{\kappa}{8}r^2,
\qquad
\gamma'(r)=-\frac{\kappa}{12}r,
\end{equation}
with
\begin{equation}
\kappa=8\pi G_5 \rho_c.
\end{equation}
Then
\begin{equation}
e^{-\Lambda(L)}=e^{-\kappa L^2/8},
\qquad
\gamma'(L)=-\frac{\kappa L^2}{12L}.
\end{equation}

For general $\sigma$, one finds
\begin{equation}
e^{\psi(L)}
=
e^{-\kappa L^2/8}\left(1-\frac{\kappa L^2}{24}\right)-4\pi G L\,\sigma ,
\end{equation}
and, to first order in $\kappa L^2$,
\begin{equation}
M_{\rm bulk}
\simeq
\frac{(2\pi R_{\mathrm{KK}})L}{G}\,
\frac{\kappa L^2}{3}\bigl(1-4\pi G L\,\sigma\bigr),
\end{equation}
\begin{equation}
M
\simeq
\frac{(2\pi R_{\mathrm{KK}})L}{G}
\left[
4\pi G L\,\sigma
-8\pi^2 G^2 L^2\sigma^2
+\frac{\kappa L^2}{6}\bigl(1-4\pi G L\,\sigma\bigr)
\right],
\end{equation}
and therefore
\begin{equation}
M_{\rm shell}
\simeq
\frac{(2\pi R_{\mathrm{KK}})L}{G}
\left[
4\pi G L\,\sigma
-8\pi^2 G^2 L^2\sigma^2
-\frac{\kappa L^2}{6}\bigl(1-4\pi G L\,\sigma\bigr)
\right].
\end{equation}
The ratio is then
\begin{equation}
\frac{M_{\rm shell}}{M_{\rm bulk}}
\simeq
\frac{3}{2}\frac{\sigma}{\rho_c L}
-\frac{1}{2}.
\end{equation}

The three shell choices can be read off directly. The minimal shell $\sigma=0$ gives $M_{\rm shell}\simeq-\tfrac12 M_{\rm bulk}$. The massless shell $M_{\rm shell}=0$ requires $\sigma_{\rm crit}\simeq\tfrac13\rho_c L$. Finally the continuous shell, $\sigma_{\rm cont}=e^{-\Lambda(L)}\gamma'(L)/8\pi G\simeq-\tfrac{1}{12}\rho_c L$, gives $M_{\rm shell}/M_{\rm bulk}\simeq-\tfrac58$. Generally, any choice of shell with $|\sigma|\sim\rho_c L$ will not affect the parametric behaviour. In the main text we take the continuous shell as representative.

\section[Solution to a gravitational system with S2 x S1 symmetry]{Solution to a gravitational system with $S^2\times S^1$ symmetry}
\label{ap:proof}
We consider again the $D=4+1$ ansatz
\begin{equation}
ds^{2} = - e^{2\psi(r)} dt^{2} + e^{2\Lambda(r)} dr^{2} + r^{2} d\theta^{2} + r^{2}\sin^{2}\theta d\varphi^{2} + e^{2\gamma(r)} d\chi^{2},
\end{equation}
with $\chi\sim\chi+2\pi R_{\text{KK}}$, and an energy-momentum tensor
\begin{equation}
T^{M}_{\,\,\,\,N} = \rho(r)\,\mathrm{diag}\left(-1,\frac{1}{4},\frac{1}{4},\frac{1}{4},\frac{1}{4}\right).
\end{equation}

Defining
\begin{equation}
A(r) = e^{-2\Lambda(r)}, \qquad u(r) = 4\gamma'(r) + \psi'(r),
\end{equation}
the Einstein equations can be written as
\begin{equation}
E_1:\quad \frac{A-1}{r^2} + A\left(\gamma'' - \gamma'\Lambda' + \gamma'^2 + \frac{2\gamma'}{r} - \frac{2\Lambda'}{r}\right) = -\rho,
\end{equation}
\begin{equation}
E_2:\quad \frac{A-1}{r^2} + A\left(\gamma'\psi' + \frac{2\gamma'}{r} + \frac{2\psi'}{r}\right) = \frac{\rho}{4},
\end{equation}
\begin{equation}
E_3:\quad \frac{A-1}{r^2} + A\left(\psi'' + \psi'^2 - \Lambda'\psi' + \frac{2\psi'}{r} - \frac{2\Lambda'}{r}\right) = \frac{\rho}{4}.
\end{equation}
The conservation equation is
\begin{equation}
\rho' = -5\rho\,\psi'.
\end{equation}

One has
\begin{equation}
E_3 - E_2 = 0, \qquad E_1 + 4E_2 = 0,
\end{equation}

while subtracting $E_2$ from $E_3$ gives
\begin{equation}
\psi'' = \gamma'\psi' + \Lambda'\psi' + \frac{2\gamma'}{r} + \frac{2\Lambda'}{r} - \psi'^2,
\end{equation}
which implies
\begin{equation}
\boxed{
\Lambda' = -\gamma' + r\frac{\psi'^2 + \psi''}{2 + r\psi'}.
}
\end{equation}

Using $E_1 + 4E_2 = 0$, one can solve for $\gamma''$. Combining this with the derivative of $E_2$ and the conservation equation, one finds that $u(r)$ satisfies
\begin{equation}
u' = \left(\Lambda' - \gamma' - \psi' - \frac{2}{r}\right) u.
\end{equation}
Regularity at the origin implies $u(0)=0$, and therefore
\begin{equation}
u(r)\equiv 0.
\end{equation}
This then imposes
\begin{equation}
\boxed{
\gamma'(r) = -\frac{1}{4}\psi'(r)
},
\end{equation}
which finishes our proof.
We can generalize this to total spacetime dimension $D=d+p$, with metric
\begin{equation}
ds^{2} = - e^{2\psi(r)} dt^{2} + e^{2\Lambda(r)} dr^{2} + r^{2} d\Omega_{d-2}^{2} + e^{2\gamma(r)} dy_p^{2},
\end{equation}
where $dy_p^2$ denotes the line element on some Ricci flat compact $p$-manifold, like a Calabi-Yau or a Riemann flat manifold. Repeating the same analysis for a perfect fluid in $d+p$ dimensions with equation of state $p=w\rho$, one finds
\begin{equation}
\boxed{
\gamma' = \frac{w-1}{(d+p-3) + (d+p-1)w}\,\psi'
}\label{eq:gammamastereq}
\end{equation}
and
\begin{equation}
\boxed{
\Lambda' = -\gamma' + r\frac{(p-1)(\gamma'^2 + \gamma'') + \psi'^2 + \psi''}{d-2 + (p-1)r\gamma' + r\psi'}.
}
\label{eq:lambdamastereq}
\end{equation}

For the general case we can also write the $r\ll1/\sqrt{\rho_c}$  behaviour as
\begin{equation}
\psi(r) \simeq \frac{w(d+p-1)+d+p-3}{2 (d-1) (d+p-2)} \rho_c r^2, \qquad \rho(r) \simeq \rho_c - \frac{(d+p-2)(w+1)^2 + w^2 - 1}{2(d-1)w(d+p-2)} \rho_c^2 r^2,
\end{equation}
while for $r\gg1/\sqrt{\rho_c}$ one finds
\begin{equation}
\psi(r) \simeq \frac{2w}{1+w}\log r, \qquad \rho(r) \simeq \frac{2 (d-3) w (d+p-2)}{(w+1)^2 (d+p-3) + 4 w}\frac{1}{r^2},
\end{equation}
and
\begin{equation}
    e^{2\Lambda(r)}\simeq
\end{equation}
\begin{equation*}
\frac{\bigl((d+p-3)(1+w)^2+4w\bigr)\bigl((d-3)(d+p-3)(1+w)^2+4w(d-3+(d+p-2)w)\bigr)}
{(d-3)(1+w)^2\bigl((d+p-3)(1+w)+2w\bigr)^2}.
\end{equation*}

These results also extend to $p=0$, one can use $\eqref{eq:gammamastereq}$ to remove any dependence on $\gamma(r)$ leading to
\begin{equation}
\Lambda' = \frac{-(w-1)\psi' + (1+w)\,r\,(\psi'^2 + \psi'')} {(d-3) + (d-1)w + (1+w)\,r\,\psi'},
\end{equation}
which one can verify holds for a system with spherical symmetry in $d$ dimensions.

Finally, we would like to point out that for $w=0$, regular static solutions require $\rho(r)=0$ for $r>0$, but with $\rho(0)$ possibly singular. The only solutions then are either flat space or black $p$-branes.

\section{Bordism groups of products of spheres}
\label{ap:bordism.products.spheres}
In this appendix we collect the topological facts that underlie the charge lattice computations of section \ref{sec:bordism} and derive eq. \eqref{eq:bordism.product}. We mainly follow \cite{Hatcher:478079}, and analogous computations of bordism groups of string backgrounds, for single spheres, tori, K3 surfaces and other Calabi--Yau manifolds can be found in \cite{Blumenhagen:2022bvh}.\footnote{The products of spheres that appear here are the topologies of the backgrounds whose charge lattices we compute. Not to be confused with the products of spheres that appear as solutions in the context of dynamical cobordisms, as in e.g. \cite{Angius:2023uqk}.}

All spaces we consider here are pointed, that is, they carry a marked base point, and along this appendix $\Sigma$ denotes the reduced suspension, not to be confused with a worldvolume. We first recall three standard constructions involving marked spaces. Firstly, the wedge sum $X\vee Y$, which is the disjoint union of $X$ and $Y$ with the two base points identified. A single shared point does not support any gluing of positive dimensional cycles, so reduced bordism is additive on wedge sums
\begin{equation}
\label{eq:wedge.bordism}
    \tilde{\Omega}_n^\xi(X\vee Y)= \tilde{\Omega}_n^\xi(X) \oplus \tilde{\Omega}_n^\xi(Y)\, .
\end{equation}
Secondly, the smash product $X\wedge Y=(X\times Y)/(X\vee Y)$, which is the product with axes collapsed, and importantly satisfies $S^m\wedge S^n=S^{m+n}$ for spheres. Thirdly, the reduced suspension $\Sigma X\simeq S ^1 \wedge X$, giving the double cone over $X$, and also satisfying $\Sigma^k X \simeq S^k \wedge X$. The key fact being that a  product decomposes after suspension as
\begin{equation}
\label{eq:suspension.decomposition}
    \Sigma(X\times Y)\simeq \Sigma X\vee \Sigma Y\vee \Sigma (X\wedge Y)\, .
\end{equation}
Notice that this is only true after suspension, as can be seen in the example of a torus, $T^2=S^1\times S^1\neq S^1\vee S^1\vee S^2$. After suspension, though, the product splits.

Importantly, for bordism theory we recall the suspension lemma
\begin{equation}
\label{eq:suspension.bordism}
     \tilde\Omega_{n+1}^{\xi}(\Sigma X)=\tilde\Omega_{n}^{\xi}(X),
\end{equation}
which means that suspension simply shifts the degree. Finally, we have the splitting of bordism classes
\begin{equation}
\label{eq:splitting}
 \Omega_{n}^{\xi}(X)=\Omega_{n}^{\xi}\oplus\tilde{\Omega}_{n}^{\xi}(X)\, ,
\end{equation}
which separates the charges of branes mapped to a point, and thus present in any background, from the reduced part, which depends on the topology of $X$.

Using all of this, we can split the computation of the bordism group of a product of a manifold with a sphere in five steps, each using one of the facts above
\begin{equation}
\label{eq:bordism.product.proof}
\begin{split}
\Omega_{n}^{\xi}(X\times S^k) & =    \Omega_{n}^{\xi} \oplus \tilde\Omega_{n}^{\xi} (X\times S^k)\\
 & =   \Omega_{n}^{\xi} \oplus \tilde\Omega_{n+1}^{\xi} (\Sigma(X\times S^k))\\
  & =   \Omega_{n}^{\xi} \oplus \tilde\Omega_{n+1}^{\xi} (\Sigma X)\oplus \tilde\Omega_{n+1}^{\xi} (\Sigma S^k)\oplus \tilde\Omega_{n+1}^{\xi} (\Sigma(X\wedge S^k))\\
  & = 
   \Omega_{n}^{\xi} \oplus \tilde\Omega_{n}^{\xi} (X)\oplus \Omega_{n-k}^{\xi}\oplus \tilde\Omega_{n-k}^{\xi} (X)\\
  & = 
   \Omega_{n}^{\xi}(X)\oplus \Omega_{n-k}^{\xi}(X)\, .
\end{split}
\end{equation}
The first line is simply \eqref{eq:splitting}, the second uses \eqref{eq:suspension.bordism}, the third combines \eqref{eq:suspension.decomposition} with \eqref{eq:wedge.bordism}, the fourth uses \eqref{eq:suspension.bordism} repeatedly in addition to ${\Sigma S^k=S^{k+1}}$ and ${\Sigma^k X= S^k\wedge X}$, and the last line applies \eqref{eq:splitting} twice. This provides \eqref{eq:bordism.product}, and focusing on $X=S^2$ and $k=d-2$ we get \eqref{eq:lattice.bh}. Let us end by remarking that bordism groups $\Omega_n^\xi(X)$ vanish for $n<0$, as there are no manifolds with dimension $n<0$. This is used in section \ref{sec:bordism} when it is stated that the charge that correspond to wrapping the horizon has no counterpart in the Horowitz--Polchinski/free string side.

\bibliography{refs.bib}
\bibliographystyle{JHEP}
\end{document}